\documentclass[10pt,a4paper]{article}
\usepackage[T1]{fontenc}
\usepackage[utf8]{inputenc}
\usepackage{authblk}

\title{Asymptotic approximations for Bloch waves and topological mode steering in a planar array of Neumann scatterers}
\date{\vspace{-5ex}}
\author[1]{\small Richard Wiltshaw }
\author[1,2,3]{\small Richard V. Craster}
\author[1,4]{\small Mehul P. Makwana}
\affil[1]{\footnotesize Department of Mathematics, Imperial College London, London, SW7 2AZ, UK}
\affil[2]{\footnotesize Department of Mechanical Engineering, Imperial College London, London SW7 2AZ, UK}
\affil[3]{\footnotesize UMI 2004 Abraham de Moivre-CNRS, Imperial College London, London SW7 2AZ, UK}
\affil[4]{\footnotesize Multiwave Technologies AG, 3 Chemin du Pr\^{e} Fleuri, 1228, Geneva, Switzerland}

\usepackage{lineno}

\usepackage{arydshln}
\usepackage{psfrag}
\usepackage{multirow}
\usepackage{url}
\usepackage{tikz}
\usepackage[symbol]{footmisc}
\usepackage{graphicx}

\usepackage[ruled,vlined]{algorithm2e}

\usepackage{float}
\usepackage{mathrsfs,amsmath,tensor}
\usepackage{wrapfig}
\usepackage{cases, mathtools}
\usepackage{amssymb, amsbsy} 
\usepackage[labelfont=bf]{caption}
\usepackage{subcaption}
\usepackage{mathrsfs,amsmath}
\usepackage{listings}
\usepackage{bm}

\definecolor{lightblue}{RGB}{6,69,173}
\definecolor{myRED}{rgb}{1,0,0.0745}
\definecolor{myBLUE}{rgb}{0.0039,0.3961,1}

\usepackage{hyperref}
\usepackage{xcolor}
\hypersetup{
    colorlinks,
    linkcolor={lightblue},
    citecolor={lightblue},
    urlcolor={lightblue}
}
\usepackage[normalem]{ulem}

\usetikzlibrary{decorations.pathmorphing}
\usetikzlibrary{decorations.markings}
\usetikzlibrary{decorations.text}
\usetikzlibrary{matrix}
\usetikzlibrary{angles, quotes}
\usetikzlibrary{calc}

\newcommand*\circled[1]{\tikz[baseline=(char.base)]{
            \node[shape=circle,draw,inner sep=2pt] (char) {#1};}}

\def\XXint#1#2#3{{\setbox0=\hbox{$#1{#2#3}{\int}$}
     \vcenter{\hbox{$#2#3$}}\kern-.5\wd0}}

\newcommand{\beq}{\begin{equation}}
\newcommand{\eeq}{\end{equation}}
\newcommand{\ba}{\begin{eqnarray}}
\newcommand{\ea}{\end{eqnarray}}

\usepackage{color}
 
\definecolor{codegreen}{rgb}{0,0.6,0}
\definecolor{codegray}{rgb}{0.5,0.5,0.5}
\definecolor{codepurple}{rgb}{0.58,0,0.82}
\definecolor{backcolour}{rgb}{0.95,0.95,0.92}

\usetikzlibrary{quotes,angles}
\usetikzlibrary{shapes.geometric}
\usepackage{rotating}

  \newcommand{\arcarrow}[8]{%
  \pgfmathsetmacro{\rin}{#1}
  \pgfmathsetmacro{\rmid}{#2}
  \pgfmathsetmacro{\rout}{#3}
  \pgfmathsetmacro{\astart}{#4}
  \pgfmathsetmacro{\aend}{#5}
  \pgfmathsetmacro{\atip}{#6}
  \fill[#7] (\astart:\rin) arc (\astart:\aend:\rin)
       -- (\aend+\atip:\rmid) -- (\aend:\rout) arc (\aend:\astart:\rout)
       -- (\astart+\atip:\rmid) -- cycle;
  \path[font = \sffamily, decoration = {text along path, text = {|\mytextstyle|#8},
    text align = {align = center}, raise = -0.5ex}, decorate]
    (\astart+\atip:\rmid) arc (\astart+\atip:\aend+\atip:\rmid);
}

\begin{document}
\maketitle

\begin{abstract}
We study the canonical problem of wave scattering by periodic arrays, either of infinite or finite extent, of Neumann scatterers in the plane; the characteristic lengthscale of the  scatterers is  considered small relative to the lattice period. We utilise the method of matched asymptotic expansions, together with Fourier series representations, to create an efficient and accurate  numerical approach for finding the  
dispersion curves associated with Floquet-Bloch waves through an infinite array of scatterers. The approach lends itself to direct scattering problems for finite arrays and we illustrate the flexibility of these asymptotic representations on topical examples from topological wave physics.
\end{abstract}

\section{Introduction}
\label{sec:intro}
A fundamental understanding of wave propagation though periodic media underpins several areas of modern wave physics particularly photonic, and phononic, crystal devices \cite{zolla05a,joannopoulos08a} and topological photonics \cite{ozawa_topological_2019} such as in valleytronics \cite{behnia_polarized_2012} - the latter relying upon the detailed orientations of multiple inclusions within the cell that is repeated.
 Although the precise setting varies between electromagnetism or acoustics, many of these periodic problems reduce to the study of the wave equation, and in the frequency domain this becomes the Helmholtz equation with periodic arrangements of inclusions. The essential computation becomes that of dispersion curves that characterise essential details of the wave spectrum such as band-gaps of forbidden frequencies, flat-bands for slow-light or slow-sound, or Dirac points exhibiting locally dispersionless waves. The overwhelming approach in engineering and physics is to compute these curves with finite elements such as the commercial package Comsol \cite{comsol}, although \textcolor{black}{there are} numerous numerical  alternatives such as the plane wave expansion method \cite{johnson01a} \textcolor{black}{that} are also highly effective. However, such numerical methods can become a distraction particularly when dealing with topological effects, where it is the geometrical arrangement of the scatterers that matters, whereby faster or more flexible simulation methods are valuable for optimisation. For flexural waves in elastic plates (that are modelled using the Kirchhoff-Love equations \cite{graff1975wave}, fourth-order partial differential equations and, unlike for the Helmholtz equation here, have non-singular Green's function \cite{evans07a}) 
 very rapid numerical methods for dispersion curve evaluation \cite{xiao12a} are created that are  well suited to studies of topological media \cite{torrent13a,makwana18a, makwana18b, makwana19b, makwana19a, tang19, proctor19}. Our aim here is to extend this rapid solution methodology to the Helmholtz system, with its singular Green's function, by using matched asymptotic expansions to build in the presence of the small Neumann inclusions so we again arrive at a eigenvalue problem. This setting also enables rapid scattering simulations for finite crystals as an extension of \textcolor{black}{Foldy's classical method} \cite{martin2006multiple}.    

Matched asymptotic expansions are the natural mathematical language in which to  couch wave scattering problems involving a small parameter; the technique in the context of waves is neatly summarised in \cite{crighton1992matched}, more extensively in \cite{varadan1986}, and in a more modern context in \cite{martin2006multiple}. The aim is to take advantage of the small parameter, the ratio of, say, defect size to wavelength or other natural lengthscale, and then create an inner problem valid in the neighbourhood of the scatterer that is matched to an outer problem; these inner and outer problems being, hopefully, relatively straightforward to determine, such that rapid, accurate, and insightful solutions can follow. There are two limiting situations to consider, Dirichlet or Neumann (sound-soft or sound-hard) inclusions and, for periodic media, these were considered by McIver and co-workers \cite{krynkin09a,mciver2007approximations} with the inner following from Laplacian or Poisson equations and complex variable methods; the outer constructed using a doubly-periodic Green's function based around multipole methods and Bessel functions. Although effective in generating limited dispersion relations these lack the flexibility to easily treat multiple inclusions within a cell or to be extended to scattering by finite arrays.  As outlined for the Dirichlet case in \cite{schnitzer2017bloch} modifying the outer solution to one based around a conditionally convergent Fourier series representation of the Green's function, and subsequent manipulation, yields  a \textcolor{black}{generalised} eigenvalue problem; this is the natural way to proceed, avoids { \color{black} convergence acceleration for lattice sums of the Helmholtz equation \cite{linton2010lattice} and the use of Graff's addition theorem \cite{mciver2007approximations,abramowitz_handbook_1964} entirely}, and the matching between the inner and outer problems tie together very neatly. The Dirichlet problem only requires a monopole source, at leading order, whereas the weak scattering by Neumann inclusions requires further analysis including additional dipole source terms, see \cite{martin2006multiple,martinscattering}, and this is the case treated here.

Importantly, for practical purposes a plane wave expansion approach, modelled around that used in flexural waves \cite{xiao12a,makwana18a,torrent13a},
leads to a highly effective semi-analytical numerical method posed in reciprocal, i.e. Fourier, space for  extracting dispersion curves; herein we require asymptotic matching to be explicitly built into that formulation. Such matching removes any singularities observed within the wavefield from the numerics, subsequently we are not constrained by any convergence issues created by  singularities. Another practical benefit is that this leads naturally into a Foldy-like approach for scattering.
Foldy's approach \cite{foldy1945multiple} was initially derived for isotropic scatterers, an implementation via matching is in \cite{schnitzer2017bloch}. For Neumann scatterers, additional information about the gradients of the scattered field are required to close the system; this extension to account for anisotropy was introduced by Martin \cite{martin2006multiple,martinscattering} and is the \textcolor{black}{generalised} or extended Foldy approach.

 These fast Foldy-like schemes complement the finite element schemes typically used in the physics and engineering literature for scattering calculations dependent upon some incident field, additionally it can also be used to generate eigensolutions by setting the incident field to zero and analysing the homogeneous Foldy problem whose solutions then cleanly identify the modes that form the scattered field.  

Whilst the primary thrust of this article focuses on the method of matched asymptotics to provide a general and systematic approach to handle scattering by small Neumann defects, there are other semi-analytical alternatives. In the present context, for the special case of circular inclusions, typically a single inclusion that resides within a cell that repeats; multipole expansions \cite{nicorovici95b} provide a route to  dispersion relation calculations, requiring knowledge of the convergence of the various lattice sums that appear in the generalised Rayleigh identity, 
and have been approximated in the dilute limit for doubly-periodic media  \cite{zalipaev2002elastic,movchan02a}. However, this approach becomes cumbersome for multiple inclusions within a cell and the settings in, say, topological photonics that have delicate dependence upon the inclusions, and their relative orientation within each periodic cell. Multipole methods are also feasible and popular for scattering problems involving circular cylinders, \textcolor{black}{see for instance } \cite{linton90a}, yielding systems of linear algebraic equations as an extension of Foldy's approach; it is nonetheless instructive to arrive at the system for small scatterers from matching and allow for non-circular scatterers.

The outline of the article is such that we first mathematically model the problem in section \ref{sec:Formulation}. Once the inclusions have been approximated by a series of monopoles and dipoles in the outer region, we are in a position to perform a traditional matching procedure about an inclusion where the singular Green's function is matched to the solution in the inner region - the solution  satisfies the Neumann condition exactly. Section \ref{sec:solnrecpsp} utilises a divergent Fourier series to represent the solution - numerically we truncate the divergent sum where the matched asymptotic analysis allows us to determine the error in doing so, this allows \textcolor{black}{a generalised} eigenvalue problem to be written down determining the dispersion relation detailing the dispersive properties of the constructed media. 

Once these properties are known, we demonstrate these effects in physical space, this is examined in section \ref{Sec:solnInPhysGen} where scattering coefficients and matrices are determined for the \textcolor{black}{generalised} Foldy approach, using the inner solution in a similar vein to \cite{schnitzer2017bloch}. {\color{black}Usually the forced problem is considered where the scattered field is determined due to interactions between some incident source and a structure. 

In Section \ref{sec:solnInPhysHomog} we take Foldy's method one step further, introducing the unforced Foldy problem by setting the incident field to zero. The result is a homogeneous system whose non-trivial solution extracts dormant modes residing within the structure (at a given frequency) awaiting excitation}. 

Our scheme determining the dispersion relation is cross-checked against full finite element computations in section \ref{sec:results}. Finally, in section \ref{sec:topological} we demonstrate the utility and efficiency of our succinct formulae by applying them to a few topical examples in topological physics.

\section{Formulation}
\label{sec:Formulation}
Assuming harmonic waves, with $\exp(-i\Omega t)$ dependence being understood (and suppressed henceforth) where $\Omega$ is the frequency, we consider the 
 dimensionless  Helmholtz equation 
\begin{equation}
    \left( \nabla^{2} + \Omega^{2} \right) \phi (\textbf{x}) = 0 \label{eqn::Helmholtz}
\end{equation}
for a two dimensional wavefield $\phi(\textbf{x})$. We \textcolor{black}{consider a finite collection of $N$ cells each containing $P$ inclusions, for a total of $NP$ inclusions enumerated by the introduction of $I = 1, \ldots, N$ and $J=1, \ldots, P$; quantities belonging to the $IJ$th inclusion are denoted by subscript $IJ$.} We denote $L$ as some characteristic length scale that we \textcolor{black}{choose to } base upon the lattice period (defined later in (\ref{eqn::Ldefn})), our attention is restricted to the case of small circular inclusions, whose radius \textcolor{black}{$\epsilon_{IJ} \ll L = \mathcal{O}(1)$.  We denote the center of each inclusion by $\textbf{x} = \textbf{X}_{IJ}$ and apply the Neumann condition on the boundary of each inclusion, that is we set}
\begin{equation}
\frac{\partial \phi}{\partial r_{IJ}} \Big|_{r_{IJ} = \epsilon_{IJ}} = 0, \quad \mbox{where $r_{IJ} = |\textbf{r}_{IJ}| = |\textbf{x} - \textbf{X}_{IJ}|$}. \label{eqn::Inclusions}
\end{equation}
 
The earlier analyses for small inclusions, \cite{crighton1992matched,martin2006multiple,mciver2007approximations},  \textcolor{black}{indicates} that the $IJ$th Neumann inclusion, being weak, should act to perturb the wavefield at \textcolor{black}{relative} order $\epsilon_{IJ}^{2}$. We further deduce that the \textcolor{black}{$IJ$th inclusion will} act as a combination of line monopoles and dipoles, \textcolor{black}{
 placed at $\textbf{X}_{IJ}$}, with \textcolor{black}{coefficients} $a_{IJ}, \textbf{b}_{IJ}$ to be determined; therefore, (\ref{eqn::Helmholtz}) subject to (\ref{eqn::Inclusions}) is approximated by
\begin{equation}
\left( \nabla^{2} + \Omega^{2} \right) \phi = 4i \sum_{I=1}^{\textcolor{black}{N}} \sum_{J=1}^{P} \epsilon^{2}_{IJ} \left\lbrace a_{IJ} \delta(\textbf{x} - \textbf{X}_{IJ}) - \textbf{b}_{IJ} \cdot \nabla \delta( \textbf{x} - \textbf{X}_{IJ} ) \right\rbrace  \label{HelmholtzOuterEase}
\end{equation}
 as the \textcolor{black}{leading outer governing equation}, where $\delta(\textbf{x})$ denotes the Dirac delta function; in the limit as the inclusions get small they are idealised as points that scatter with monopolar and dipolar contributions. \textcolor{black}{In section \ref{Sec:solnInPhysGen} we solve for scattering by finite arrays using a generalisation of Foldy's method conventionally used for Dirichlet (sound-soft) scatterers, but here we consider Neumann scatterers.} 
 
\begin{figure}[h!]
\centering
\begin{tikzpicture}[scale=0.45, transform shape]
\draw (-4.625, -3.5) node[inner sep=0] {\includegraphics[scale=0.5]{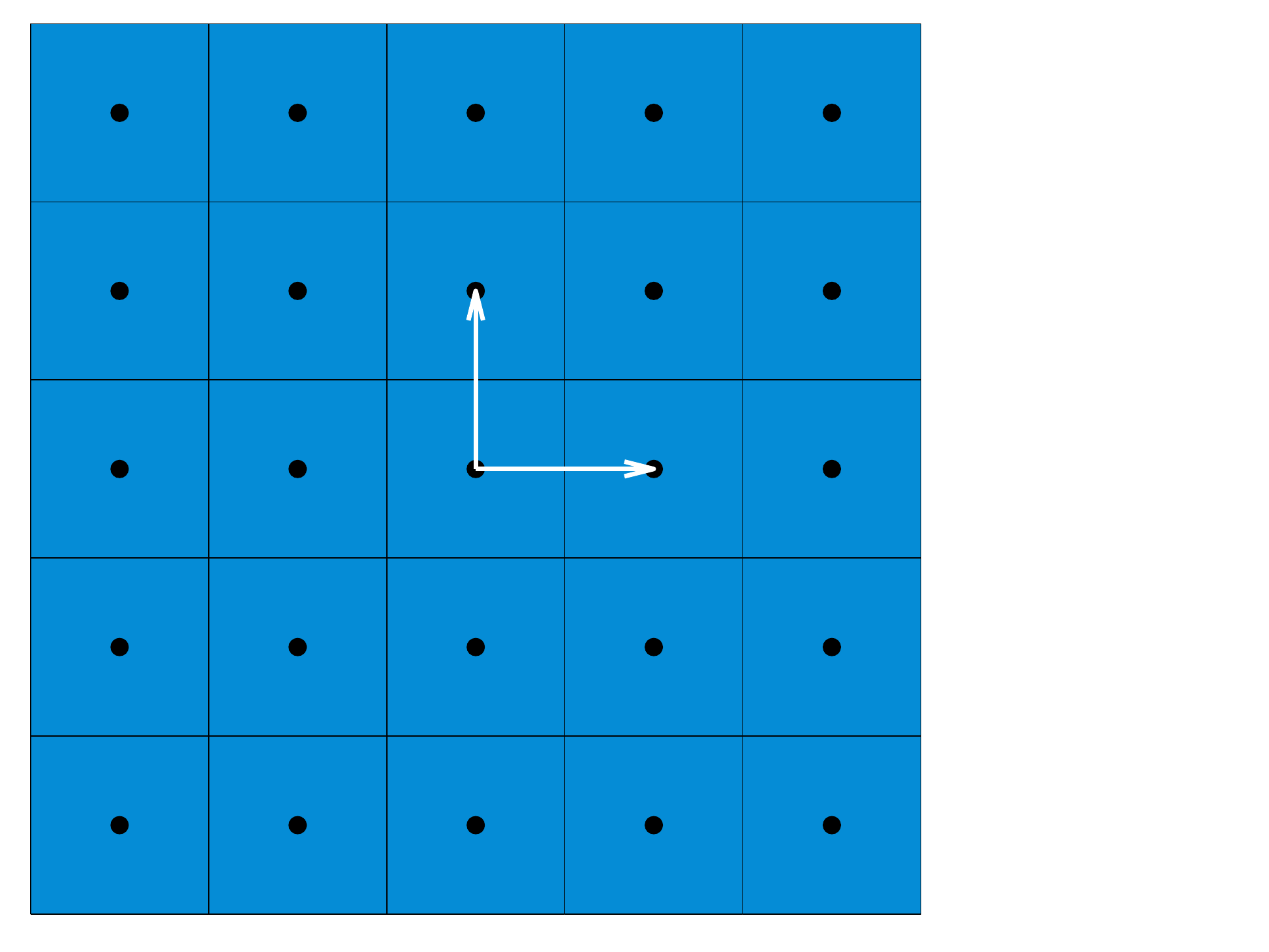}};
\node[below, scale=2,white] at (-3.85,-3.1) {$\displaystyle  \boldsymbol{\alpha}_{1}$}; 
\node[below, scale=2,white] at (-5.9,-3.1+1.9) {$\displaystyle  \boldsymbol{\alpha}_{2}$}; 
\node[below, scale=1.75,] at (-5.9+4.3,-3.2) {$\displaystyle  \ldots$}; 
\node[below, scale=1.75,] at (-5.9-4.3,-3.2) {$\displaystyle  \ldots$}; 
\node[below, scale=1.75,] at (-5.9,-3.2+5.0) {$\displaystyle  \vdots$}; 
\node[below, scale=1.75,] at (-5.9,-3.2-4.3+0.35) {$\displaystyle  \vdots$}; 
\node[below, scale=1.75,] at (-5.9,-3.2-4.3-1) {$\displaystyle  (i)$}; 
\draw (2.7, -1) node[inner sep=0] {\includegraphics[scale=0.25]{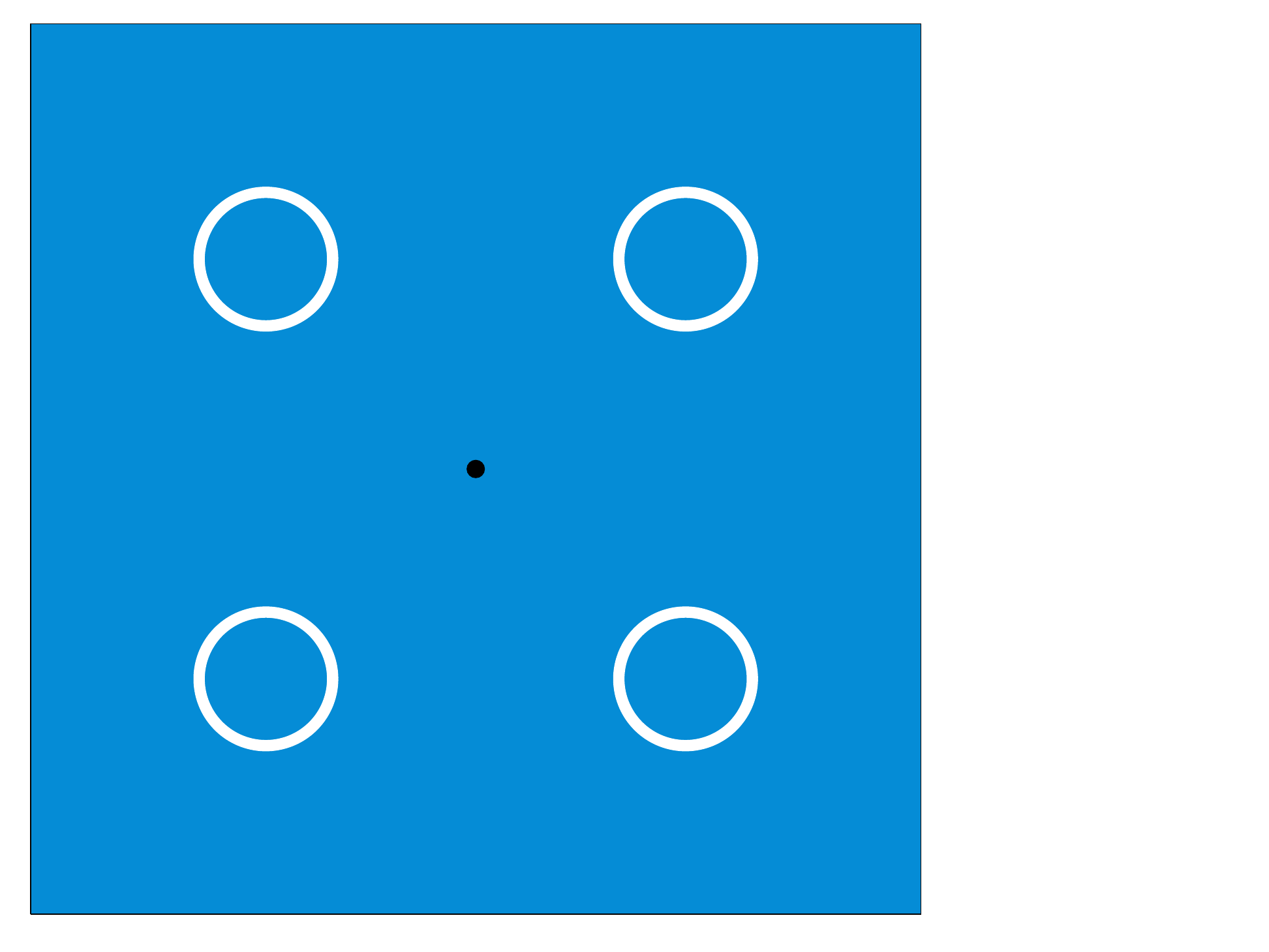}};
\node[below, scale=1.95] at (2.0,-2.75) {$\displaystyle  (ii)$};
\begin{scope}[shift = {(16,0)}]
\draw (-4.625, -3.5) node[inner sep=0] {\includegraphics[scale=0.5]{simple-eps-converted-to.pdf}};
\node[below, scale=2,white] at (-3.85,-3.1) {$\displaystyle  \boldsymbol{\beta}_{1}$}; 
\node[below, scale=2,white] at (-5.9,-3.1+2.1) {$\displaystyle  \boldsymbol{\beta}_{2}$}; 
\node[below, scale=1.75,] at (-5.9+4.3,-3.2) {$\displaystyle  \ldots$}; 
\node[below, scale=1.75,] at (-5.9-4.3,-3.2) {$\displaystyle  \ldots$}; 
\node[below, scale=1.75,] at (-5.9,-3.2+5.0) {$\displaystyle  \vdots$}; 
\node[below, scale=1.75,] at (-5.9,-3.2-4.3+0.35) {$\displaystyle  \vdots$}; 
\node[below, scale=1.75,] at (-5.9,-3.2-4.3-1) {$\displaystyle  (iv)$}; 
\end{scope}
\begin{scope}[transform canvas={scale=1.1}, shift={(1.4+0.5,-2)}]
\node[regular polygon, regular polygon sides=4, draw, inner sep=0.5*6.28*10.0 pt,rotate=0] at (0 pt,-6.28*10.0*1.5 pt) {};
\draw[line width=0.5mm,gray,-] (0pt,-6.28*10.0*1.5 pt) -- (0+ 0.5*6.28*10*1.41 pt,-6.28*10.0*1.5 pt);
\draw[line width=0.5mm,gray,-] (0+ 0.5*6.28*10*1.41 pt,-6.28*10.0*1.5 pt) -- (0+ 0.5*6.28*10*1.41  pt, 0.5*6.28*10*1.41 -6.28*10.0*1.5 pt);
\draw[line width=0.5mm,gray,-] (0+ 0.5*6.28*10*1.41  pt, 0.5*6.28*10*1.41 -6.28*10.0*1.5 pt) -- (0pt,-6.28*10.0*1.5 pt);
\node[below,left,scale=1.75] at (0pt,-6.28*10.0*1.5 pt) {$\displaystyle  \Gamma$}; 
\node[below,right,scale=1.75] at (0+ 0.5*6.28*10*1.41 pt,-6.28*10.0*1.5 pt) {$\displaystyle  X$};
\node[above,right,scale=1.75] at (0+ 0.5*6.28*10*1.41  pt, 0.5*6.28*10*1.41 -6.28*10.0*1.5 pt) {$\displaystyle M$};
\node[regular polygon, circle, draw, inner sep=1.25pt,rotate=0,line width=0.5mm,shading=fill,outer color=gray,gray] at (0pt,-6.28*10.0*1.5 pt)  {};
\node[regular polygon, circle, draw, inner sep=1.25pt,rotate=0,line width=0.5mm,shading=fill,outer color=gray,gray] at (0+ 0.5*6.28*10*1.41 pt,-6.28*10.0*1.5 pt)  {};
\node[regular polygon, circle, draw, inner sep=1.25pt,rotate=0,line width=0.5mm,shading=fill,outer color=gray,gray] at (0+ 0.5*6.28*10*1.41  pt, 0.5*6.28*10*1.41 -6.28*10.0*1.5 pt)  {};
\node[below, scale=1.75] at (0pt,-0.5*6.28*10*1.41 -6.28*10.0*1.5 pt) {$\displaystyle  (iii)$}; 
\end{scope}
\end{tikzpicture}

\caption{A doubly periodic crystalline structure of Neumann inclusions, with  primitive cells represented by black squares with centroids ($\bullet$). 
$(i)$ The crystal in physical space, spanned by \textcolor{black}{$\boldsymbol{\alpha}_{1}$ and  $\boldsymbol{\alpha}_{2}$}.
$(ii)$ An arrangement of four Neumann inclusions (white circles) per primitive cell. 
$(iii)$ The first Brillouin zone (a primitive cell in reciprocal space) and irreducible Brillouin zone ($\Gamma X M \Gamma$), in reciprocal space.
$(iv)$ The crystal in reciprocal space, spanned by \textcolor{black}{ $\boldsymbol{\beta}_{ 1}$ and $\boldsymbol{\beta}_{ 2}$.} } 
\label{fig:brillouinFormation}
\end{figure}

 \textcolor{black}{We also consider infinite arrays (with no forcing) in section \ref{sec:solnrecpsp}, where  \textcolor{black}{we assume a periodic arrangement cells each of which contains $P$ inclusions; this allows} us to concentrate upon a single cell, with Bloch-Floquet conditions, and construct dispersion relations relating phase-shift across the cell to frequency. These periodic eigensolutions act to inform choices of frequency for the excitation of finite arrays.}
 When \textcolor{black} {extending \eqref{HelmholtzOuterEase} to} an infinite, doubly periodic, arrangement of Neumann inclusions we \textcolor{black}{then} consider primitive cells in physical space that are spanned by lattice vectors \textcolor{black}{$\boldsymbol{\alpha}_{1}$, $\boldsymbol{\alpha}_{2}$}, as in Fig.  \ref{fig:brillouinFormation}. {\color{black} The centroids of each cell form a two dimensional  Bravais lattice exemplified with physical position vector }
\begin{equation}
\textbf{R} = n \boldsymbol{\alpha}_{1} + m \boldsymbol{\alpha}_{2} \quad \mbox{for some $n,m \in \mathbb{Z} $.} \label{R=nmeqn}
\end{equation}
 
  \textcolor{black}{The solution is found in a single cell so}
 the double sum within (\ref{HelmholtzOuterEase}) is 
 \textcolor{black}{no longer required and we } consider the $I$th primitive cell; \textcolor{black}{ without loss of generality, we set $I=1$ and refer to this cell as the fundamental cell.}  
 
 
The aforementioned characteristic length $L$ is defined to be that of the lattice period
\begin{equation}
    L = \min(|\boldsymbol{\alpha}_{1}|, |\boldsymbol{\alpha}_{2}| ); \label{eqn::Ldefn}  
\end{equation}
\textcolor{black}{ implicitly this assumes that the length of one lattice vector is order one and the other is of the same order or larger.}
The periodic nature of the material allows us to utilise Bloch's theorem \cite{kittel1996introduction}, \textcolor{black}{wherein, applying the quasi-periodic Bloch-Floquet conditions we require}
\begin{equation}
\phi(\textbf{x}) = \Phi( \textbf{x} ) \exp (i \boldsymbol{\kappa} \cdot \textbf{x}), \quad {\rm and} \quad \Phi(\textbf{x} + \textbf{R}) = \Phi(\textbf{x}). \label{BlochPhaseShift}
\end{equation}
Here $\boldsymbol{\kappa}$ denotes the Bloch-wave vector and $\Phi$ inherits the  periodicity of the lattice. Subsequently, we express $\Phi(\textbf{x})$ by means of a Fourier series \textcolor{black}{and seek solutions of the form} 
\begin{equation}
\phi(\textbf{x}) = \sum_{\textbf{G}} \Phi_{\textbf{G}} \exp (i \textbf{K}_{\textbf{G}} \cdot \textbf{x}) \quad \mbox{where $\textbf{K}_{\textbf{G}} = \boldsymbol{\kappa} + \textbf{G}$}. \label{eqn::BlochExp}
\end{equation}
\textcolor{black}{Here $\textbf G$ is the reciprocal lattice vector and }
$\Phi_{\textbf{G}}$ denotes the Fourier coefficients representing the amplitude of the $\textbf{G}$th excited mode characterising incoming and outgoing Bloch waves, propagating throughout the primitive cells. The reciprocal space is spanned by the lattice vectors \textcolor{black}{ $\boldsymbol{\beta}_{1}$ and $\boldsymbol{\beta}_{2}$} satisfying  the following orthogonality conditions with the physical lattice vectors 
\begin{equation}
    \boldsymbol{\alpha}_{i} \cdot \boldsymbol{\beta}_{j} = 2 \pi \widetilde{\delta}_{ij},
\end{equation}
where $\widetilde{\delta}_{ij}$ denotes the Kronecker delta function. The  reciprocal lattice vector is explicitly given by
\begin{equation}
    \textbf{G} = n \boldsymbol{\beta}_{1} + m \boldsymbol{\beta}_{2}, \quad \mbox{for some $n,m \in \mathbb{Z} $}. \label{egn::G}
\end{equation}
Due to the underlying periodicity  we consider the solution throughout the reduced $\boldsymbol{\kappa}$ space known as the first Brillouin zone, symmetry  allows us to reduce this further to the irreducible Brillouin zone as detailed in \cite{brillouin1953wave}.

{\color{black}
Once the $\Omega (\boldsymbol{\kappa})$ dependent solution for $\phi$ (and hence coefficients $a_{1J}$, $\textbf{b}_{1J}$) are known in the fundamental cell, a phase shift using \eqref{BlochPhaseShift}  deduces $\phi$ (and $a_{IJ} = a_{1J} \exp(i \boldsymbol{\kappa} \cdot (\textbf{X}_{IJ} - \textbf{X}_{1J} ))$ similarly for $\textbf{b}_{IJ}$) from cell-to-cell. Note equations (\ref{R=nmeqn}) - (\ref{egn::G}) are only valid when the media is periodic and of infinite extent; these assumptions form the basis of the analysis in section \ref{sec:solnrecpsp}. 
}

\section{A scheme in Fourier space} \label{sec:solnrecpsp}

\textcolor{black}{We begin by considering the periodic infinite array problem, considering the fundamental cell and constructing a numerical method from the asymptotics that generates dispersion relations between the phase-shift across the cell and frequency.}

The solution in reciprocal space utilises Fourier series expansions {\color{black} in} \eqref{HelmholtzOuterEase}. However the solution is singular at the location of each of the line monopole or dipole source terms, therefore the series expansions must be divergent within its inner limits. Unfortunately the coefficients of the monopole and dipole terms are determined \textit{a posteriori} and require inspection within the inner limit, where we expect the series expansions to diverge. \textcolor{black}{ Singularities are removed by considering the truncation error in truncating the divergent series; the singularities present within the} sum coincide with the singularities obtained from the outer limit of the inner solution. 


\textcolor{black}{We consider solutions in the fundamental cell - for clarity of exposition we take $P=1$ although the analysis is easily generalised for multiple inclusions per cell. These considerations mean any summations over $IJ=11$ are dropped and we drop subscripts $11$ throughout the remainder of this section.}

 Applying the Bloch expansion  \eqref{eqn::BlochExp}  to 
 \eqref{HelmholtzOuterEase}  we obtain 
\begin{equation}
\Phi_{\textbf{G}} = \frac{4}{i \mathscr{A}} \frac{\epsilon^{2} \left\lbrace a - i \textbf{b} \cdot \textbf{K}_{\textbf{G}} \right\rbrace \exp \left( - i  \textbf{K}_{\textbf{G}} \cdot \textbf{X} \right)}{\left( \textbf{K}_{\textbf{G}} \cdot \textbf{K}_{\textbf{G}} - \Omega^{2} \right)}, \label{G<Rcont} 
\end{equation}
where $\mathscr{A}$ denotes the area of the primitive cell. 
Throughout the text, $\textbf{b}$ and the Bloch wave vector $\boldsymbol{\kappa}$ are  two dimensional vectors  with components in $\textbf{e}_{x}$ and $\textbf{e}_{y}$ directions.

The series representation of $\phi$ is subsequently 
\begin{equation}
    \phi = \frac{4}{i \mathscr{A}} \sum_{\textbf{G}} \frac{\epsilon^{2} \left\lbrace a - i \textbf{b} \cdot \textbf{K}_{\textbf{G}} \right\rbrace \exp \left[  i  \textbf{K}_{\textbf{G}} \cdot \textbf{r} \right]}{\left( \textbf{K}_{\textbf{G}} \cdot \textbf{K}_{\textbf{G}} - \Omega^{2} \right)}. \label{eqn::series}
\end{equation}
The isotropic (monopolar) term in \eqref{eqn::series} is, other than its relative order, identical to that in \cite{schnitzer2017bloch}. {\color{black} This series is conditionally convergent when $r \neq 0$ and diverges  as $r \to 0$. The monopolar and dipolar terms introduce logarithmic and algebraic singularities respectively, observed when considering the outer limit of the inner solution (derived in \ref{Appendix::matching})}  
\begin{equation}
\begin{split}
\phi {\color{black} \sim} \frac{4 i}{\pi} \frac{a}{\Omega^{2}} \left[ 1 - \frac{\Omega^{2} r^{2}}{4} \right] + \frac{4}{i \pi} \frac{ \textbf{b} \cdot \hat{\textbf{r}} }{\Omega}  \left( \frac{r \Omega}{2} - \frac{\Omega^{3} r^{3}}{16} \right)  + \\
+ \epsilon^{2} \left\lbrace \frac{2ia}{\pi} \left[ \log \frac{r}{\epsilon} + \frac{3}{4} \right] + \frac{\textbf{b} \cdot \hat{\textbf{r}}}{i \pi} \left(\frac{2}{r} - \Omega^{2} r \left[ \log \frac{r}{\epsilon} - \frac{7}{4} \right] \right) \right\rbrace,
\end{split} \label{NAP::HelmholtzInnerLim} \quad \mbox{{\color{black} $\quad$ as $r \to 0$}}.
\end{equation}
 Here $\hat{\textbf{r}} = {\textbf{r}}/{|\textbf{r}|}$. Equation \eqref{NAP::HelmholtzInnerLim} implies the series given in \eqref{eqn::series} diverges as ${1}/{r}$ as $r \to 0$. \textcolor{black}{In comparison the Dirichlet case \cite{schnitzer2017bloch} only contains logarithmic singularities due to $a$, and the presence of $\textbf{b}$ here provides algebraic singularities and $2P$ more unknowns; therefore, $2P$ more equations are required to close the system. We require the consideration of both $\phi$ (Dirichlet and Neumann) and $\nabla \phi$ (Neumann) as $r \to 0$ to solve the Neumann case}. The gradient of (\ref{NAP::HelmholtzInnerLim}) implies that  $\nabla$(\ref{eqn::series}) diverges as ${1}/{r^{2}}$ as $r \to 0$.
 This poses an intriguing numerical issue revolving around performing a series expansion to evaluate all unknowns of a system, whilst the series {\color{black}itself \eqref{eqn::series}} does not converge.

\subsection{ \textcolor{black}{Utilizing the truncation error as $r \to 0$}}
The solution in the outer region is given by the series expansion (\ref{eqn::series}), {\color{black} whose divergent behaviour is accounted for by splitting the series in two, inside and outside a truncation radius $R'$; the asymptotic regime is identical to \cite{schnitzer2017bloch}, i.e. $1 \ll R' \ll {1}/{r}$ as $r \to 0$. We consider 


%

\begin{equation}
\lim_{r \to 0} \phi = \lim_{r \to 0} \left\lbrace \overbrace{ \sum_{\substack{\textbf{G} \\ |\textbf{G}| < R'}}}^{\phi_{\mathrm{tr}}} + \overbrace{\sum_{\substack{\textbf{G} \\ |\textbf{G}| > R'}}}^{\phi_{\mathrm{res}}} \right\rbrace \Phi_{\textbf{G}} \exp(i \textbf{K}_{\textbf{G}} \cdot \textbf{x}).  \label{HelmCondConvSer}
\end{equation}
Here:
\begin{itemize}
    \item $\phi_{\mathrm{tr}}$ represents the truncated portion of the series expansion, where we take  $\Phi_{\textbf{G}}$ from \eqref{G<Rcont},
    \item $\phi_{\mathrm{res}}$ denotes the residual portion of the sum, i.e. the leftover piece beyond truncation. The truncation error $\phi - \phi_{\mathrm{tr}}$ is given by $\phi_{\mathrm{res}}$, whose approximation is determined in \ref{sec:Residual}.
\end{itemize}
Simple numerical analysis exists in convergent (finite) problems, where $\phi_{\mathrm{res}}$ tends to zero for increasing $R'$. In the case treated here, singularities exist as $r \to 0$ which complicates the analysis; here both $\phi$ and $\phi_{\mathrm{res}}$ are singular when $1 \ll R' \ll {1}/{r}$ as $r \to 0$.

In the limit as $r \to 0$ the left hand side of (\ref{HelmCondConvSer}) is given by the inner solution (\ref{NAP::HelmholtzInnerLim}), and is independent of $R'$. The right hand side is also independent of $R'$; the apparent 
dependence on $R'$ cancels when considering $\phi_{\mathrm{tr}} + \phi_{\mathrm{res}}$. $\phi_{\mathrm{res}}$ is asymptotically approximated and contains terms which are singular with respect to $R'$ and $r$, and the summation in $\phi_{\mathrm{tr}}$ contains $R'$ terms that perfectly cancel those in $\phi_{\mathrm{res}}$.

To
treat (\ref{HelmCondConvSer}) 
 we choose an arbitrary large value for $R'$, as ultimately we construct a numerical method that uses truncation, 
 and rearrange (\ref{HelmCondConvSer}) as 
\begin{equation}
\phi_{\mathrm{tr}} = \phi - \phi_{\mathrm{res}} \quad \mbox{as $r \to 0$ }. \label{eqn::residualError}
\end{equation}
The singularities with respect to $r$ in $\phi$ and   $\phi_{\mathrm{res}}$ cancel, leaving behind the truncation error dependent upon $R'$. 

The analysis allows us to create a generalised matrix eigenvalue problem, where the unknowns $\Phi_{|\textbf{G}|<R'}$, $a_{1J}$, $\textbf{b}_{1J}$ form the eigenvector and the frequency $\Omega^{2}$ is the eigenvalue. The matrices required depend explicitly on the Bloch wavevector $\boldsymbol{\kappa}$, and solving the generalised eigenvalue problem determines all the unknowns for any wavevector.
}

\textcolor{black}{\subsection{The generalised eigenvalue problem}}
The derivation of $\phi_{\mathrm{res}}$ is outlined in \ref{sec:Residual}, {\color{black}and we obtain} 
\begin{equation}
\begin{split}
& \phi_{\mathrm{res}} \sim \epsilon^{2} \frac{\exp \left( i  \boldsymbol{\kappa} \cdot \textbf{r} \right)}{i \pi} \Big\lbrace \frac{2 \textbf{b} \cdot \hat{\textbf{r}}}{r} J_{0}(R'r) + Ji_{0} (R'r) \Big[ i r (i \textbf{b} \Omega^{2} + 2 a \boldsymbol{\kappa}) \cdot \hat{\textbf{r}} - 2  a  \Big] - \\
& \quad - \frac{J_{1}(R'r)}{R'r} \Big[ 2 i \Big( \textbf{b} \cdot \boldsymbol{\kappa}\cos(2 \varphi - 2 \theta_{\kappa}) - (\textbf{b} \times \boldsymbol{\kappa})\cdot \textbf{e}_{z} \sin(2 \varphi - 2 \theta_{\kappa}) \Big) + \\
&\quad \quad \quad  \quad \quad \quad \quad \quad \quad  \quad \quad \quad \quad \quad \quad  \quad \quad \quad  + i r(i \textbf{b} \Omega^{2} + 2 a \boldsymbol{\kappa}) \cdot \hat{\textbf{r}} \Big] - \\
& \quad -  2 \kappa \Big[ \textbf{b} \cdot \boldsymbol{\kappa} \cos( 3 \theta_{\kappa} - 3 \varphi) + (\textbf{b} \times \boldsymbol{\kappa})\cdot \textbf{e}_{z} \sin(3 \theta_{\kappa} - 3 \varphi)  \Big] \frac{J_{2} (R'r)}{R'^{2} r}  \Big\rbrace 
\end{split} \label{NAP::G>Rcont}
\end{equation}
here $Ji_{0}$ refers to Van der Pol's Bessel-integral function of zero order\footnote{When constructing the eigensolution in the fundamental cell, in physical space, a useful identity is \begin{equation}
    Ji_{0}(x) = -\frac{x^{2}}{8} \,  \tensor[_2]{F}{_3}(1,1;2,2,2;-\frac{x^{2}}{2}) + \log \frac{x}{2} + \gamma_{E},
\end{equation} where $\tensor[_t]{F}{_u}(v_{1}, \ldots , v_{t}; w_{1} , \ldots, w_{u}; z)$ is the generalised hypergeometric function.} and $J_0, J_1, J_2$ are Bessel functions.
 The behaviour of \textcolor{black}{the Bessel functions  for small arguments are well known \cite{abramowitz1965handbook}, and Humbert \cite{humbert1933bessel} deduces}
\begin{equation}
    Ji_{0}(x) = Ci(x) - \log 2 = \log \frac{x}{2} + \gamma_{E} -\frac{x^{2}}{4} + \mathcal{O}(x^{4}) \quad \mbox{as $x \to 0$}, \label{eqn::Ji0small}
\end{equation}
\textcolor{black}{where the Euler-Mascheroni constant, denoted by $\gamma_{E}$, is present and 
h}ere $Ci(x)$ is the cosine integral function.

We return to  (\ref{eqn::residualError}) and use the inner limit of the outer 
 (\ref{NAP::HelmholtzInnerLim}) and the residual (\ref{NAP::G>Rcont}) to obtain 
\begin{equation}
 \lim_{r \to 0} \phi_{\mathrm{tr}} = \frac{4ia}{\pi \Omega^{2}} + \epsilon^{2} \left\lbrace \frac{2ai}{\pi} \left[ \log \frac{2}{\epsilon R'} + \frac{3}{4} - \gamma_{E} \right]  -  \frac{\boldsymbol{\kappa} \cdot \textbf{b}}{\pi} \right\rbrace +  \mathcal{O}(r). \label{HelmholtzEigeAsymp}
 \end{equation}

We require two further equations to close the system; \textcolor{black}{since we seek real eigenvalues, the matrices forming the generalised eigenvalue problem must be Hermitian. Therefore we must consider $\nabla \phi$ to close the problem, so we require}

\begin{equation}
    \nabla \phi_{\mathrm{tr}} = \nabla \phi - \nabla \phi_{\mathrm{res}} \quad \mbox{as $r \to 0$}.
\end{equation}
For multiple objects it is easiest to consider the gradients, in the above, in polar coordinates localized about the center of each inclusion; then transforming to a global Cartesian system we find the following equations \textcolor{black}{for the $\textbf{e}_{x}$ and $\textbf{e}_{y}$ components}:
\begin{equation}
\begin{split}
\textbf{e}_{x}: \quad \lim_{r \to 0} \sum_{G < R'} i K_{1 \, \textbf{G}}  \Phi_{\textbf{G}} \exp(i \textbf{K}_{\textbf{G}} \cdot \textbf{x}) & = \quad \quad \quad \quad \quad \quad \quad \quad \quad \quad \quad \quad \\
= \Big[ \frac{2}{i \pi} + i \frac{\epsilon^{2}}{\pi} \Omega^{2} \Big( \log \frac{2}{\epsilon R'} & - \frac{5}{4} - \gamma_{E} \Big)  + \frac{\epsilon^{2}}{2i \pi } R'^{2} \Big] b_{1} + \\
+ \epsilon^{2} \frac{a}{\pi} \kappa_{1}  & -  \frac{\epsilon^{2} i}{4 \pi} \Big[  \kappa_{1}  \textbf{b} \cdot \boldsymbol{\kappa} - \kappa_{2} (\textbf{b} \times \boldsymbol{\kappa} ) \cdot \textbf{e}_{z} \Big] {\color{black}+ \mathcal{O}(r)}, 
\end{split} \label{HelmholtzGradExEigeAsymp} 
\end{equation}

\begin{equation}
\begin{split}
\textbf{e}_{y}: \quad \lim_{r \to 0} \sum_{G < R'} i K_{2 \, \textbf{G}}  \Phi_{\textbf{G}} \exp(i \textbf{K}_{\textbf{G}} \cdot \textbf{x}) & = \quad \quad \quad \quad \quad \quad \quad \quad \quad \quad \quad \quad \\
= \Big[ \frac{2}{i \pi} + i \frac{\epsilon^{2}}{\pi} \Omega^{2} \Big( \log \frac{2}{\epsilon R'} & - \frac{5}{4} - \gamma_{E} \Big)+ 
\frac{\epsilon^{2}}{2i \pi } R'^{2} \Big] b_{2} + \\
+\epsilon^{2} \frac{a}{\pi} \kappa_{2} 
& -  \frac{\epsilon^{2} i}{4 \pi} \Big[ \kappa_{2} \textbf{b} \cdot \boldsymbol{\kappa} + \kappa_{1} (\textbf{b} \times \boldsymbol{\kappa} ) \cdot \textbf{e}_{z} \Big] {\color{black}+ \mathcal{O}(r)} .  
\end{split} \label{HelmholtzGradEyEigeAsymp} 
\end{equation}
The complete generalised eigenvalue problem is formed from equations $(\textbf{K}_{\textbf{G}} \cdot \textbf{K}_{\textbf{G}} - \Omega^{2}) \cdot (\ref{G<Rcont})$, $\Omega^{2} \cdot (\ref{HelmholtzEigeAsymp})$, (\ref{HelmholtzGradExEigeAsymp}) and (\ref{HelmholtzGradEyEigeAsymp}) {\color{black}as follows}
\begin{equation}
\left( \mathcal{A}\textcolor{black}{(\boldsymbol{\kappa})} - \Omega^{2} \mathcal{B }\textcolor{black}{(\boldsymbol{\kappa})}  \right) \boldsymbol{\Phi} = \textbf{0}. \label{HelmholtzHardScheme_1}
\end{equation}
The matrices $\mathcal{A}$, $\mathcal{B}$ are lengthy, and given in \ref{AllOfTheComponentsEigen}. The numerical scheme is efficient and dispersion curves are found by looping through each of the required values of $\boldsymbol{\kappa}$, to extract the frequencies-squared, $\Omega^2$, as the eigenvalues of \eqref{HelmholtzHardScheme_1}. {\color{black}The the eigenvector $\boldsymbol{\Phi}$ gives the monopolar and dipolar coefficents as well as the $\Phi_{\mathbf{G}}$ components. The solution $\phi$ in physical space, utilising (\ref{eqn::series}), is reconstructed from $\boldsymbol{\Phi}$; analytical expressions for $\nabla \phi$, as well as the flux, can be found -  some examples are given and computed from our schemes} in section \ref{sec:results}. 
\textcolor{black}{Although the analysis has been presented for a single object, the extension to consider multiple objects is simple provided that the inclusions are placed far enough apart such that the matching from \ref{Appendix::matching} remains valid. The components in (\ref{HelmholtzHardScheme_1}) are shown for general $P$ inclusions per cell in \ref{AllOfTheComponentsEigen}.}



\section{Generalised Foldy solution}
\label{Sec:solnInPhysGen}
\textcolor{black}{We now complement the dispersion curve analysis, of section \ref{sec:solnrecpsp}, by considering scattering from a finite array of Neumann inclusions in the physical domain.} 
\textcolor{black}{In this section we drop the requirement that the media has to be periodic, so equations (\ref{R=nmeqn}) - (\ref{egn::G}) no longer apply. We no longer consider the $IJ$th scatterer but instead we consider the $n$th scatterer in some arbitrary arrangement - subscript $IJ$ is replaced accordingly. The solutions of the preceding section are still valid provided one constructs comparable media (like in Fig. (\ref{SchematicDZwithEigenFoldy})) for the scattering simulation.}

 Foldy's method \cite{foldy1945multiple} is popular for modelling scattering from small Dirichlet {\color{black}inclusions}, \cite{martinscattering}, as to leading order they have an isotropic monopolar behaviour; an implementation is in \cite{schnitzer2017bloch}. We now extend this to the Neumann case and draw upon the discussion in  \cite{martinscattering, martin2006multiple}. Foldy's method considers the effect of an incident field interacting with multiple isotropic scatterers. \textcolor{black}{ Singularities at the $n$th scatterer are resolved by considering  Foldy's hypothesis based on use of the external field. The external field, at the $n$th scatterer, is defined as the total field minus the contribution of the $n$th scatterer. In our case}
\begin{equation}
\phi_{n}(\textbf{X}_{n}) = \lim_{\textbf{x} \to \textbf{X}_{n}} \left\lbrace \phi - \epsilon_{n}^{2} \Big( a_{n} H_{0}^{(1)} (\Omega | \textbf{x} - \textbf{X}_{n} |) + \textbf{b}_{n} \cdot \hat{\textbf{r}} \Omega H_{1}^{(1)} ( \Omega |\textbf{x} - \textbf{X}_{n}|) \Big) \right\rbrace,\label{externalField} 
\end{equation}
 where, through the Green's function \footnote{{\color{black}The introduction of arbitrary monopole ($a_{\mathrm{arb}}$) or dipole ($\textbf{b}_{\mathrm{arb}}$) point sources, placed at $\textbf{x} = \textbf{X}$, are considered by
\begin{equation}
    \left( \nabla^{2} + \Omega^{2} \right) \phi = \begin{cases}
a_{\mathrm{arb}} \delta( \textbf{x} - \textbf{X}) \quad \mbox{monopole source}, \\
\textbf{b}_{\mathrm{arb}} \cdot \nabla \delta( \textbf{x} - \textbf{X}) \quad \mbox{dipole source},
\end{cases} \label{eqn::Assoc}
\end{equation}
within the outer field. The corresponding Green's functions are
\begin{equation}
\phi = \frac{1}{4i} \cdot \begin{cases}
a_{\mathrm{arb}} H_{0} (\Omega r) \quad \mbox{monopole source}, \\
- \textbf{b}_{\mathrm{arb}} \cdot \hat{\textbf{r}} \Omega H_{1} (\Omega r) \quad \mbox{dipole source}.
\end{cases} \label{eqn::AssocGreens}
\end{equation}
The Green's functions may be calculated utilizing the Fourier transform, the inversion of which requires integrals between combinations of Bessel functions and powers \cite{gradshteyn2014table} - the monopole Green's function is derived in Graff \cite{graff1975wave} pp. 284-285. Extending this analysis for the dipole source is simple with the aid of Fig.  \ref{Xixandb} (i). }},  
Hankel functions of the first kind are present.  We adopt the notation that
\[
    H_{\nu}^{(1)}(x) =  H_{\nu}(x)
\] 
for the Hankel function of the first kind and $\nu$th order, and do not use the superscript henceforth. 

{\color{black}Foldy's \cite{foldy1945multiple} hypothesis states that the strength of the $n$th isotropic scatterer, $a_{n}$, will be proportional to the external field incident upon the $n$th scatterer \eqref{externalField}}. The proportionality constant, the monopole scattering coefficient denoted $\tau_{n}$, is given by
\begin{equation}
a_{n} = \tau_{n} \phi_{n}(\textbf{X}_{n}). \label{SoundHardMonopoleScatterCoeff}
\end{equation}
\textcolor{black}{
As noted in \cite{linton2004semi}, dipolar (anisotropic) behaviour  can be incorporated by generalising Foldy's approach as outlined in Martin \cite{martinscattering} \cite{martin2006multiple} - by considering gradients,  \textcolor{black}{in a similar fashion to equation (\ref{HelmholtzEigeAsymp})}, the dipole scattering coefficient is related to $\nabla \phi_{n}(\textbf{X}_{n})$ by:
}
\begin{equation}
\textbf{b}_{n} = \mathrm{T}_{n}\cdot \nabla \phi_{n}(\textbf{X}_{n}). \label{SoundHardDipoleScatterCoeff}
\end{equation}
Here $\mathrm{T}_{n}$ is the dipole scattering coefficient matrix, a $2 \times 2$ matrix fully encapsulating any anisotropy introduced by the dipole contribution of the sound-hard scatterer.

The $\lim_{r_{n} \to 0} \phi$ is given by equation (\ref{HelmholtzInnerLim}), therefore by  (\ref{SoundHardMonopoleScatterCoeff}) and (\ref{SoundHardDipoleScatterCoeff}) we find that
\begin{equation}
\begin{rcases}
\tau_{n} {\color{black}=} \frac{1}{\frac{4i}{\pi \Omega^{2}} - \epsilon_{n}^{2} \left[ 1 - \frac{2i}{\pi} \left( \log \frac{2}{\epsilon_{n} \Omega} + \frac{3}{4} - \gamma_{E}  \right) \right] + {\color{black}\mathcal{O}(r_{n})} } \\
\mathrm{T}_{n}  {\color{black}=}  \frac{1}{\frac{2}{i \pi} + \frac{\epsilon_{n}^{2} i \Omega^{2}}{\pi} \left( \log \frac{2}{\epsilon_{n} \Omega} - \frac{5}{4} - \gamma_{E} \right) - \frac{\epsilon_{n}^{2} \Omega^{2}}{2} + {\color{black}\mathcal{O}(r_{n})} }   \mathrm{I}
\end{rcases} \quad \mbox{as $r_{n} \to 0$}, \label{MonoDiPoleCoeffs}
\end{equation}
where $\mathrm{I}$ is the 2-by-2 identity matrix.

We take the incident field to be a line source placed at $\textbf{\textbf{x}} = \textbf{X}_{\mathrm{inc}}$, with a strength comparable to that of the scatterers; subsequently 
\begin{equation}
(\nabla^{2} + \Omega^{2}) \phi =  \epsilon_{\mathrm{min}}^{2} \widetilde{\phi}_{\mathrm{inc}} +4i \sum_{j} \epsilon^{2}_{j}  \left\lbrace  a_{j} \delta(\textbf{x} - \textbf{X}_{j})- \textbf{b}_{j} \cdot \nabla \delta(\textbf{x} - \textbf{X}_{j})  \right\rbrace. \label{eqn::FoldyPDE}
\end{equation}
Here $\epsilon_{\mathrm{min}}$ denotes the smallest inclusion present within the field and $\widetilde{\phi}_{\mathrm{inc}}$ denotes the incident field source term. It is rather natural for us to consider two sources, we set $\widetilde{\phi}_{\mathrm{inc}}$ to some arbitrary incident values as defined in the right hand side of (\ref{eqn::Assoc}). Here $a_{\mathrm{arb}}$ and $\textbf{b}_{\mathrm{arb}}$ are some order unity constants altering the strength (monopole and dipole) and alignment (dipole) of the incident field. Subsequently, by (\ref{eqn::AssocGreens}), considering $m$ scatterers within the field and denoting the Green's function associated with $\widetilde{\phi}_{\mathrm{inc}}$ by $\phi_{\mathrm{inc}}$, it follows
\begin{equation}
\phi = \epsilon^{2}_{\mathrm{min}} \phi_{inc} (\textbf{x}) + \sum_{j=1}^{m} \epsilon^{2}_{j} \left\lbrace a_{j} H_{0}(\Omega |\textbf{x} - \textbf{X}_{j}|) + \textbf{b}_{j} \cdot \hat{\textbf{r}} \Omega H_{1} (\Omega |\textbf{x} - \textbf{X}_{j}|) \right\rbrace . \label{HelmholtzTotalField}
\end{equation}
Substituting (\ref{HelmholtzTotalField}) into (\ref{SoundHardMonopoleScatterCoeff}) and $\nabla$(\ref{HelmholtzTotalField}) into (\ref{SoundHardDipoleScatterCoeff}), one finds
\begin{equation}
\begin{split}
a_{n} \left\lbrace \frac{4i}{\pi \Omega^{2}} - \epsilon^{2}_{n} \left[ 1 - \frac{2i}{\pi} \left( \log \frac{2}{\epsilon \Omega} + \frac{3}{4} - \gamma_{E}  \right) \right] \right\rbrace - & \\
- \sum_{\substack{j=1 \\ j \neq n}}^{m} \epsilon_{j}^{2} \Big\lbrace a_{j} H_{0}(\Omega r_{nj}) + \textbf{b}_{j} \cdot \hat{\textbf{r}} \Omega & H_{1} (\Omega r_{nj} ) \Big\rbrace = \epsilon_{\mathrm{min}}^{2} \phi_{\mathrm{inc}},
\end{split}
\label{ScatterHelmMonoCoeff}
\end{equation}
\begin{equation}
\begin{split}
\textbf{b}_{n} \left\lbrace \frac{2}{i \pi} + \frac{\epsilon_{n}^{2} i \Omega^{2}}{\pi} \left( \log \frac{2}{\epsilon_{n} \Omega} - \frac{5}{4} - \gamma_{E} \right) - \frac{\epsilon_{n}^{2} \Omega^{2}}{2}  \right\rbrace & - \sum_{\substack{j=1 \\ j \neq n}}^{m} \epsilon_{j}^{2} \Big\lbrace \Big( - a_{j} \Omega H_{1}(\Omega r_{nj}) +  \\ 
+ \frac{\textbf{b}_{j} \cdot \textbf{e}_{r \,j} }{2} \Omega^{2} \Big[ H_{0} (\Omega r_{nj}) - H_{2} (\Omega r_{nj}) \Big] \Big)  \textbf{e}_{r \, j} + & \\
+ \Big( \frac{\textbf{b}_{j} \cdot \textbf{e}_{\varphi \,j}}{r_{nj}} & \Omega H_{1} (\Omega r_{nj}) \Big) \textbf{e}_{\varphi \, j} \Big\rbrace = \epsilon_{\mathrm{min}}^{2} \nabla \phi_{\mathrm{inc}},
\end{split} \label{ScatterHelmDipoleCoeff}
\end{equation}
where $r_{nj} = |\textbf{X}_{n} - \textbf{X}_{j}|$. Comparing (\ref{MonoDiPoleCoeffs}) with the heuristic argument of \cite{martinscattering}, the `good choices' for the monopole and dipole scattering coefficients  are $\tau_{n}$ and $\mathrm{T}_{n}$ to leading order. Herein the solution is correct up to order $\epsilon^{2}$, the order where our singular Green's functions act to induce the inherently weak scattered field. 


To avoid confusion between local radial coordinate systems, around each $j$th scatterer, we define a global Cartesian basis in which
\begin{equation}
\textbf{e}_{r \, j} = \cos \varphi_{ij} \textbf{e}_{x} + \sin \varphi_{ij} \textbf{e}_{y}, \quad \quad \textbf{e}_{\varphi \, j} = - \sin \varphi_{ij} \textbf{e}_{x} + \cos \varphi_{ij} \textbf{e}_{y}.
\end{equation}
Here $\varphi_{ij}$ is the polar angle of $\textbf{X}_{i} - \textbf{X}_{j}$, the angle from the $j$th to the $i$th scatterer, centred on the $j$th. {\color{black}Therefore, considering equations} (\ref{ScatterHelmMonoCoeff}),  (\ref{ScatterHelmDipoleCoeff})$\cdot \textbf{e}_{x}$ and (\ref{ScatterHelmDipoleCoeff})$\cdot \textbf{e}_{y}$ we form the following matrix system. 
\begin{equation}
\underbrace{ \begin{pmatrix}
\circled{1} & \circled{2} & \circled{3} \\
\circled{4} & \circled{5} & \circled{6} \\
\circled{7} & \circled{8} & \circled{9} 
\end{pmatrix}}_{\circledast} \begin{pmatrix}
\textbf{a}_{s} \\
\textbf{b}_{1s} \\
\textbf{b}_{2s}
\end{pmatrix} = \begin{pmatrix}
\boldsymbol{\phi}_{\mathrm{inc} \,s} \\
\nabla \boldsymbol{\phi}_{\mathrm{inc} 1 \,s} \\
\nabla \boldsymbol{\phi}_{\mathrm{inc} 2 \,s} 
\end{pmatrix}, \label{FoldyMatrix}
\end{equation} 
where
\begin{equation}
\textbf{a}_{s}^{\dagger}  = \begin{pmatrix}
a_{1} & a_{2} & \ldots & a_{m} 
\end{pmatrix}, \quad  \textbf{b}_{i \, s}^{\dagger}  = \begin{pmatrix}
b_{i \, 1} & b_{i \, 2} & \ldots & b_{i \, m}
\end{pmatrix}, \quad \mbox{\textcolor{black}{for $i=1,2$}}.
\end{equation}
The superscript $\dagger$ denoting the transpose. Again, the details in (\ref{FoldyMatrix}) are lengthy and given in \ref{AllOfTheComponentsFoldy}. Eq. (\ref{FoldyMatrix}) is easily solved numerically and the scattered field by a collection of small Neumann inclusions is rapidly extracted.

\begin{figure}[t]
\begin{tikzpicture}[scale=0.3, transform shape]
\draw (14 - 1.5, -0.2) node[inner sep=0] {\includegraphics[scale=1]{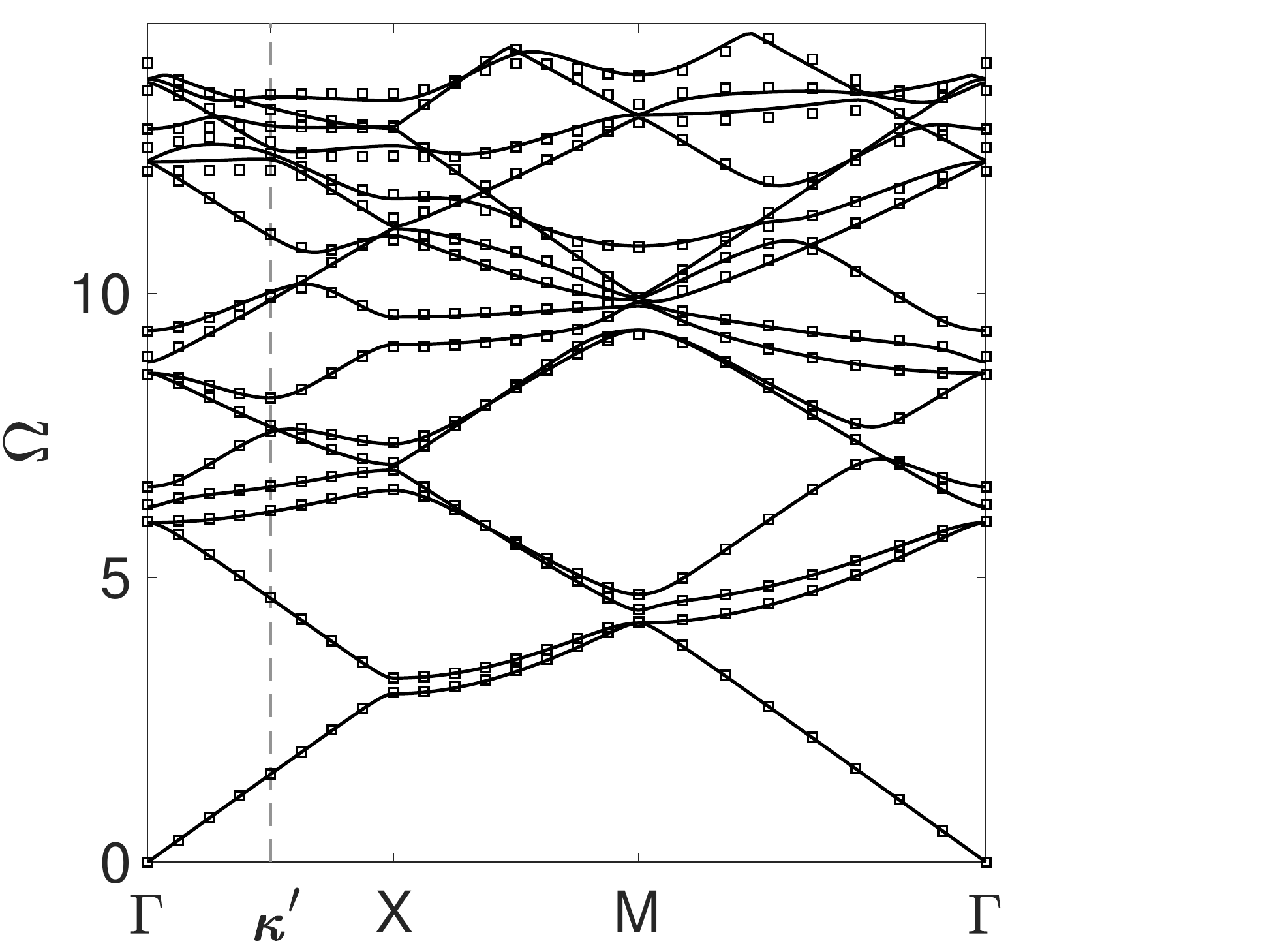}};
\draw (30.35-0.03 - 1.5, 3.5) node[inner sep=0] {\includegraphics[scale=1]{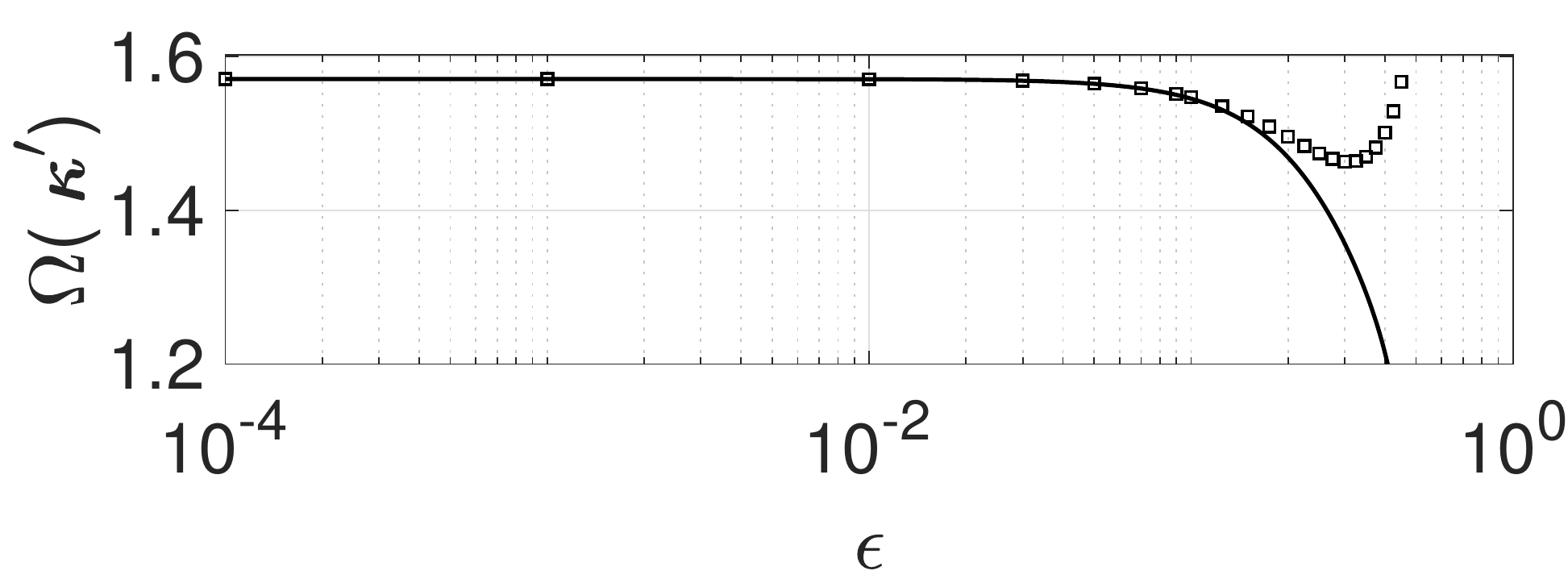}};
\draw (30.35-0.03 - 1.5, -3.9) node[inner sep=0] {\includegraphics[scale=1]{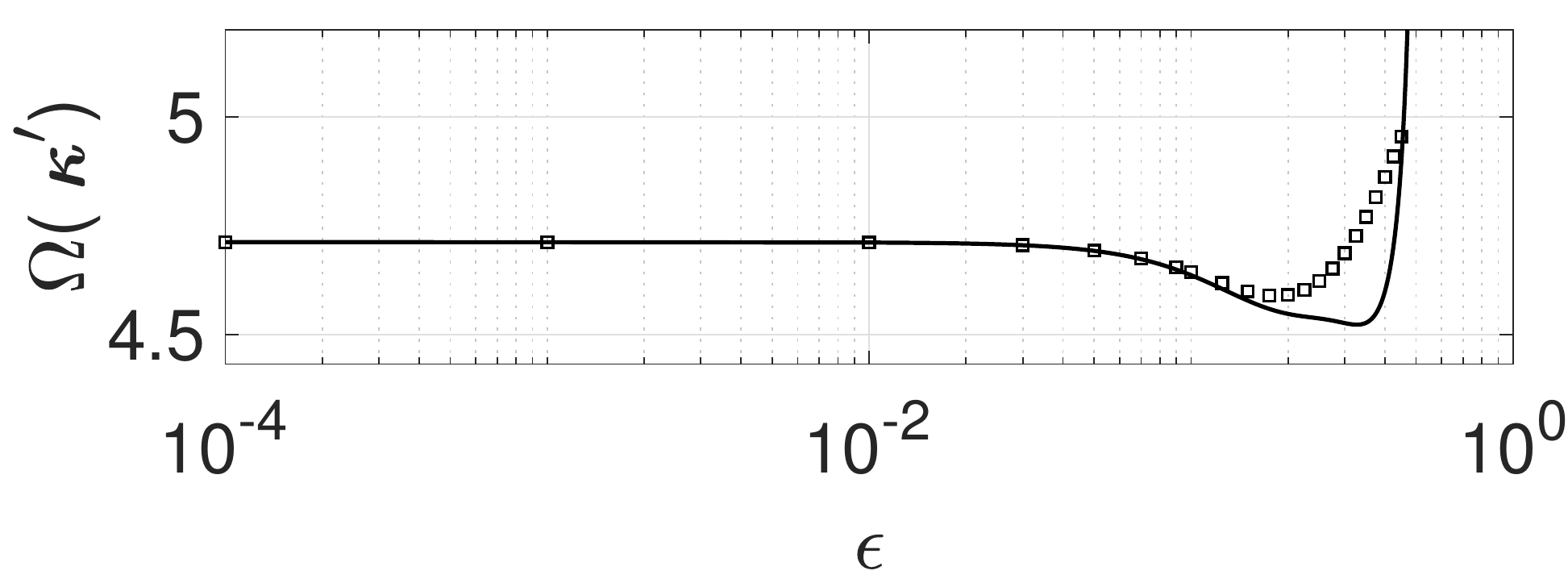}};
\node[below, scale=2.5] at (14 - 1.5 - 7.5 + 0.5+0.2+0.5,-0.1 + 7+0.2-0.2+0.2) {$\displaystyle  (iii)$}; 
\node[below, scale=2.5] at (14 - 1.5 - 7.5 +32.5-0.1-0.2,-0.2 + 7-0.3-2.5) {$\displaystyle  (iv)$}; 
\node[below, scale=2.5] at (14 - 1.5 - 7.5 +32.5-0.2,-0.2 + 7 - 7.4-2.5) {$\displaystyle  (v)$};

\node[regular polygon, regular polygon sides=4, draw, inner sep=0.5*6.28*10.0 pt,rotate=-90] at (0,6.28*10.0*1.5 pt) {};
\node[regular polygon, circle, draw, inner sep= 0.5*6.28*10.0*0.1*2 pt,rotate=0] at (0,6.28*10.0*1.5 pt) {};

\node[regular polygon, regular polygon sides=4, draw, inner sep=0.5*6.28*10.0 pt,rotate=0] at (0 pt,-6.28*10.0*1.5 pt) {};
\draw[line width=0.5mm,gray,-] (0pt,-6.28*10.0*1.5 pt) -- (0+ 0.5*6.28*10*1.41 pt,-6.28*10.0*1.5 pt);
\draw[line width=0.5mm,gray,-] (0+ 0.5*6.28*10*1.41 pt,-6.28*10.0*1.5 pt) -- (0+ 0.5*6.28*10*1.41  pt, 0.5*6.28*10*1.41 -6.28*10.0*1.5 pt);
\draw[line width=0.5mm,gray,-] (0+ 0.5*6.28*10*1.41  pt, 0.5*6.28*10*1.41 -6.28*10.0*1.5 pt) -- (0pt,-6.28*10.0*1.5 pt);
\node[below,left,scale=2.5] at (0pt,-6.28*10.0*1.5 pt) {$\displaystyle  \Gamma$}; 
\node[below,right,scale=2.5] at (0+ 0.5*6.28*10*1.41 pt,-6.28*10.0*1.5 pt) {$\displaystyle  X$};
\node[above,right,scale=2.5] at (0+ 0.5*6.28*10*1.41  pt, 0.5*6.28*10*1.41 -6.28*10.0*1.5 pt) {$\displaystyle M$};
\node[regular polygon, circle, draw, inner sep=1.25pt,rotate=0,line width=0.5mm,shading=fill,outer color=gray,gray] at (0pt,-6.28*10.0*1.5 pt)  {};
\node[regular polygon, circle, draw, inner sep=1.25pt,rotate=0,line width=0.5mm,shading=fill,outer color=gray,gray] at (0+ 0.5*6.28*10*1.41 pt,-6.28*10.0*1.5 pt)  {};
\node[regular polygon, circle, draw, inner sep=1.25pt,rotate=0,line width=0.5mm,shading=fill,outer color=gray,gray] at (0+ 0.5*6.28*10*1.41  pt, 0.5*6.28*10*1.41 -6.28*10.0*1.5 pt)  {};
\node[below, scale=2.5] at (0pt, -0.5*6.28*10*1.41+6.28*10.0*1.5 pt) {$\displaystyle  (i)$}; 
\node[below, scale=2.5] at (0pt,-0.5*6.28*10*1.41 -6.28*10.0*1.5 pt) {$\displaystyle  (ii)$};
\end{tikzpicture}
\begin{subfigure}[t]{.4\textwidth}
\begin{tikzpicture}[scale=0.3, transform shape]
\draw (14 - 1.5 - 0.25, -0.2) node[inner sep=0] {\includegraphics[scale=1]{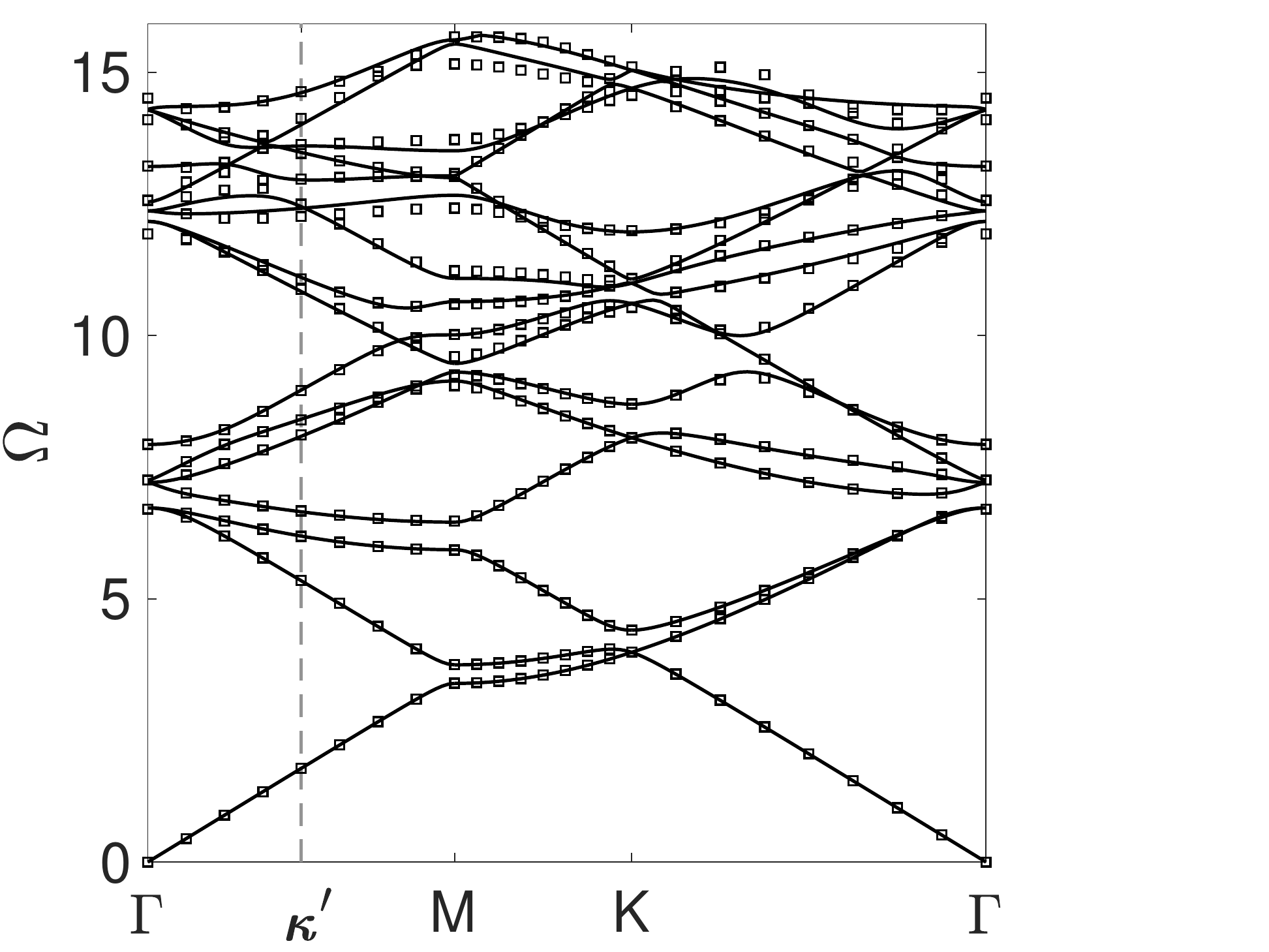}};
\draw (30.35-0.03 - 1.5 - 0.25, 3.5) node[inner sep=0] {\includegraphics[scale=1]{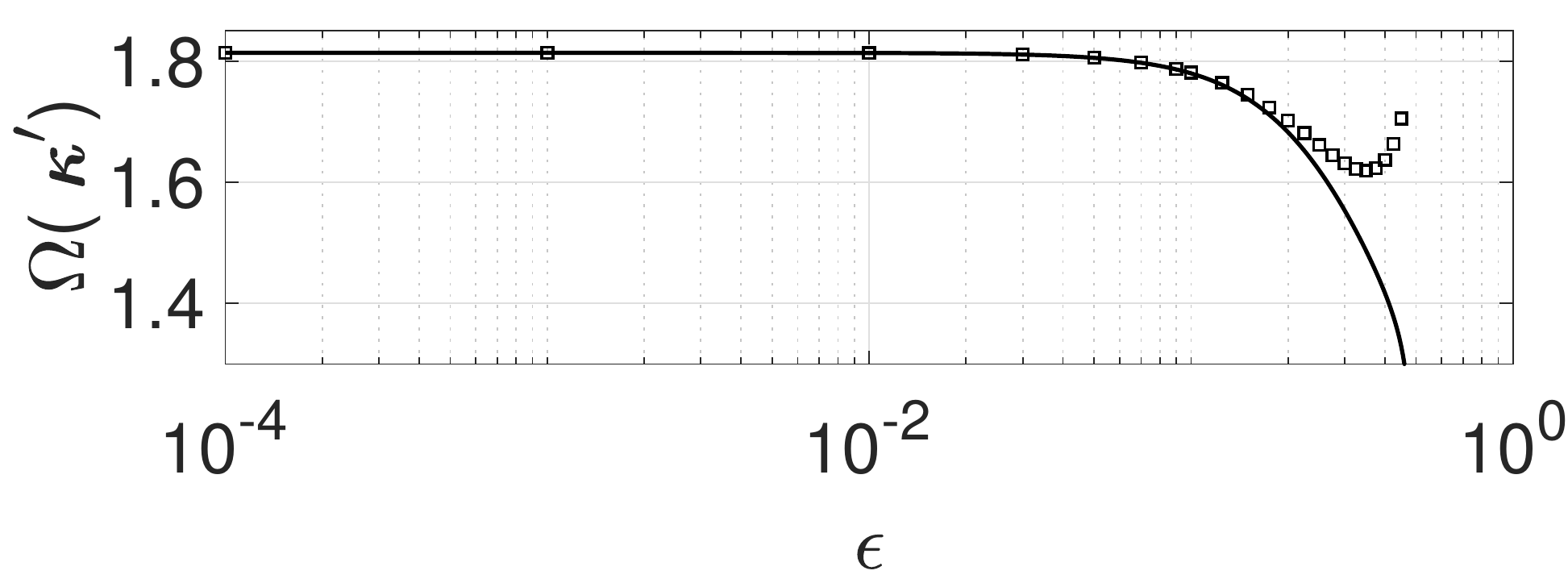}};
\draw (30.35-0.03 - 1.5 -0.25, -3.9) node[inner sep=0] {\includegraphics[scale=1]{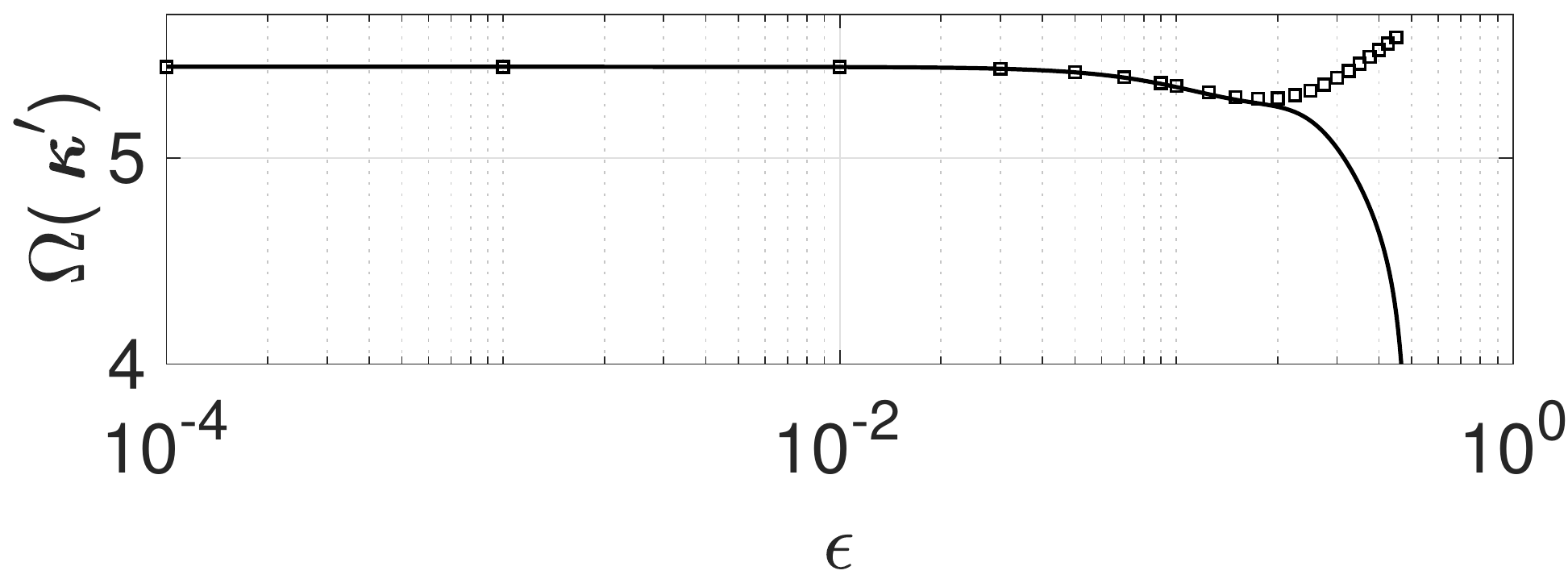}};
\node[below, scale=2.5] at (14 - 1.5 - 7.5 + 0.5+0.2,-0.1 + 7+0.2-0.2) {$\displaystyle  (iii')$}; 
\node[below, scale=2.5] at (14 - 1.5 - 7.5 +32.5 - 0.25-0.2-0.2,-0.2 + 7-2.5) {$\displaystyle  (iv')$}; 
\node[below, scale=2.5] at (14 - 1.5 - 7.5 +32.5-0.25-0.1-0.2,-0.2 + 7 - 7.4+0.2-2.5) {$\displaystyle  (v')$};

\node[regular polygon, regular polygon sides=6, draw, inner sep=0.5*6.28*10.0 pt,rotate=0] at (0,6.28*10.0*1.5 pt) {};
\node[regular polygon, circle, draw, inner sep= 0.5*6.28*10.0*0.1*2 pt,rotate=0] at (0,6.28*10.0*1.5 pt) {};
\node[regular polygon, regular polygon sides=6, draw, inner sep=0.5*6.28*10.0 pt,rotate=90] at (0 pt,-6.28*10.0*1.5 pt) {};
\draw[line width=0.5mm,gray,-] (0pt,-6.28*10.0*1.5 pt) -- (0+ 0.5*6.28*10*1.41 pt,-6.28*10.0*1.5 pt);
\draw[line width=0.5mm,gray,-] (0+ 0.5*6.28*10*1.41 pt,-6.28*10.0*1.5 pt) -- (0+ 0.5*6.28*10*1.41  pt, 0.577*0.5*6.28*10*1.41 -6.28*10.0*1.5 pt);
\draw[line width=0.5mm,gray,-] (0+ 0.5*6.28*10*1.41  pt, 0.577*0.5*6.28*10*1.41 -6.28*10.0*1.5 pt) -- (0pt,-6.28*10.0*1.5 pt);
\node[below,left,scale=2.5] at (0pt,-6.28*10.0*1.5 pt) {$\displaystyle  \Gamma$}; 
\node[below,right,scale=2.5] at (0+ 0.5*6.28*10*1.41 pt,-6.28*10.0*1.5 pt) {$\displaystyle  M$};
\node[above,right,scale=2.5] at (0+ 0.5*6.28*10*1.41  pt, 0.577*0.5*6.28*10*1.41 -6.28*10.0*1.5 pt) {$\displaystyle  K$};
\node[regular polygon, circle, draw, inner sep=1.25pt,rotate=0,line width=0.5mm,shading=fill,outer color=gray,gray] at (0pt,-6.28*10.0*1.5 pt)  {};
\node[regular polygon, circle, draw, inner sep=1.25pt,rotate=0,line width=0.5mm,shading=fill,outer color=gray,gray] at (0+ 0.5*6.28*10*1.41 pt,-6.28*10.0*1.5 pt)  {};
\node[regular polygon, circle, draw, inner sep=1.25pt,rotate=0,line width=0.5mm,shading=fill,outer color=gray,gray] at (0+ 0.5*6.28*10*1.41  pt, 0.577*0.5*6.28*10*1.41 -6.28*10.0*1.5 pt)  {};
\node[below,scale=2.5] at (0pt, -0.5*6.28*10*1.41+6.28*10.0*1.5 pt) {$\displaystyle  (i')$}; 
\node[below,scale=2.5] at (0pt,-0.5*6.28*10*1.41 -6.28*10.0*1.5 -5pt) {$\displaystyle  (ii')$}; 

\end{tikzpicture}
\end{subfigure}
\caption{Comparison between eigenvalues from (\ref{HelmholtzHardScheme_1})  and the  FE generated dispersion relation for a single inclusion in either square or hexagonal primitive cells. The full dispersion diagram is shown in $(iii)$ for $\epsilon = 0.1L$, {\color{black}where $L=1$}, with asymptotic (squares) and FE (solid lines). For a fixed wavenumber, $\kappa'$, we show the frequency variation with $\epsilon$ on the first branch $(iv)$ and second branch $(v)$. The regular and dashed Roman enumerated quantities denote a square and hexagonal primitive cell configuration respectively.}
\label{HelmholtzNeumannErrorAnalysis}
\end{figure}
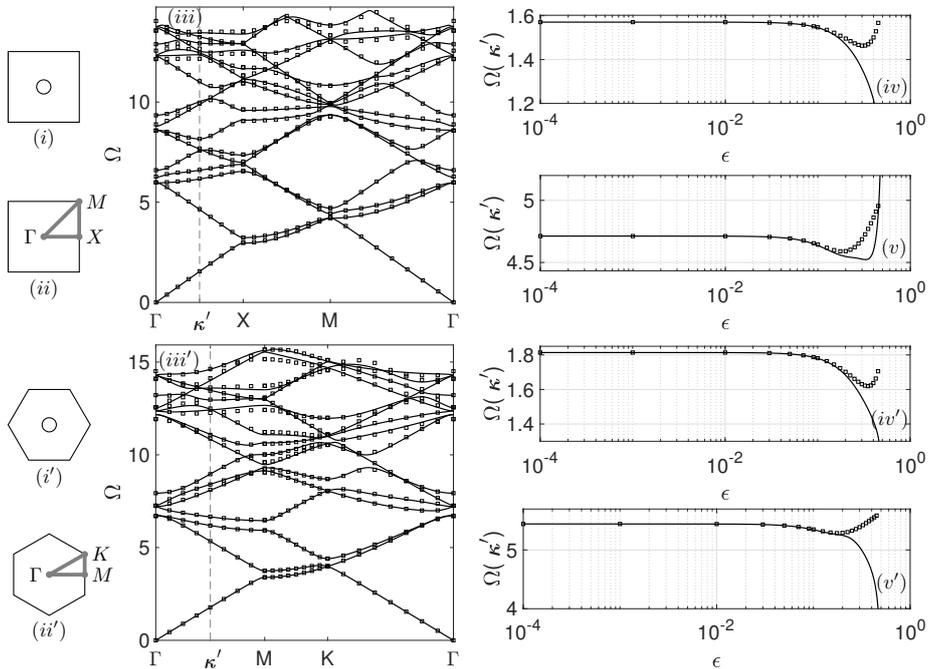

\subsection{Solutions to the \textcolor{black}{unforced} (homogeneous) Foldy problem} 
\label{sec:solnInPhysHomog}

Setting the incident field {\color{black}to zero in \eqref{FoldyMatrix}} leads to an interesting practical result where we can extract \textcolor{black}{dormant ``modes" for unforced problems - these modes await excitation}. The rows and columns of $\circledast$ in (\ref{FoldyMatrix}), given in \ref{AllOfTheComponentsFoldy},  are Hankel functions evaluated at the {\color{black}centers of the inclusions throughout the structure; they are linearly independent. Therefore $\circledast$ is a matrix of full rank and, by the rank-nullity theorem, its non-trivial null-space is empty - we are} doomed to never find a perfect non-trivial solution to our homogeneous problem.

However, we can apply the singular value decomposition to $\circledast$. {\color{black}We choose the right-singular vector, corresponding to the singular values of the smallest magnitude, for the column vector containing $\textbf{a}_{s}$, $\textbf{b}_{1 \, s}$ and $\textbf{b}_{2 \, s}$ in equation \eqref{FoldyMatrix}. Provided the chosen} singular value is small and the total number of scatterers considered (hence dimension of the matrix) large, each row of $\circledast$ multiplied by the column vector {\color{black}containing $\textbf{a}_{s}$, $\textbf{b}_{1 \, s}$ and $\textbf{b}_{2 \, s}$} is negligible. The error involved in considering the right-singular vector as a valid non-trivial solution would introduce errors, in many cases, smaller than the asymptotic error in considering the solution of the extended Foldy problem accurate to order $\epsilon^{2}$. {\color{black}We demonstrate the benefit of this approach, in approximating the solution to unforced problems, by showing it is} capable of finding the edge modes existing between the two media in section \ref{sec:topological}, refer to Fig.  \ref{SchematicDZwithEigenFoldy}.
 This approach is not limited to Foldy's method and could be used in a finite element scheme to rapidly extract the dominant ``modes" for a large lattice system.



\section{Dispersion curves}
\label{sec:results}

Having developed the asymptotic technique, \textcolor{black}{and outlined the numerical  methodology for dispersion curves via a generalised eigenvalue problem}, we now compare and contrast with dispersion curves obtained from full numerical simulations using the open source finite element (FE) package FreeFEM++ 
 \cite{hecht12a}. We begin, as shown in Fig.  \ref{HelmholtzNeumannErrorAnalysis}, by considering a single inclusion within a square or hexagonal fundamental cell. The dispersion curves are shown for both FE (solid), and from the asymptotics (squares). We choose $\epsilon=0.1$ (i.e. relatively large for such an asymptotic scheme) and note that there is still a pleasing agreement, even for the higher branches in the dispersion diagram; the discrepancy as $\epsilon$ increases is illustrated in Fig.  \ref{HelmholtzNeumannErrorAnalysis}  for a typical wavevector. As expected, from the matching procedure, both (\ref{InnerOuter}) and (\ref{HelmholtzInnerLim}) lose their validity as $\epsilon \Omega$ approaches order unity, the asymptotic scheme ultimately breaks down.
 
\begin{figure}[h!]
\centering
\begin{tikzpicture}[scale=0.3, transform shape]
\draw (14 - 1.5, -0.2) node[inner sep=0] {\includegraphics[scale=1]{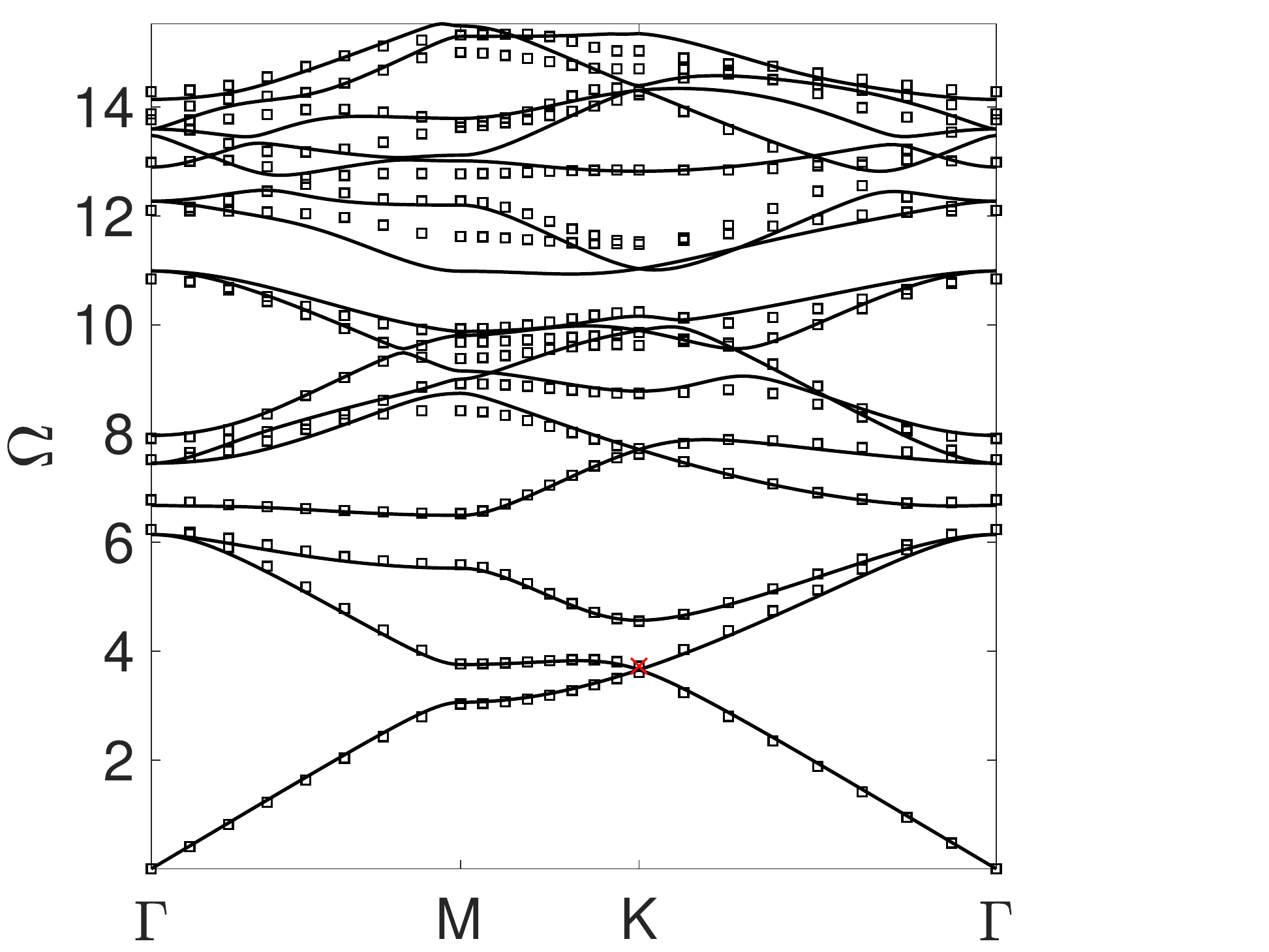}};
\draw (30.35-0.03 - 1.5, -0.2) node[inner sep=0] {\includegraphics[scale=1]{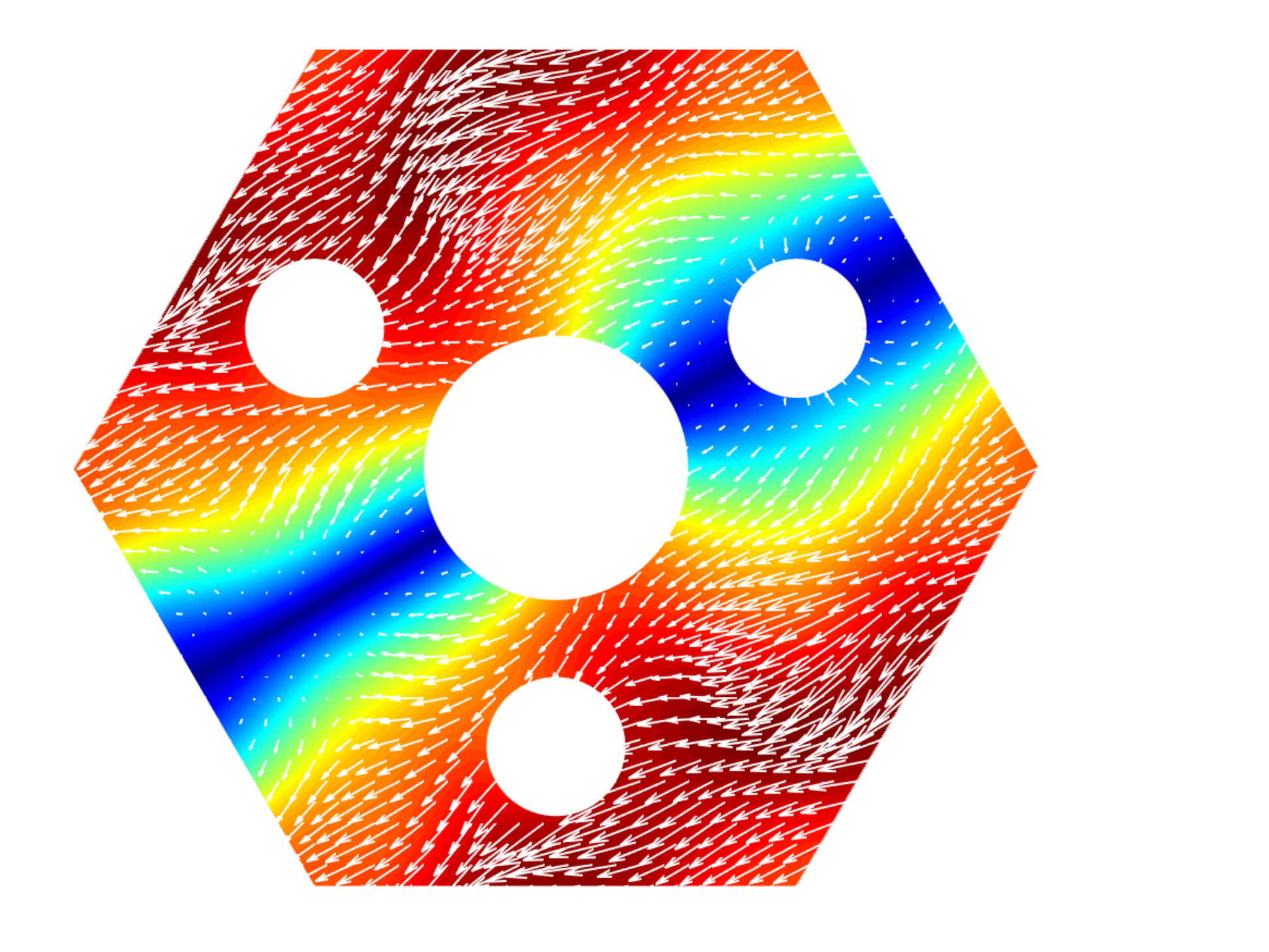}};
\node[below,scale=2.5] at (14 - 1.5 - 1.25,-0.2 - 7 - 0.5) {$\displaystyle  (iii)$}; 
\node[below,scale=2.5] at (30.35-0.03 - 1.5 - 1.25,-0.2 - 7 - 0.5) {$\displaystyle  (iv)$}; 

\node[regular polygon, regular polygon sides=6, draw, inner sep=1.1 cm,rotate=0] at (0,3.3) {};
\foreach \num in {0, 1,...,2}
	{\node (A) at ({1.4142*1.25*0.55*cos((2*pi*\num/3 + pi/6) r)}, {3.3+1.4142*1.25*0.55*sin((2*pi*\num/3+ pi/6) r)}) {};
 	\node[regular polygon, circle, draw, inner sep= 0.5*6.28*10.0*0.075*2 pt,rotate=0] at (A) {};}
 	
 	\node (B) at (0,3.3){};
 	\node[regular polygon, circle, draw, inner sep= 0.5*6.28*10.0*0.15*2 pt,rotate=0] at (B) {};
\node[regular polygon, regular polygon sides=6, draw, inner sep=0.5*6.28*10.0 pt,rotate=90] at (0 pt,-6.28*10.0*1.5 pt) {};
\draw[line width=0.5mm,gray,-] (0pt,-6.28*10.0*1.5 pt) -- (0+ 0.5*6.28*10*1.41 pt,-6.28*10.0*1.5 pt);
\draw[line width=0.5mm,gray,-] (0+ 0.5*6.28*10*1.41 pt,-6.28*10.0*1.5 pt) -- (0+ 0.5*6.28*10*1.41  pt, 0.577*0.5*6.28*10*1.41 -6.28*10.0*1.5 pt);
\draw[line width=0.5mm,gray,-] (0+ 0.5*6.28*10*1.41  pt, 0.577*0.5*6.28*10*1.41 -6.28*10.0*1.5 pt) -- (0pt,-6.28*10.0*1.5 pt);
\node[below,left,scale=2.5] at (0pt,-6.28*10.0*1.5 pt) {$\displaystyle  \Gamma$}; 
\node[below,right,scale=2.5] at (0+ 0.5*6.28*10*1.41 pt,-6.28*10.0*1.5 pt) {$\displaystyle  M$};
\node[above,right,scale=2.5] at (0+ 0.5*6.28*10*1.41  pt, 0.577*0.5*6.28*10*1.41 -6.28*10.0*1.5 pt) {$\displaystyle  K$};
\node[regular polygon, circle, draw, inner sep=1.25pt,rotate=0,line width=0.5mm,shading=fill,outer color=gray,gray] at (0pt,-6.28*10.0*1.5 pt)  {};
\node[regular polygon, circle, draw, inner sep=1.25pt,rotate=0,line width=0.5mm,shading=fill,outer color=gray,gray] at (0+ 0.5*6.28*10*1.41 pt,-6.28*10.0*1.5 pt)  {};
\node[regular polygon, circle, draw, inner sep=1.25pt,rotate=0,line width=0.5mm,shading=fill,outer color=gray,gray] at (0+ 0.5*6.28*10*1.41  pt, 0.577*0.5*6.28*10*1.41 -6.28*10.0*1.5 pt)  {};

\node[below,scale=2.5] at (0pt, -0.5*6.28*10*1.41+6.28*10.0*1.5 pt) {$\displaystyle  (i)$}; 
\node[below,scale=2.5] at (0pt,-0.5*6.28*10*1.41 -6.28*10.0*1.5 -7.5pt) {$\displaystyle  (ii)$}; 
\end{tikzpicture}
\caption{The dispersion diagram $\Omega = \Omega(\boldsymbol{\kappa})$ in panel $(iii)$ for a primitive cell $(i)$ containing 4 Neumann inclusions of radius $\epsilon_{{\color{black}1}1}=0.15 {\color{black}L}$ at $\textbf{X}_{11} = \textbf{0}$ and $\epsilon_{{\color{black}1}J}=0.075 {\color{black}L}$ placed at $\textbf{X}_{1J} = \frac{L}{3}\left[\cos \left( \frac{2(J-2) \pi}{3} + \frac{\pi}{6} \right), \sin \left( \frac{2(J-2) \pi}{3} + \frac{\pi}{6} \right) \right] $ for $J = 2,3,4$ {\color{black}and $L=1$}. The asymptotics are square symbols and FE simulations are solid lines. 
The $\boldsymbol{\kappa}$ space $\Gamma M K \Gamma$ is the irreducible Brillouin zone shown in $(ii)$. Panel $(iv)$ shows the eigensolution $\phi = \phi(\textbf{x})$ constructed over the primitive cell in physical space at frequency and wavevector at the red cross in $(iii)$, with the white arrows representing the time-averaged energy flux (\ref{TimeAvEnFlux}). } 
\label{HelmholtzNeumannOurArrangement}
\end{figure}

Although such agreement is pleasing, our primary aim is to employ the asymptotic scheme for clusters of inclusions within a primitive cell, and in particular use the scheme as a rapid route for prototyping and optimising arrangements of scatterers to obtain specific physical effects. Fig.  \ref{HelmholtzNeumannOurArrangement} shows an arrangement of inclusions, chosen to have specific symmetries such that a symmetry-induced Dirac point occurs at the $K$ point in the dispersion diagram; this underlies  so-called valley-Hall edge states \cite{khanikaev_two-dimensional_2017}, 
  and we use this geometrical arrangement to illustrate that the asymptotic scheme is capable of generating these, along with the underlying numerics required to interpret them. \cite{makwana18a} used group theoretic arguments to demonstrate that by having point group symmetries of $C_{3v}$, at both $\Gamma$ and $K$, it would guarantee the presence of a Dirac cone; the geometry chosen here is case (ii) of \cite{makwana18a}. The topological effects occur due to the breaking of the mirror symmetry by rotating the system of inclusions, lowering the point group symmetry to $C_3$, and gapping the Dirac point to open a band-gap. For the purposes of the asymptotic scheme it is interesting to note that in Fig.  \ref{HelmholtzNeumannOurArrangement}, the lowest dispersion curves display a symmetry induced Dirac point, of low enough frequency to be well captured by the scheme - certainly well enough that one can explore the topological valley-Hall effect.

\begin{figure}[h!]
\begin{subfigure}[t]{.35\textwidth}
\centering
\end{subfigure} $\quad \quad$
\begin{subfigure}[t]{.35\textwidth}
\centering
\begin{tikzpicture}[scale=0.275, transform shape]
\begin{scope}[shift={(0,0)}]
\draw (-10.5 -7.5+5+5, 0.35-3.1-15) node[inner sep=0] {\includegraphics[scale=1.0]{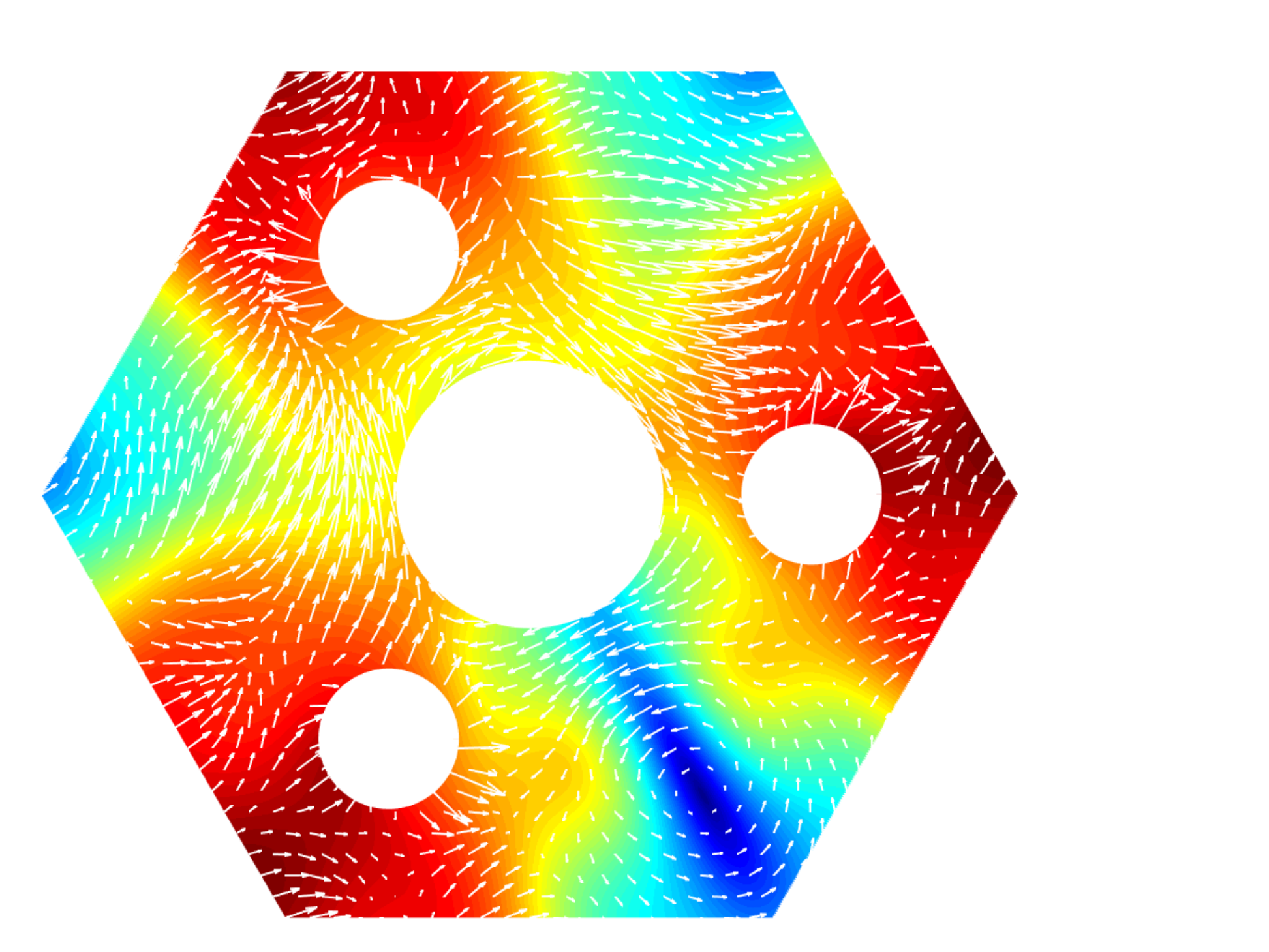}};
\node[below,scale=3.5] at (-10.5 -7.5+5+5 + 0.75-2 - 0.5,0.35-3.1-15 -0.2 - 7 - 0.5) {$\displaystyle  (iv)$};

  \begin{scope}[transform canvas={scale=0.75}, shift={(-26.2+10*0.5+6.666+6.666, -4.1 - 10*0.8660-20)},rotate=-137-60]
     \fill[even odd rule,blue!50!black,opacity=0.2] circle (3.8) circle (3.2);
  \foreach \x in {0,60,...,180} {
    \arcarrow{3}{3.5}{4}{\x+20}{\x+70}{5}{blue,
      draw = blue!50!black, very thick}{}
 }
  \end{scope}

 \begin{scope}[transform canvas={scale=0.75}, shift={(-26.2+10*0.5+6.666+6.666,-4.1 + 10*0.8660-20)},rotate=-137+60]
     \fill[even odd rule,blue!50!black,opacity=0.2] circle (3.8) circle (3.2);
  \foreach \x in {0,60,...,180} {
    \arcarrow{3}{3.5}{4}{\x+20}{\x+70}{5}{blue,
      draw = blue!50!black, very thick}{}
 }
  \end{scope}

   \begin{scope}[transform canvas={scale=0.75}, shift={(-26.2-10+6.666+6.666,-4.1-20)},rotate=-137-180]
     \fill[even odd rule,blue!50!black,opacity=0.2] circle (3.8) circle (3.2);
  \foreach \x in {0,60,...,180} {
    \arcarrow{3}{3.5}{4}{\x+20}{\x+70}{5}{blue,
      draw = blue!50!black, very thick}{}
 }
  \end{scope}
 \end{scope}

\begin{scope}[shift={(12,1.0)}]
\draw (0, 2) node[inner sep=0] {\includegraphics[scale=1.25]{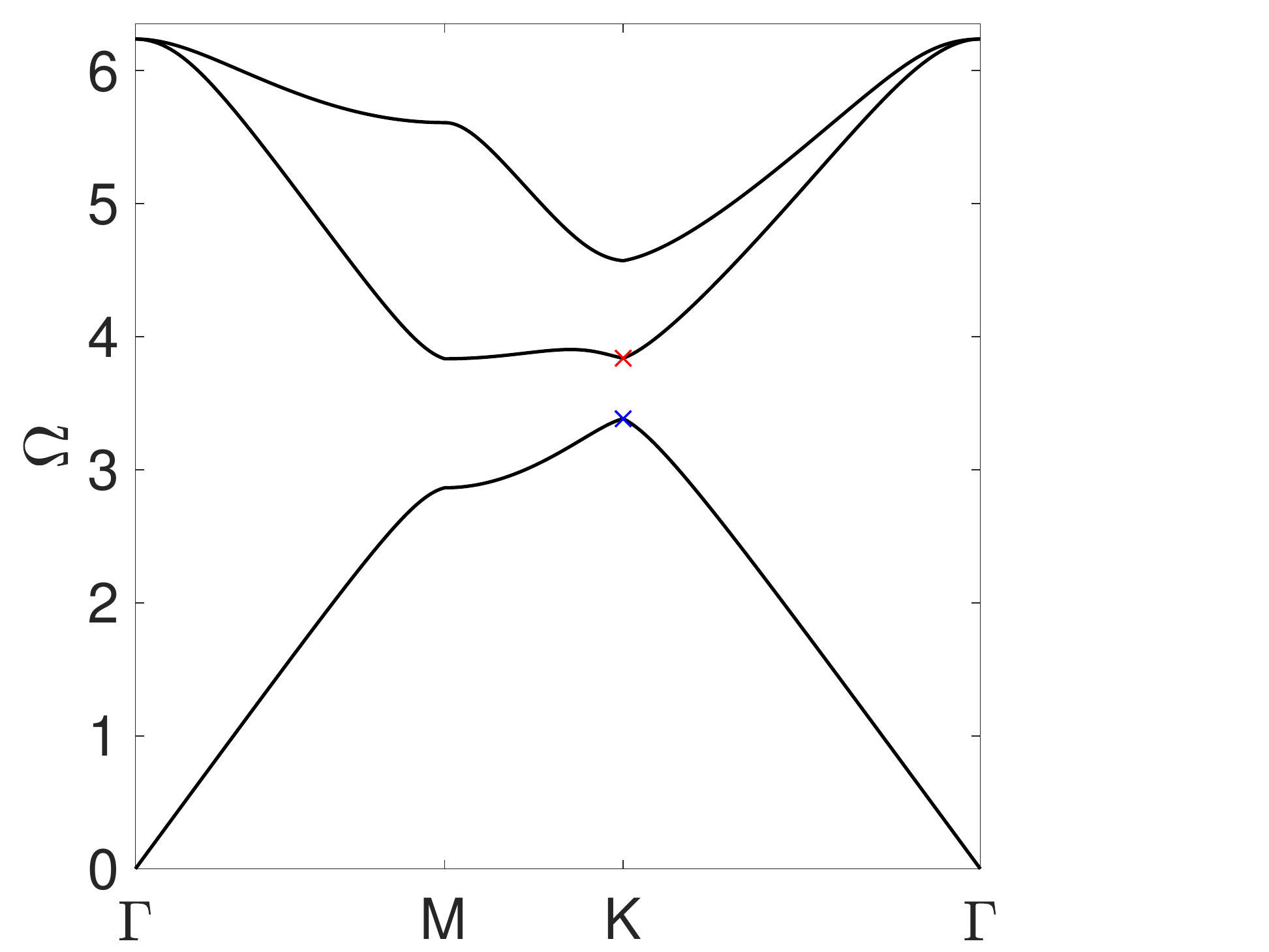}};
\node[below,scale=3.5] at (0.75-2,-0.2 - 7) {$\displaystyle  (iii)$}; 
\end{scope}

\begin{scope}[shift={(0,0)}]
\draw (-10.5 -7.5+40-3.5-5, 0.35-3.1-15) node[inner sep=0] {\includegraphics[scale=1.0]{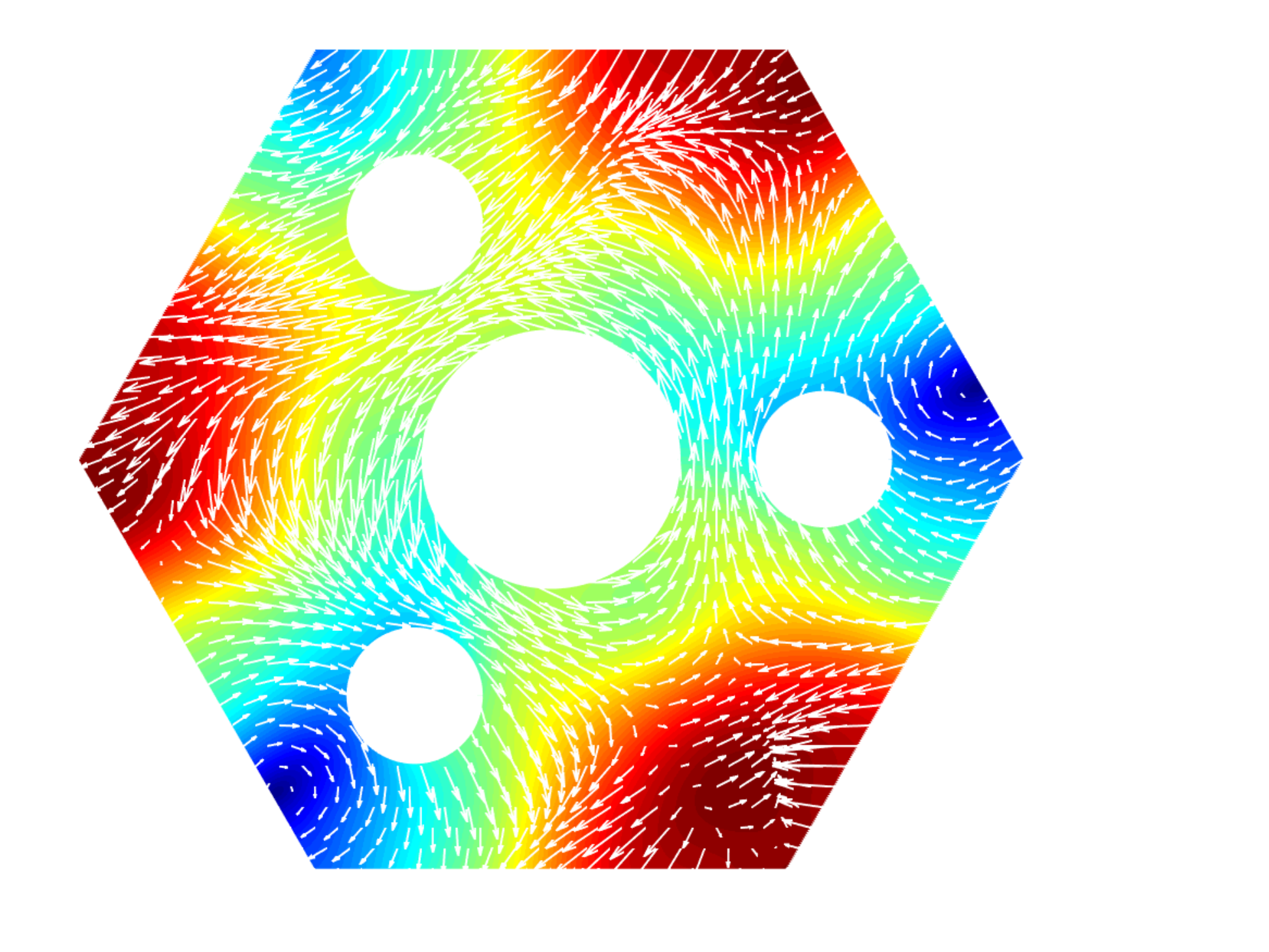}};
\node[below,scale=3.5] at (-10.5 -7.5+5+5 + 0.75-2 - 0.5+22,0.35-3.1-15 -0.2 - 7 - 0.5) {$\displaystyle  (v)$};

 \begin{scope}[yscale=-1,xscale=1, transform canvas={scale=0.75}, shift={(-25.75-10*0.5+53.3333-5-6.666,3.4 - 10*0.8660+20)},rotate=-137-120]
     \fill[even odd rule,blue!50!black,opacity=0.2] circle (3.8) circle (3.2);
  \foreach \x in {0,60,...,180} {
    \arcarrow{3}{3.5}{4}{\x+20}{\x+70}{5}{blue,
      draw = blue!50!black, very thick}{}
 }
  \end{scope}

 \begin{scope}[yscale=-1,xscale=1, transform canvas={scale=0.75}, shift={(-25.75-10*0.5+53.3333-5-6.666,3.4 + 10*0.8660+20)},rotate=-137+120]
     \fill[even odd rule,blue!50!black,opacity=0.2] circle (3.8) circle (3.2);
  \foreach \x in {0,60,...,180} {
    \arcarrow{3}{3.5}{4}{\x+20}{\x+70}{5}{blue,
      draw = blue!50!black, very thick}{}
 }
  \end{scope}

   \begin{scope}[yscale=-1,xscale=1, transform canvas={scale=0.75}, shift={(-25.75+10+53.3333-5-6.666,3.4+20)},rotate=-137]
     \fill[even odd rule,blue!50!black,opacity=0.2] circle (3.8) circle (3.2);
  \foreach \x in {0,60,...,180} {
    \arcarrow{3}{3.5}{4}{\x+20}{\x+70}{5}{blue,
      draw = blue!50!black, very thick}{}
 }
  \end{scope}

\end{scope}

\begin{scope}[transform canvas={scale=2}, shift={(-5,0.9)}]
\node[regular polygon, regular polygon sides=6, draw, inner sep=1.1 cm,rotate=0] at (0,3.3) {};
\foreach \num in {0, 1,...,2}
	{\node (A) at ({1.4142*1.25*0.55*cos((2*pi*\num/3) r)}, {3.3+1.4142*1.25*0.55*sin((2*pi*\num/3) r)}) {};
 	\node[regular polygon, circle, draw, inner sep= 0.5*6.28*10.0*0.075*2 pt,rotate=0] at (A) {};}
 	
 	\node (B) at (0,3.3){};
 	\node[regular polygon, circle, draw, inner sep= 0.5*6.28*10.0*0.15*2 pt,rotate=0] at (B) {};
 	\node[below,scale=1.75] at (0pt, -0.5*6.28*10*1.41+6.28*10.0*1.5 pt) {$\displaystyle  (i)$}; 
 \end{scope}
 
\begin{scope}[transform canvas={scale=2}, shift={(-5,2.5)}]
\node[regular polygon, regular polygon sides=6, draw, inner sep=0.5*6.28*10.0 pt,rotate=90] at (0 pt,-6.28*10.0*1.5 pt) {};
\draw[line width=0.5mm,gray,-] (0pt,-6.28*10.0*1.5 pt) -- (0+ 0.5*6.28*10*1.41 pt,-6.28*10.0*1.5 pt);
\draw[line width=0.5mm,gray,-] (0+ 0.5*6.28*10*1.41 pt,-6.28*10.0*1.5 pt) -- (0+ 0.5*6.28*10*1.41  pt, 0.577*0.5*6.28*10*1.41 -6.28*10.0*1.5 pt);
\draw[line width=0.5mm,gray,-] (0+ 0.5*6.28*10*1.41  pt, 0.577*0.5*6.28*10*1.41 -6.28*10.0*1.5 pt) -- (0pt,-6.28*10.0*1.5 pt);
\node[below,left,scale=1.75] at (0pt,-6.28*10.0*1.5 pt) {$\displaystyle  \Gamma$}; 
\node[below,right,scale=1.75] at (0+ 0.5*6.28*10*1.41 pt,-6.28*10.0*1.5 pt) {$\displaystyle  M$};
\node[above,right,scale=1.75] at (0+ 0.5*6.28*10*1.41  pt, 0.577*0.5*6.28*10*1.41 -6.28*10.0*1.5 pt) {$\displaystyle  K$};
\node[regular polygon, circle, draw, inner sep=1.25pt,rotate=0,line width=0.5mm,shading=fill,outer color=gray,gray] at (0pt,-6.28*10.0*1.5 pt)  {};
\node[regular polygon, circle, draw, inner sep=1.25pt,rotate=0,line width=0.5mm,shading=fill,outer color=gray,gray] at (0+ 0.5*6.28*10*1.41 pt,-6.28*10.0*1.5 pt)  {};
\node[regular polygon, circle, draw, inner sep=1.25pt,rotate=0,line width=0.5mm,shading=fill,outer color=gray,gray] at (0+ 0.5*6.28*10*1.41  pt, 0.577*0.5*6.28*10*1.41 -6.28*10.0*1.5 pt)  {};
\node[below,scale=1.75] at (0pt,-0.5*6.28*10*1.41 -6.28*10.0*1.5 -7.5pt) {$\displaystyle  (ii)$}; 
 \end{scope}

\end{tikzpicture}
\end{subfigure}
\caption{\textcolor{black}{The dispersion diagram $(iii)$ generated from (\ref{HelmholtzHardScheme_1}), g}apping the Dirac point at $K$ by perturbing the structure present in Fig.  \ref{HelmholtzNeumannOurArrangement}$(i)$ via a $-\frac{\pi}{6}$ rotation as shown in $(i)$. Panels $(iv)$ and $(v)$ show the eigensolution and associated flux at the frequency shown by the blue and red crosses in (iii), respectively.}
\label{HelmholtzNeumannOurArrangementPerturbed}
\end{figure}

\section{Topological mode steering in a planar array of Neumann scatterers}
\label{sec:topological}

{\color{black} There is intense activity exploring topological-like effects for wave transport that shows little sign of abating \cite{khanikaev_two-dimensional_2017}. A specific sub-class of topological insulators that are pragmatic and simple to design are those of symmetry induced edge states \cite{dong_valley_2017, ju_topological_2015, he_acoustic_2016, lu_observation_2016}. These are formed from the strategic breaking of parity symmetry which in turn reveals topologically nontrivial band-gaps in which broadband edge modes are guaranteed to reside. In this section we use the asymptotic machinery we have developed to rapidly compute solutions that pertain to the robust transport of energy around bends in partitioned media \cite{shalaev_robust_2018, makwana_geometrically_2018}.}

\subsection{Time-averaged energy flux for a symmetry-induced topological system}
The earlier sections focused on deriving asymptotically accurate formulae, dealing with the  singular behaviour present within a divergent sum, and turning this into an effective numerical tool. Here we give a topical example on how the use of these formulae allows one to expedite computations of the energy flux; this is a physically useful quantity that allows us to determine whether a state is topologically protected. The time-averaged energy flux is defined as follows,
\begin{equation}
    \left< \textbf{F} \right> = - \frac{1}{2} \Re \left\lbrace \left[ - i \Omega \phi \right] \overline{\left[ \nabla \phi \right]} \right\rbrace, \label{TimeAvEnFlux}
\end{equation}
 with the overbar denoting complex conjugate. 
By utilising (\ref{HelmCondConvSer}) and $\nabla$(\ref{HelmCondConvSer}) or  (\ref{HelmholtzTotalField}) and $\nabla$\eqref{HelmholtzTotalField} we are able to rapidly compute \eqref{TimeAvEnFlux}; as demonstrated in Fig.  \ref{HelmholtzNeumannOurArrangement}(iv). 

\begin{figure}[h!]
\centering
\begin{tikzpicture}[scale=0.4, transform shape]
\draw (9+3.5-1, 0.35-3.1) node[inner sep=0] {\includegraphics[scale=1.175]{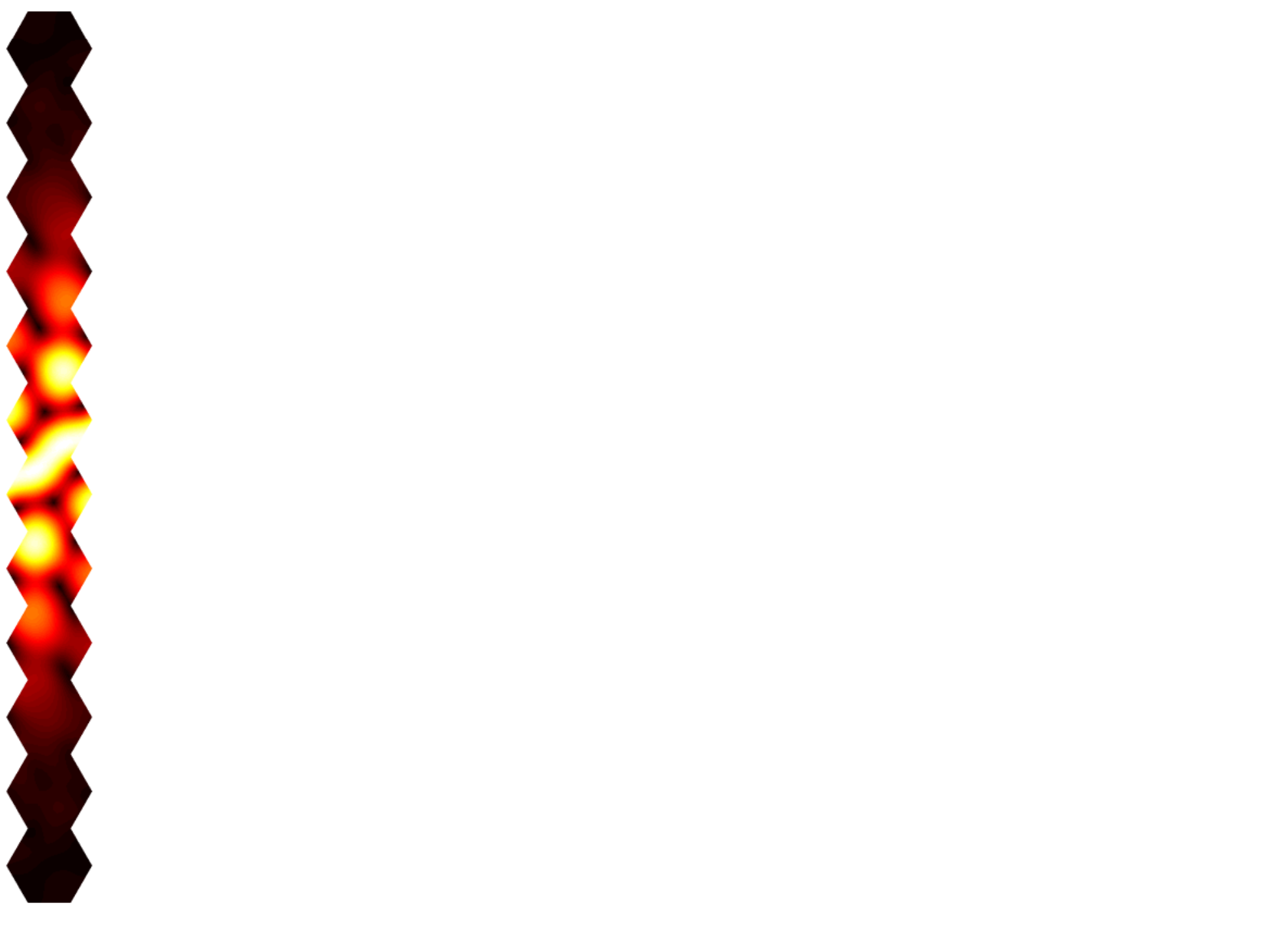}};
\draw (11.5 +3.5 +0.5-1, 0.35-3.1) node[inner sep=0] {\includegraphics[scale=1.175]{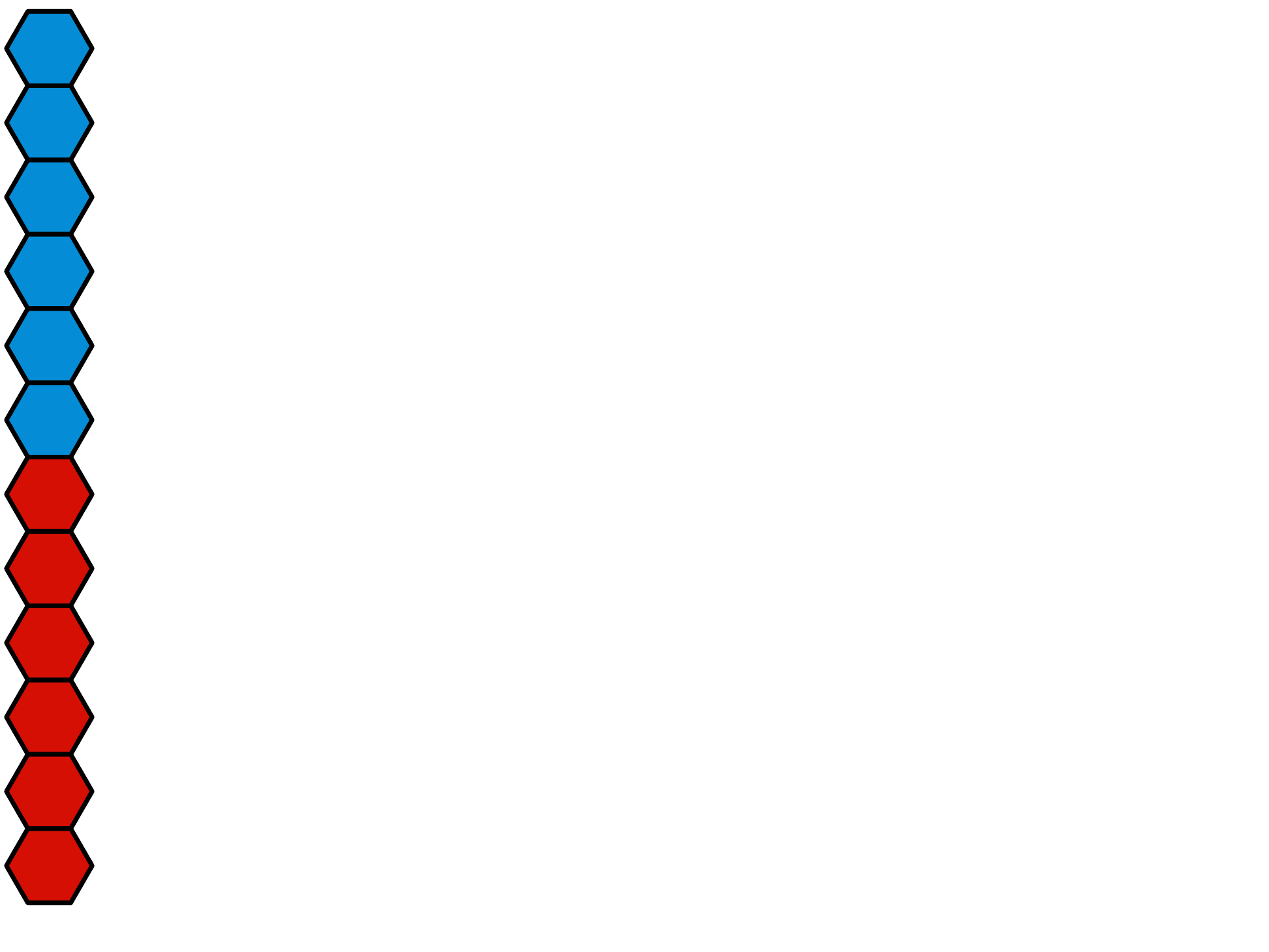}};
\draw (15.75 - 1.35, -0.750) node[inner sep=0] {\includegraphics[scale=0.85]{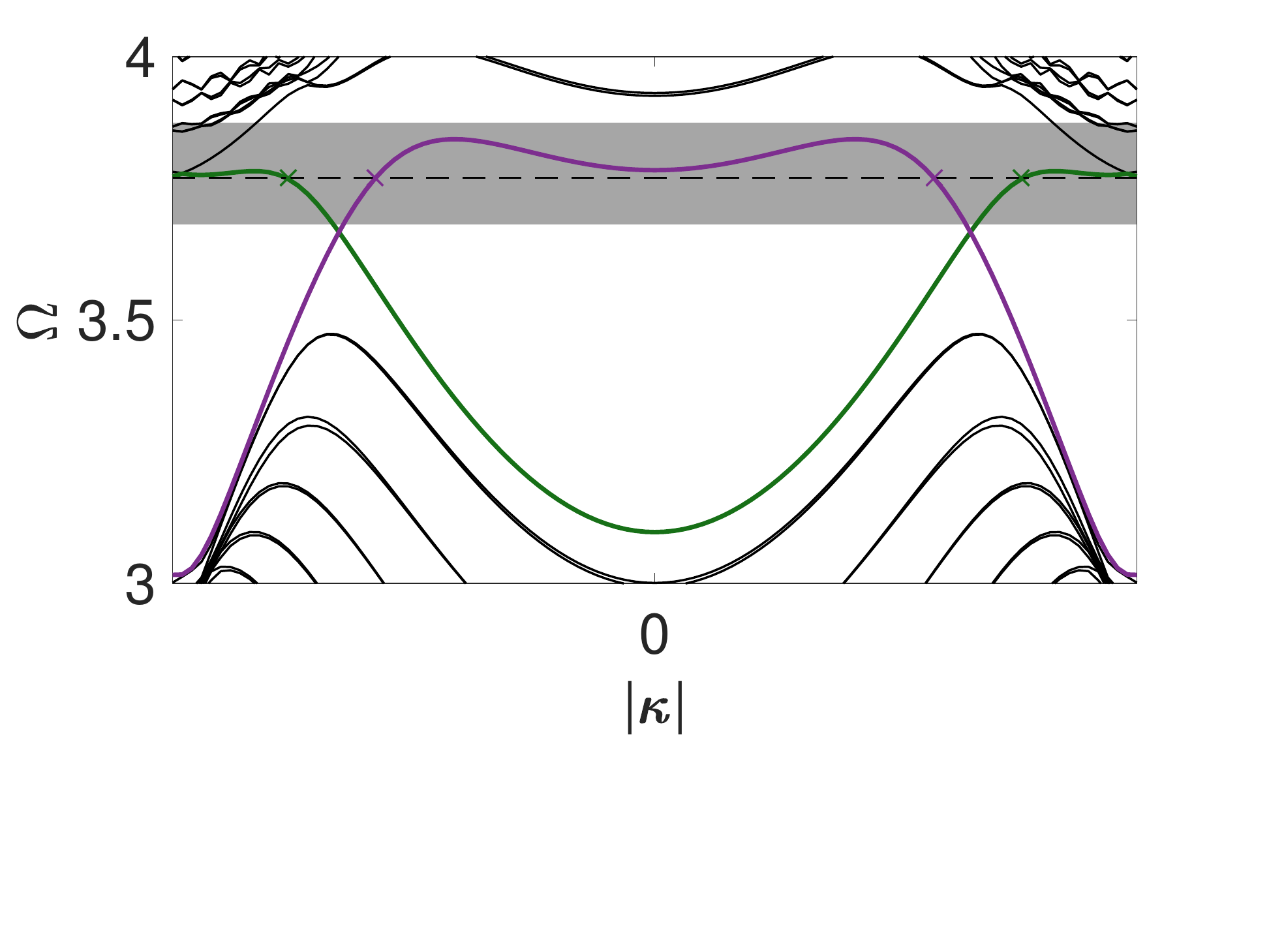}};
\draw (30.75+3.85+1, 0.35-3.1) node[inner sep=0] {\includegraphics[scale=1.175]{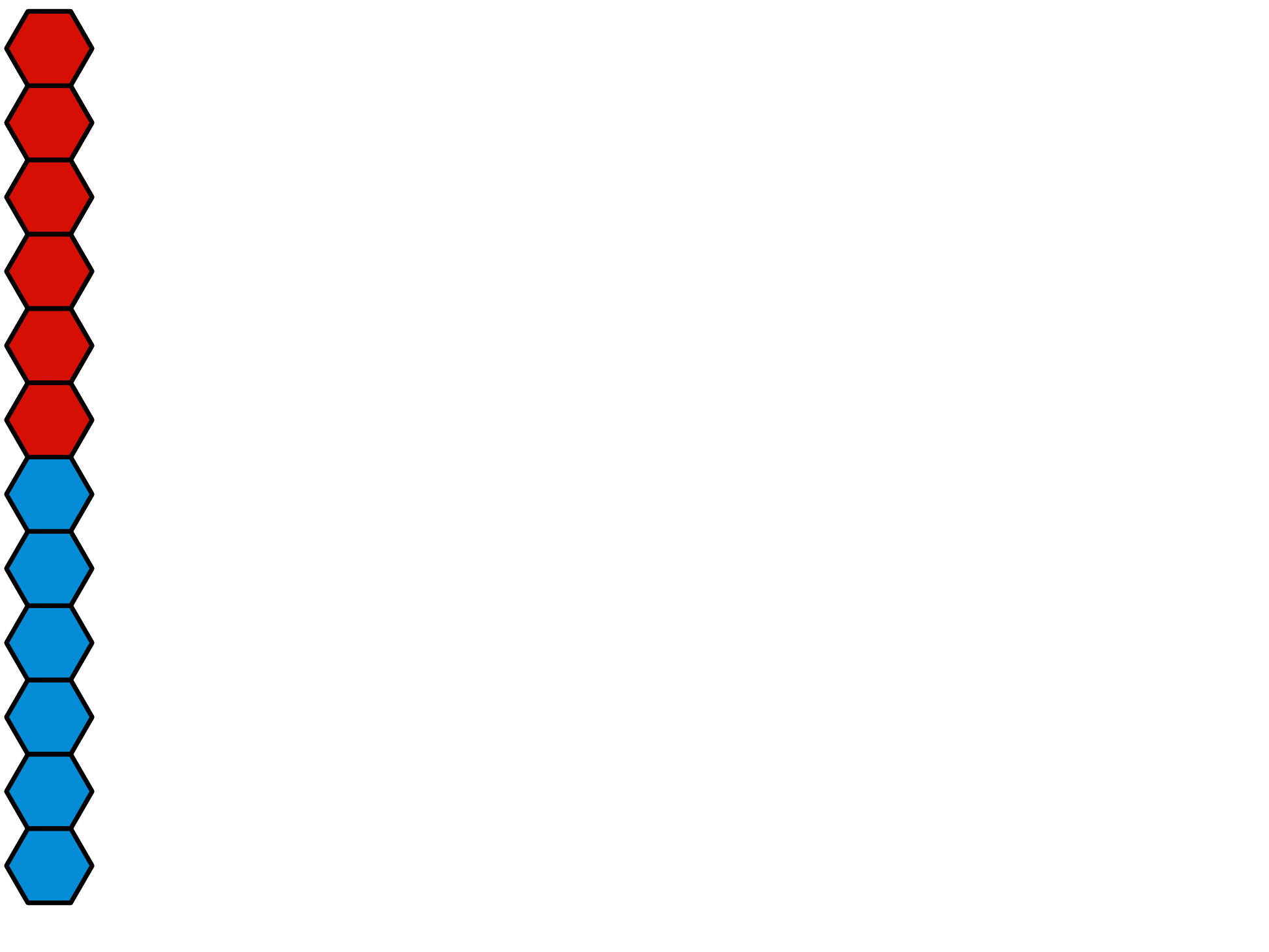}};
\draw (30.75+2.5+0.5+3.85+1, 0.35-3.1) node[inner sep=0] {\includegraphics[scale=1.175]{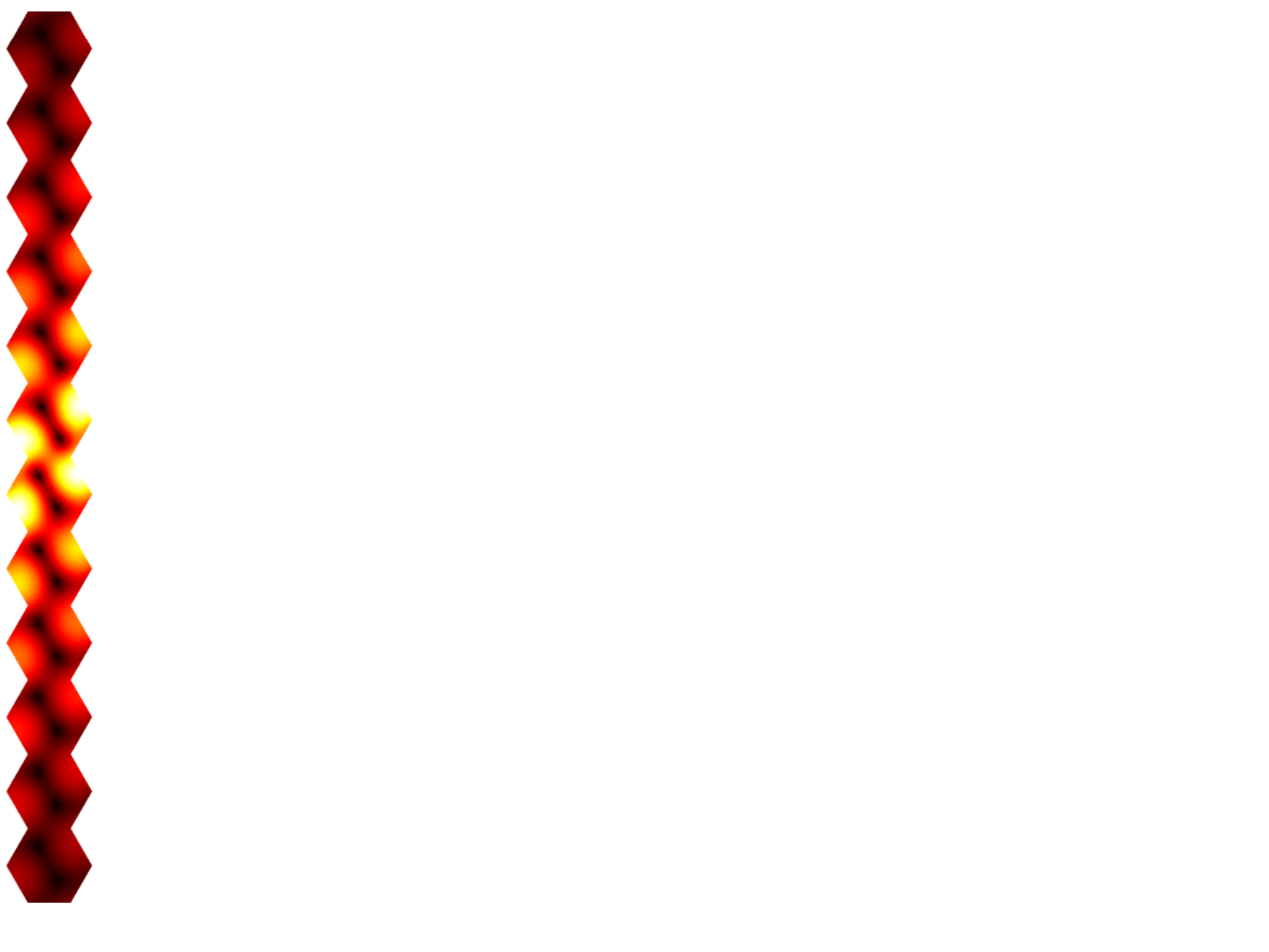}};
\draw (14-2 - 1.5, -8+0.75) node[inner sep=0] {\includegraphics[scale=0.35]{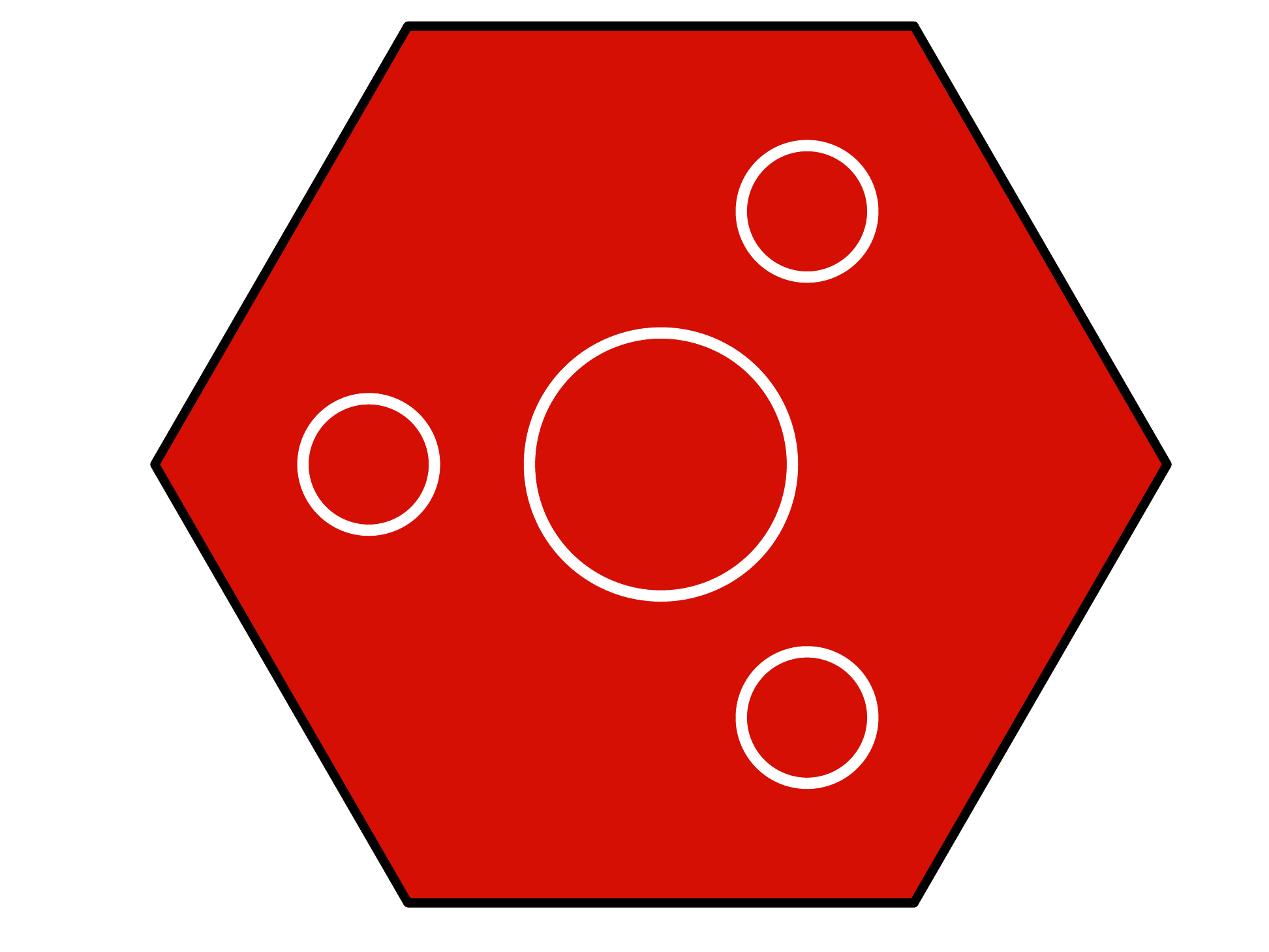}};
\draw (14+6 -1.5, -8+0.75) node[inner sep=0] {\includegraphics[scale=0.35]{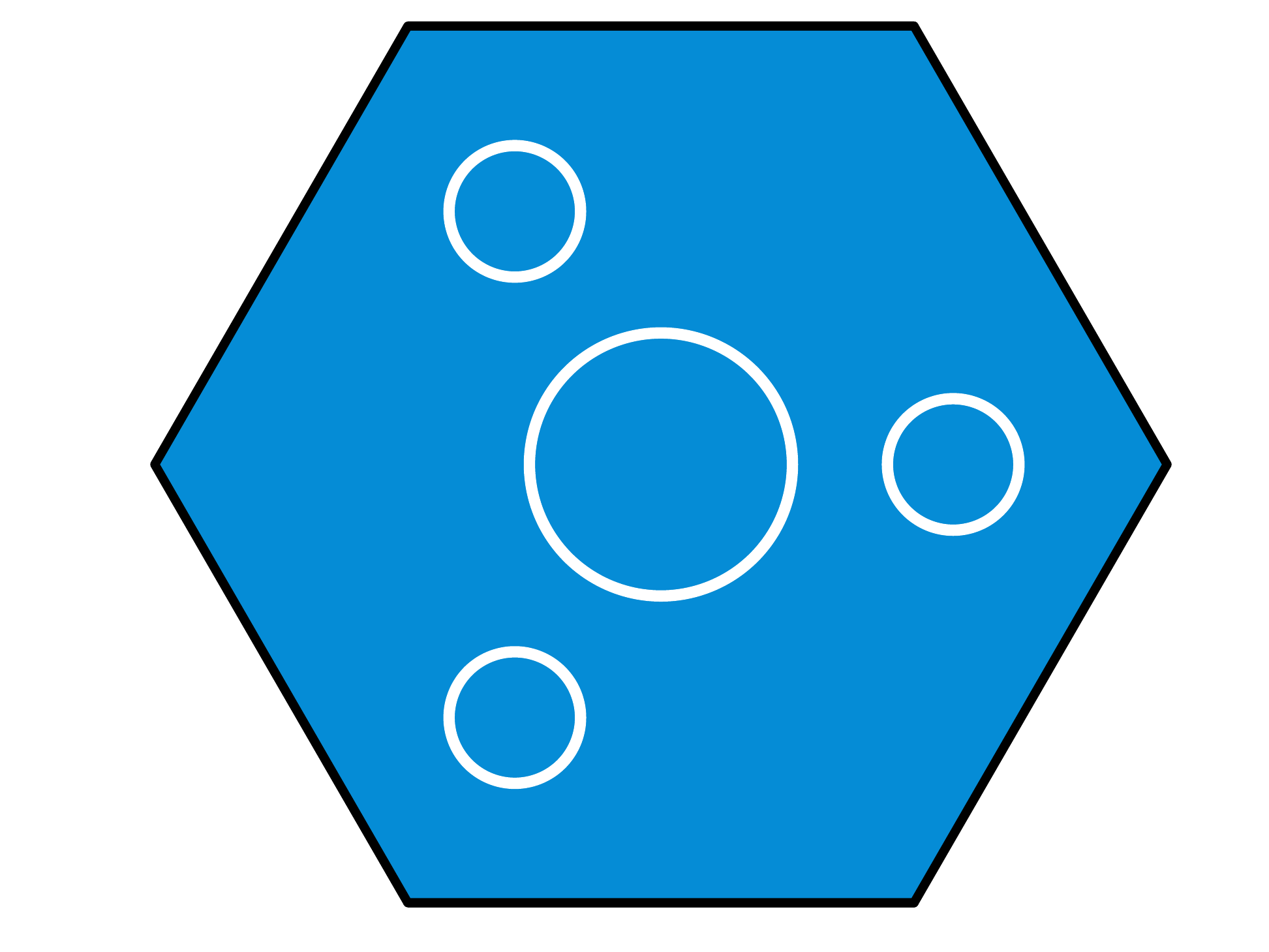}};
\end{tikzpicture}
\caption{\textcolor{black}{The dispersion diagram and eigensolutions (generated from (\ref{HelmholtzHardScheme_1})) throughout the singly periodic infinite ribbon - here $\boldsymbol{\alpha}_{1} = L( \cos \frac{\pi}{6}, \sin \frac{\pi}{6})$,  $\boldsymbol{\alpha}_{2} = ( 0, \infty)$ and $L=1$. The fundamental (super)cell here is built by stacking primitive cells (with base vectors $\boldsymbol{\alpha}_{1} = L( \cos \frac{\pi}{6}, \sin \frac{\pi}{6})$, $\boldsymbol{\alpha}_{2} = L( 0, 1)$ with $L =1$) on top of one another. We stack two sets of primative cells, $\pm \pi/6$ rotations of the arrangement Fig. \ref{HelmholtzNeumannOurArrangement}$(i)$, colour coordinated in the lower central panel. We create two media,  medium 1 and medium 2 built from stacking blue and red cells respectively. Further stacking medium 1 on top of medium 2 (or vice versa) creates the infinite ribbon and a distinct interface connecting the two media. The grey section represents the band gap in Fig. \ref{HelmholtzNeumannOurArrangementPerturbed}, where we observe two interfacial modes plotted in purple and green. These interfacial modes,  or edge states,} coexist for a certain range of frequencies. The wave field $\phi$ has been plotted for the edge states, within the extremities of the figure, next to the arrangement (medium 1 over 2 or 2 over 1) in which the edge state resides. The leftward and rightward wavefields are that of the green and purple interfacial modes respectively - specifically at the crosses where $\Omega = 3.73$ \textcolor{black}{(at the intersecting dashed line)}.} 
\label{RibbonDisp}
\end{figure}

{\color{black} The generation of symmetry induced topological modes is reliant upon there being a pair of time-reversal symmetric Dirac cones (see lowest two bands  in Fig.  \ref{HelmholtzNeumannOurArrangement}(iii)) that are well separated in Fourier space.} Upon symmetry reduction of the cellular structure, {\color{black} and by rotating the Neumann inclusion set}, we obtain the dispersion curves shown in Fig.   \ref{HelmholtzNeumannOurArrangementPerturbed}(iii). Notably, the energy fluxes of the modes that demarcate the band gap have opposite chirality (Fig. \ref{HelmholtzNeumannOurArrangementPerturbed}(iv, v)) and it is precisely this property that imbues the ensuing edge modes with their protective property {\color{black} \cite{Wong_Kagome_2020}}. The two distinct interfaces, that are constructable using the cells in Fig. \ref{HelmholtzNeumannOurArrangementPerturbed}(iv, v), are shown in Fig.  \ref{RibbonDisp}. {\color{black} The eigensolutions for the perturbed system, Fig.  \ref{HelmholtzNeumannOurArrangementPerturbed}(iv, v), are related to the ribbon edge modes in Fig.  \ref{RibbonDisp} via the zone-folding bulk-boundary correspondence \cite{qian_theory_2018}; this states that if the two media, either side of an interface, are related by mirror symmetry then we are guaranteed a pair of counterpropagating modes for both stackings (i.e. positively rotated set of Neumann inclusions over a negatively rotated set and vice versa).}

\begin{figure}[h!]
\begin{subfigure}[t]{.35\textwidth}
\centering
\end{subfigure} $\quad \quad \quad$
\begin{subfigure}[t]{.35\textwidth}
\begin{tikzpicture}[scale=0.3, transform shape]
\draw (-10.5 -7.5+ 21 - 10, 0.35-3.1) node[inner sep=0] {\includegraphics[scale=1.0]{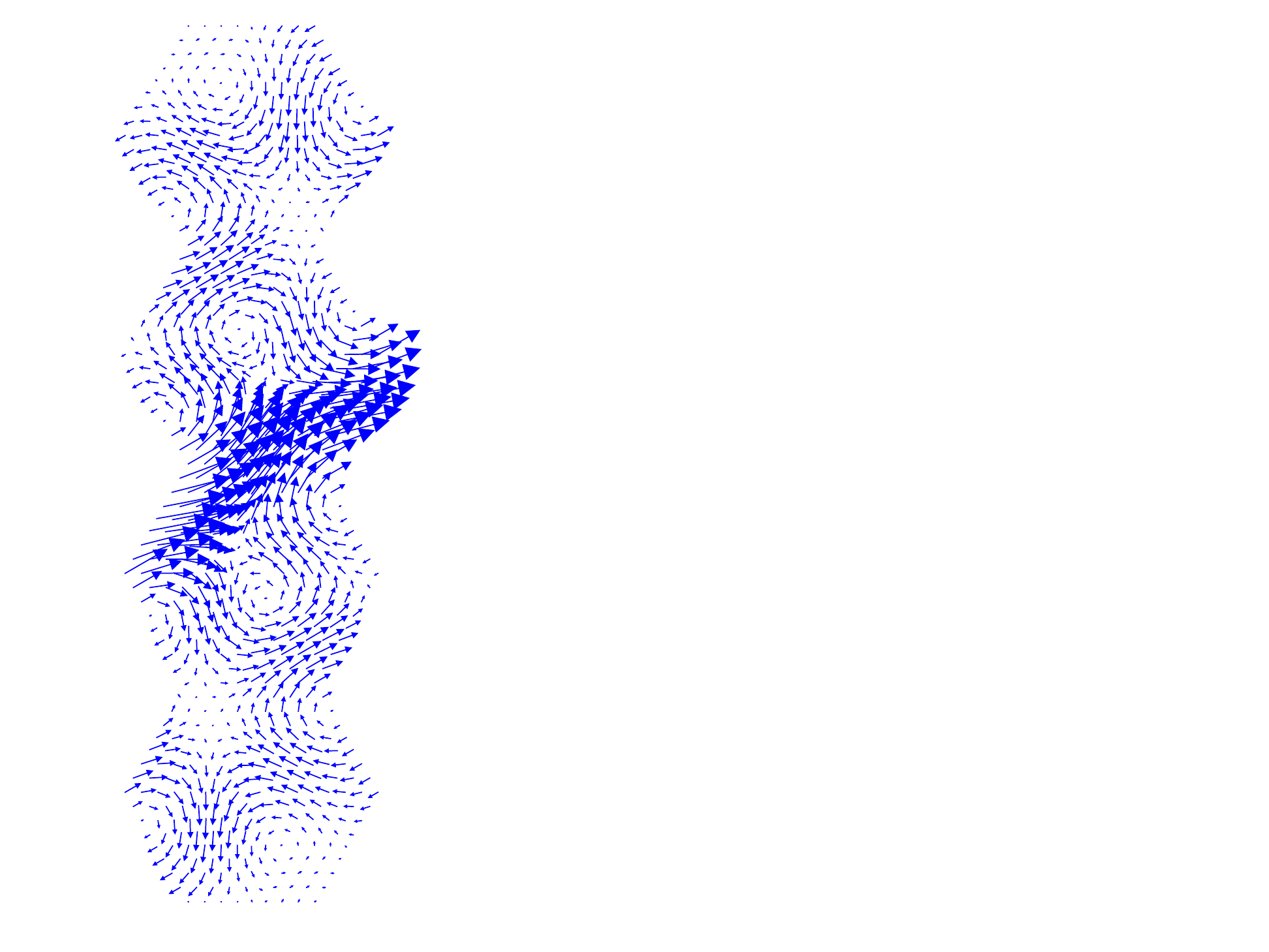}};
\draw (-10.5 + 21 - 10, 0.35-3.1) node[inner sep=0] {\includegraphics[scale=1.0]{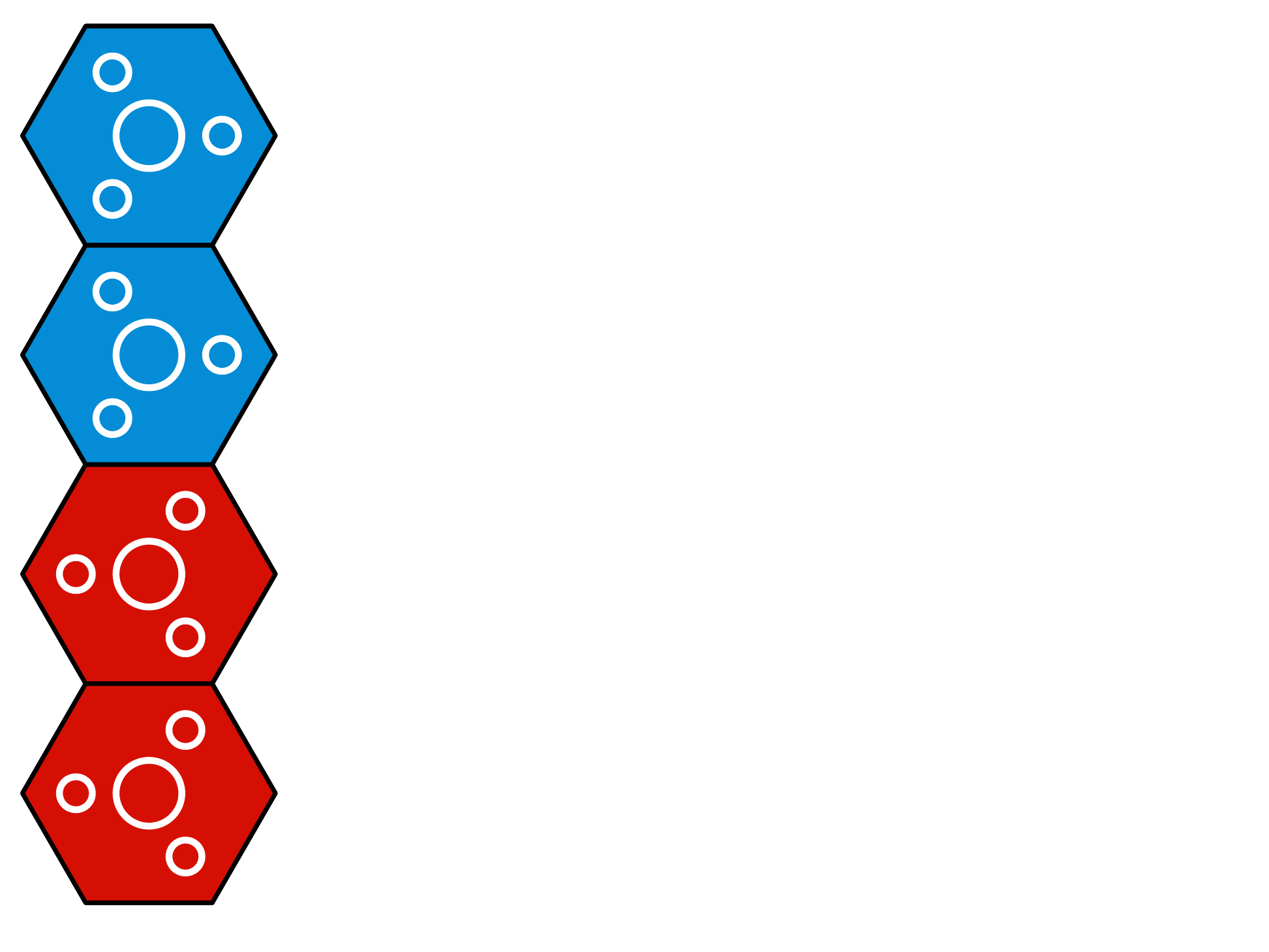}};

\draw (10.5, 0.35-3.1) node[inner sep=0] {\includegraphics[scale=1.0]{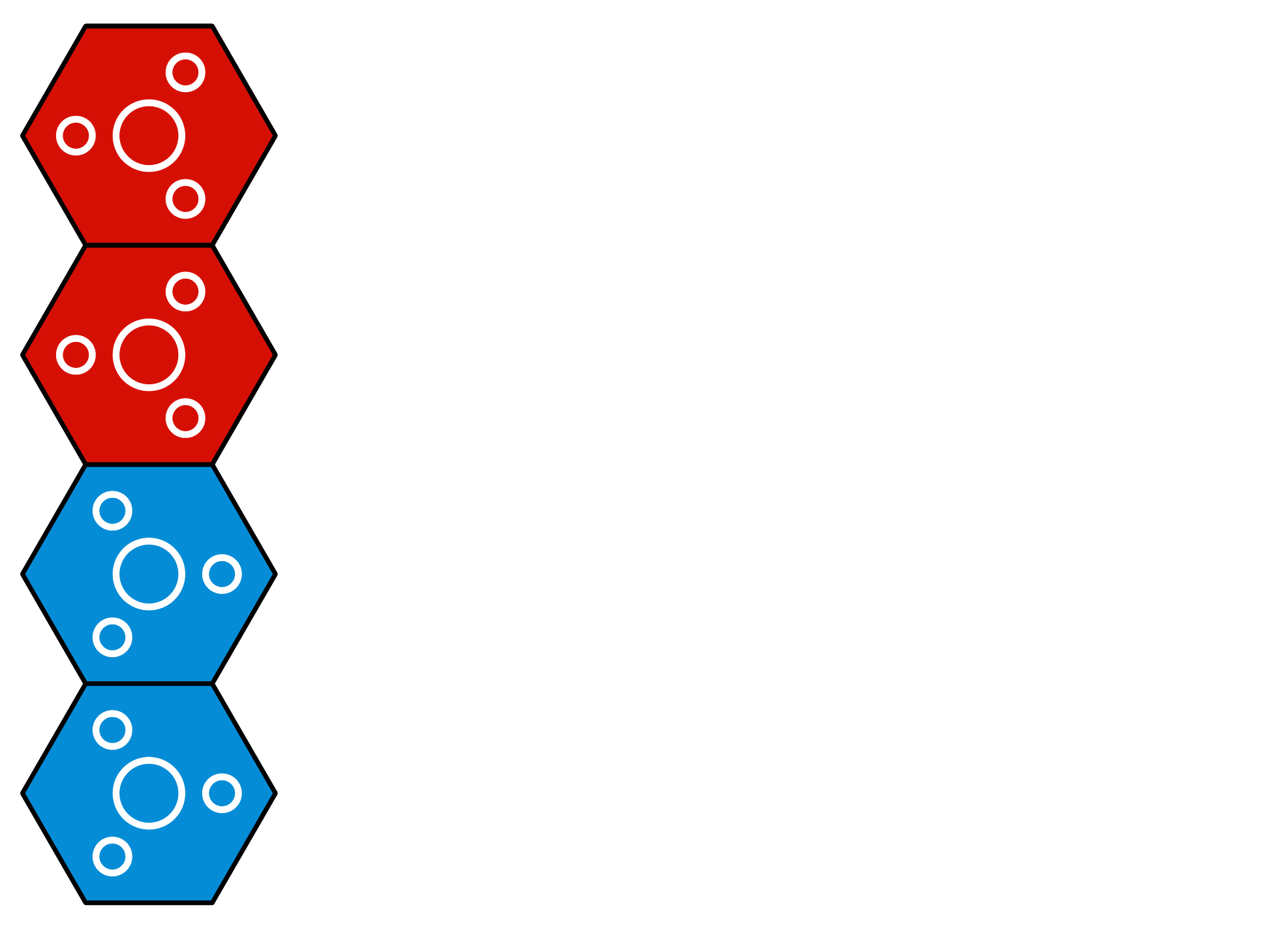}};
\draw (10.5 +4.5, 0.35-3.1) node[inner sep=0] {\includegraphics[scale=1.0]{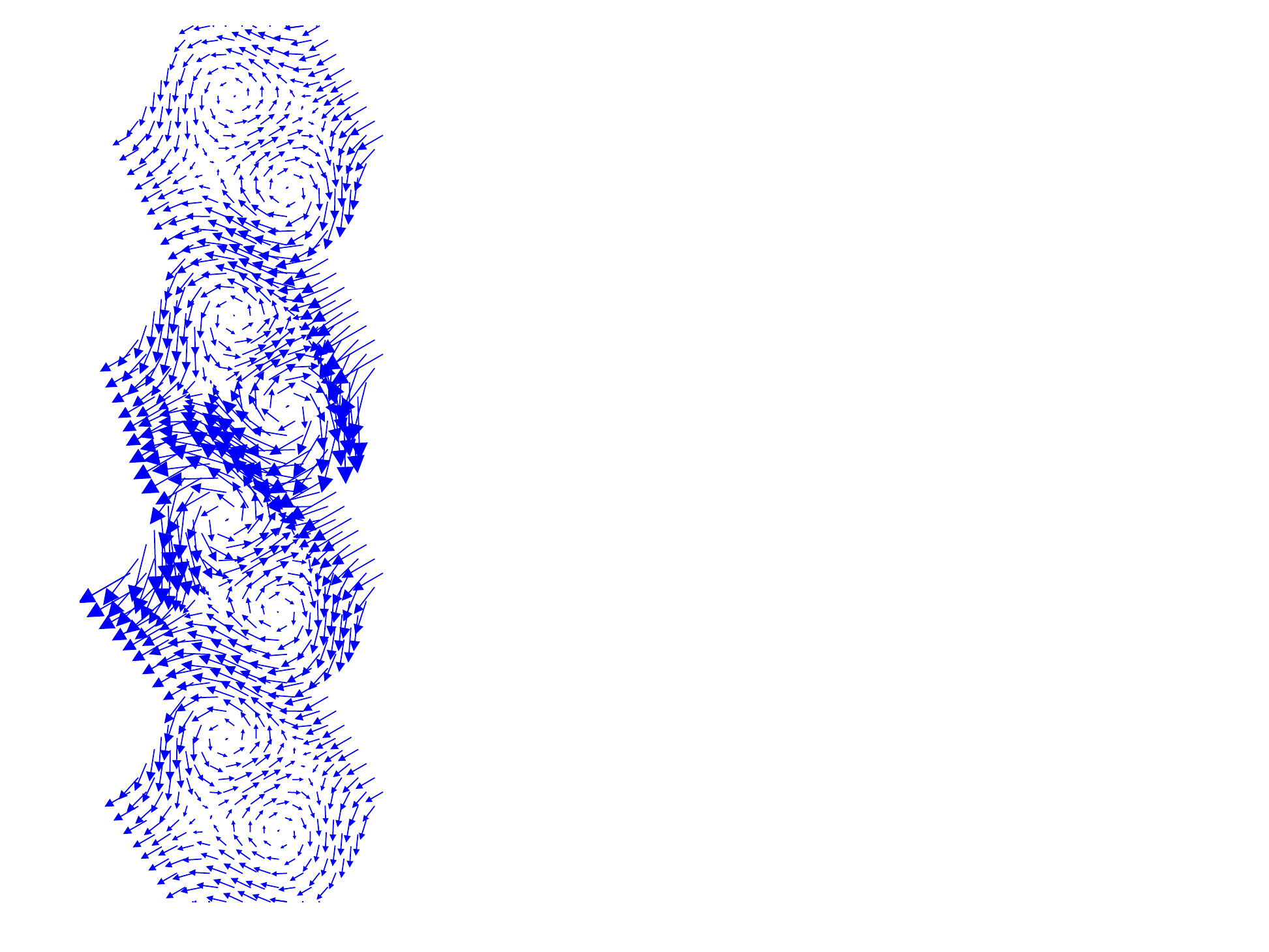}};
\end{tikzpicture}
\end{subfigure}
\caption{Fluxes corresponding to \textcolor{black}{zoomed in sections of the interfacial modes} shown in Fig.  \ref{RibbonDisp}. Rightward ($+\kappa$) and leftward ($-\kappa$) propagating modes for both of the geometrically distinct interfaces are shown.}

\label{FluxyMcFlux}
\end{figure}

The resulting pair of concave and convex dispersion curves (figure \ref{RibbonDisp}) yield modes that are of either even or odd-parity, and are hence physically distinguishable. The coupling between the even and odd-parity modes around different angled bends has been explored in \cite{makwana18a, tang19} and also in the context of more complicated topological domains in \cite{makwana18b}. The time-averaged energy flux of the two distinct edge modes is shown in Fig.  \ref{FluxyMcFlux} where the accuracy of our numerical scheme is exemplified by the clarity of the orbital motion in these figures. The right/left propagating modes shown are often said to have right/left chiral pseudospins. The near orthogonality of the forward and backwards propagating pseudospin modes (Fig.  \ref{FluxyMcFlux}) is inherited from the bulk solutions (Fig.  \ref{HelmholtzNeumannOurArrangementPerturbed}) via the bulk-boundary correspondence \cite{qian_theory_2018}. Protection against backscattering depends upon the orthogonality of these opposite pseudospin states and it has been shown to be approximately valid for small band gaps \cite{fefferman_honeycomb_2012}.

\subsection{Modal coupling between topologically distinct domains}

\begin{figure}[h!]
\centering
\begin{tikzpicture}[scale=0.315, transform shape]
\draw (0.75+2, 0.75) node[inner sep=0] {\includegraphics[scale=1.1]{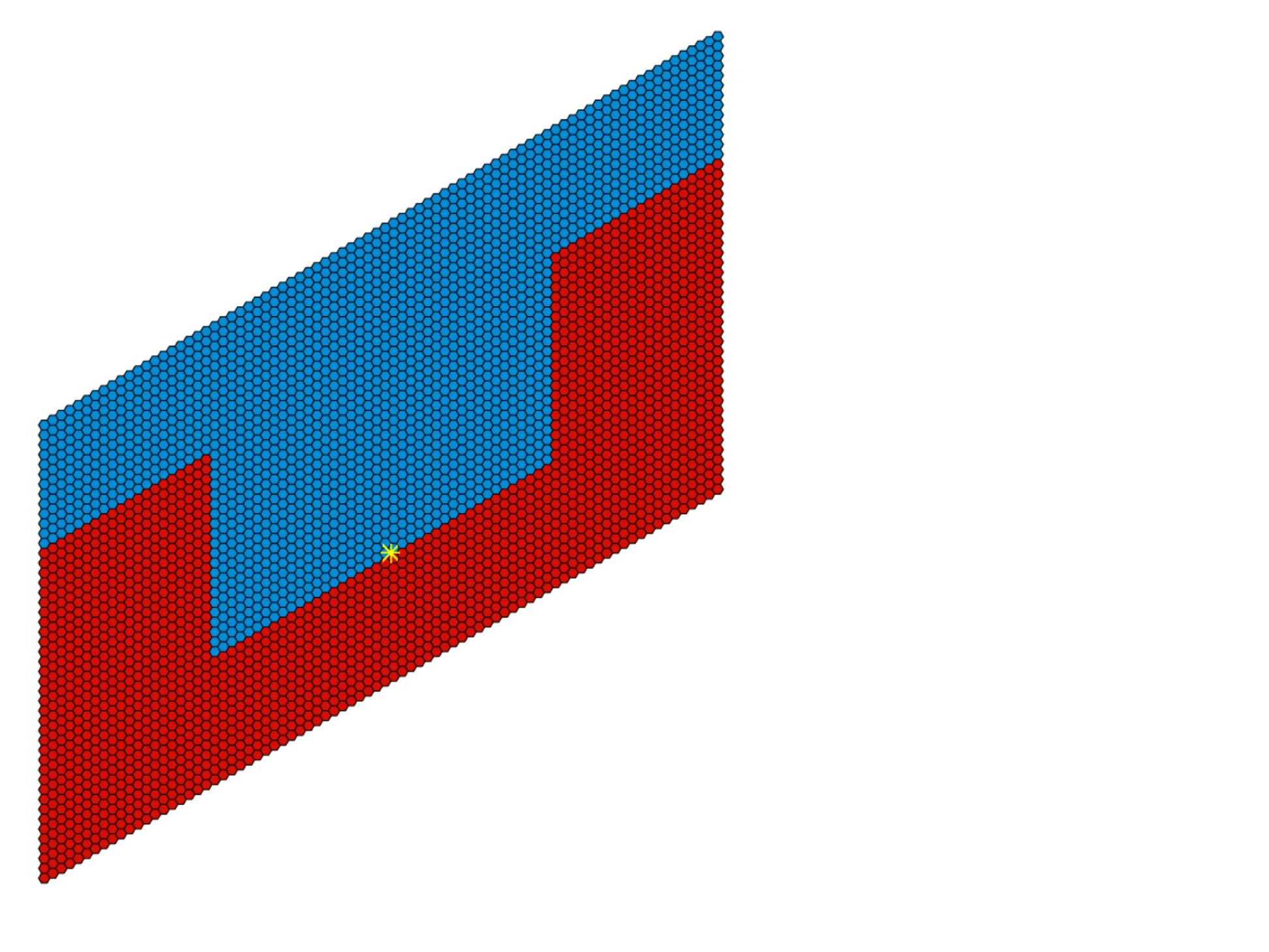}};

\draw (-1.10, -14.8-0.80) node[inner sep=0] {\includegraphics[scale=0.510]{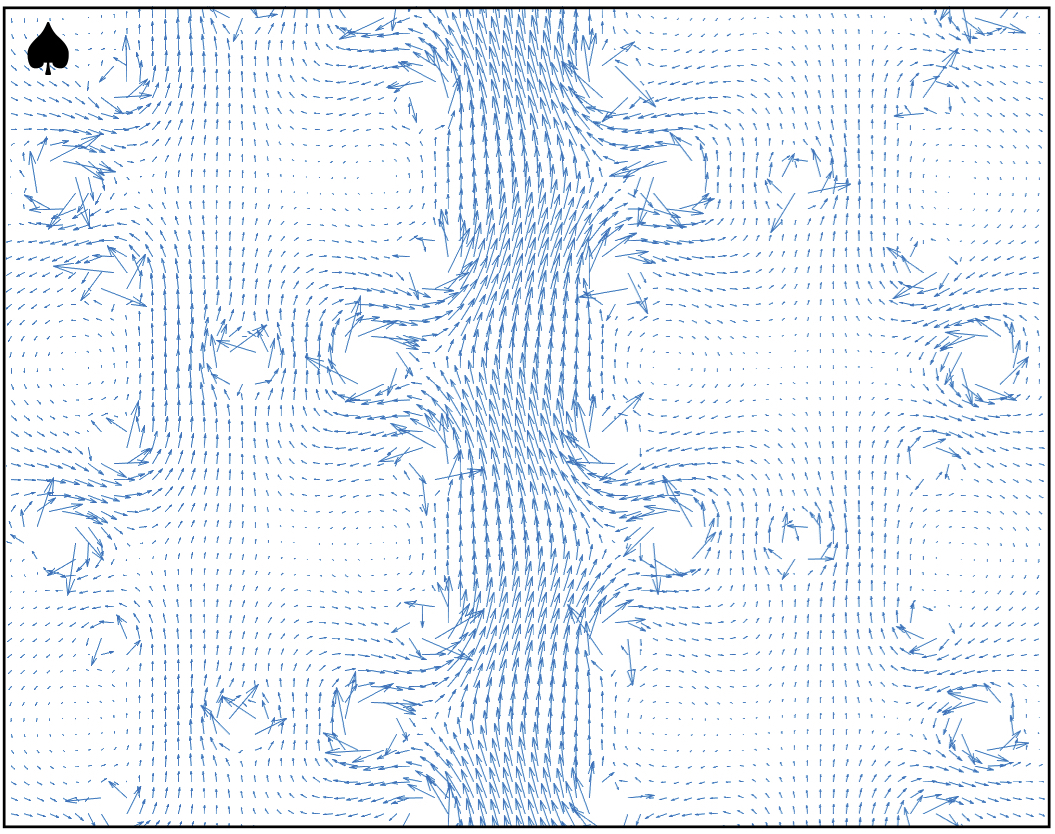}};
\draw (18.05, -14.8-0.8) node[inner sep=0] {\includegraphics[scale=0.515]{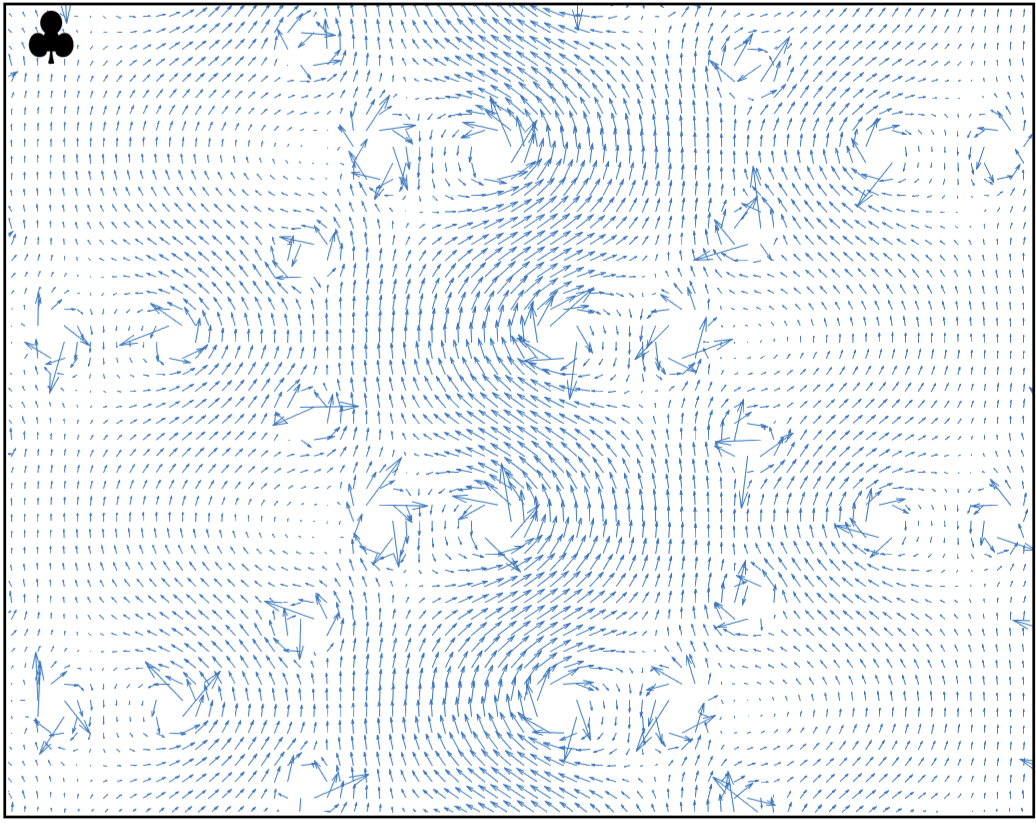}};
\draw (18.5+1.25, 0.75) node[inner sep=0] {\includegraphics[scale=1.175]{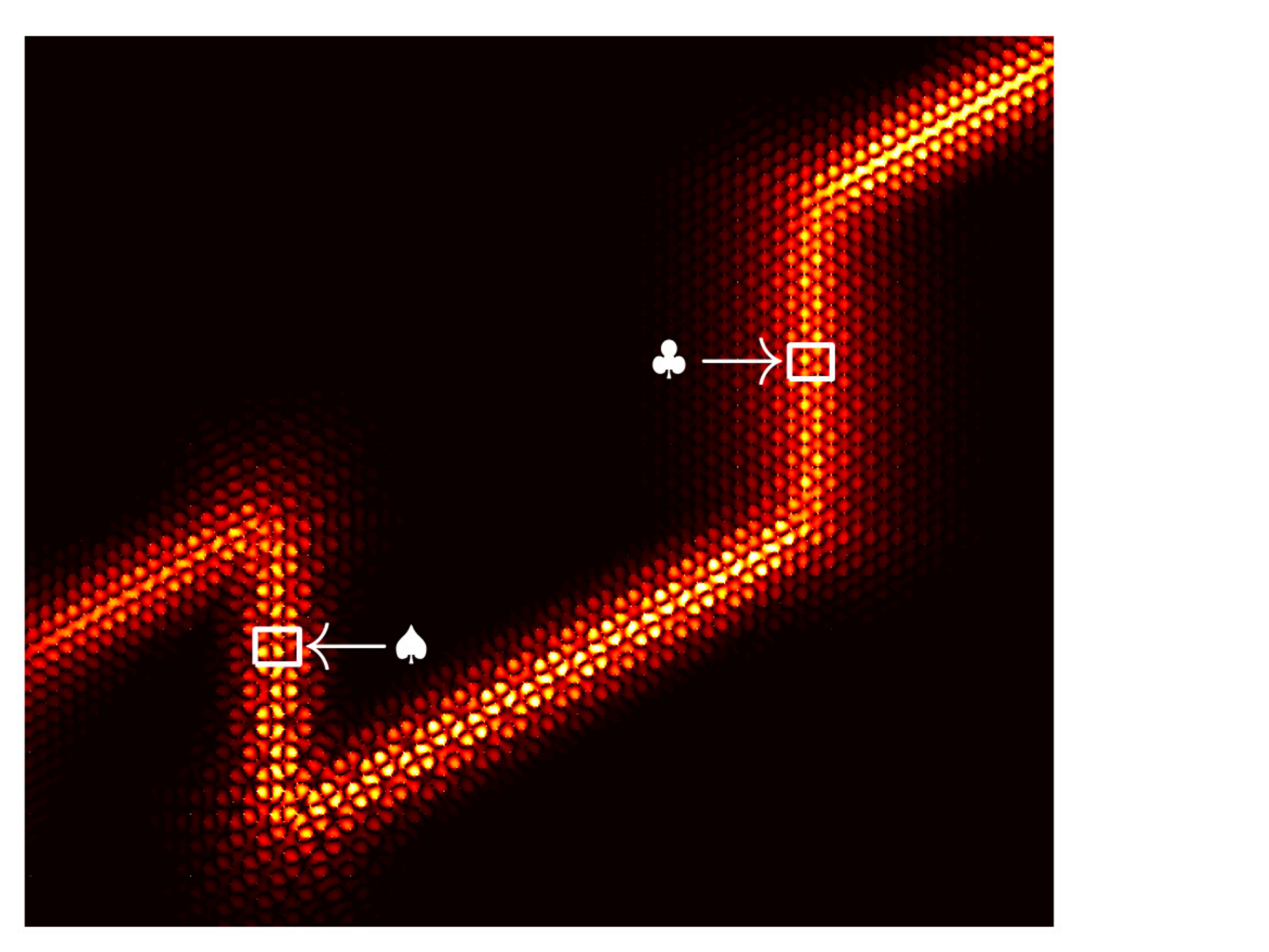}};
\end{tikzpicture}
\caption{Double-Z schematic containing 3760 cells (half blue {\color{black}the other} red) forming a structure with a total of 15040 scatterers (top left). The homogeneous-Foldy solution (top right), with enlarged sections $\spadesuit$ and $\clubsuit$ {\color{black}in which we plot the associated flux (blue arrows). Sections $\spadesuit$ and $\clubsuit$ show the different interfacial modes (same frequency different $\boldsymbol{\kappa}$) as in Fig.  \ref{FluxyMcFlux}. In this simulation $\Omega = 3.73$, and the corresponding singular value was $0.0013$; small enough to provide a good approximation to the homogeneous-Foldy solution}.}
\label{SchematicDZwithEigenFoldy}
\end{figure}

To demonstrate the utility of our {\color{black}generalised} Foldy (section \ref{Sec:solnInPhysGen}), and the {\color{black}unforced (homogeneous) Foldy solutions} we show how the
distinct modes in Fig.  \ref{RibbonDisp} couple around different angled bends. Transport of energy around corners in structured media is of inherent interest across wave physics \cite{mekis_high_1996, chutinan_wider_2002, ma_guiding_2015}. The modal symmetries are indispensable for determining whether or not energy will couple around a bend or along parallel interfaces \cite{makwana18b, tang19}. The majority of the valley-Hall literature, to name but a few \cite{zhang_topological_2018, zhang_manipulation_2018, Shalaev:18, lu_observation_2016, liu_experimental_2018, jung_active_2018, gao_valley_2017, chen_tunable_2018}, uses a Z-shaped interface to demonstrate robustness of the modes. However this design, that solely contains $\pi/3$ bends, does not result in modal conversion between the even and odd-parity edge modes (Fig.  \ref{RibbonDisp}). To clearly demonstrate  both the modal conversion and modal preservation cases, by the use of one all encompassing figure, we use a double Z configuration (Fig.  \ref{SchematicDZwithEigenFoldy}). The displacement pattern, shown here, uses the homogeneous-Foldy method. 
 Interestingly it is solely along the gentle $2\pi/3$ bend  ($\clubsuit$) in which the edge state undergoes modal conversion; along the left-sided interface ($\spadesuit$) there is modal preservation as the energy traverses a $\pi/3$ bend. 
The modal differences between the edge states along the two vertical interfaces ($\clubsuit$ and $\spadesuit$) is further exemplified by the fluxes shown. 

\begin{figure}[h!]
\centering
$\quad \quad \quad \quad$
\begin{tikzpicture}[scale=0.5, transform shape]
\draw (0, 0) node[inner sep=0] {\includegraphics[scale=0.40]{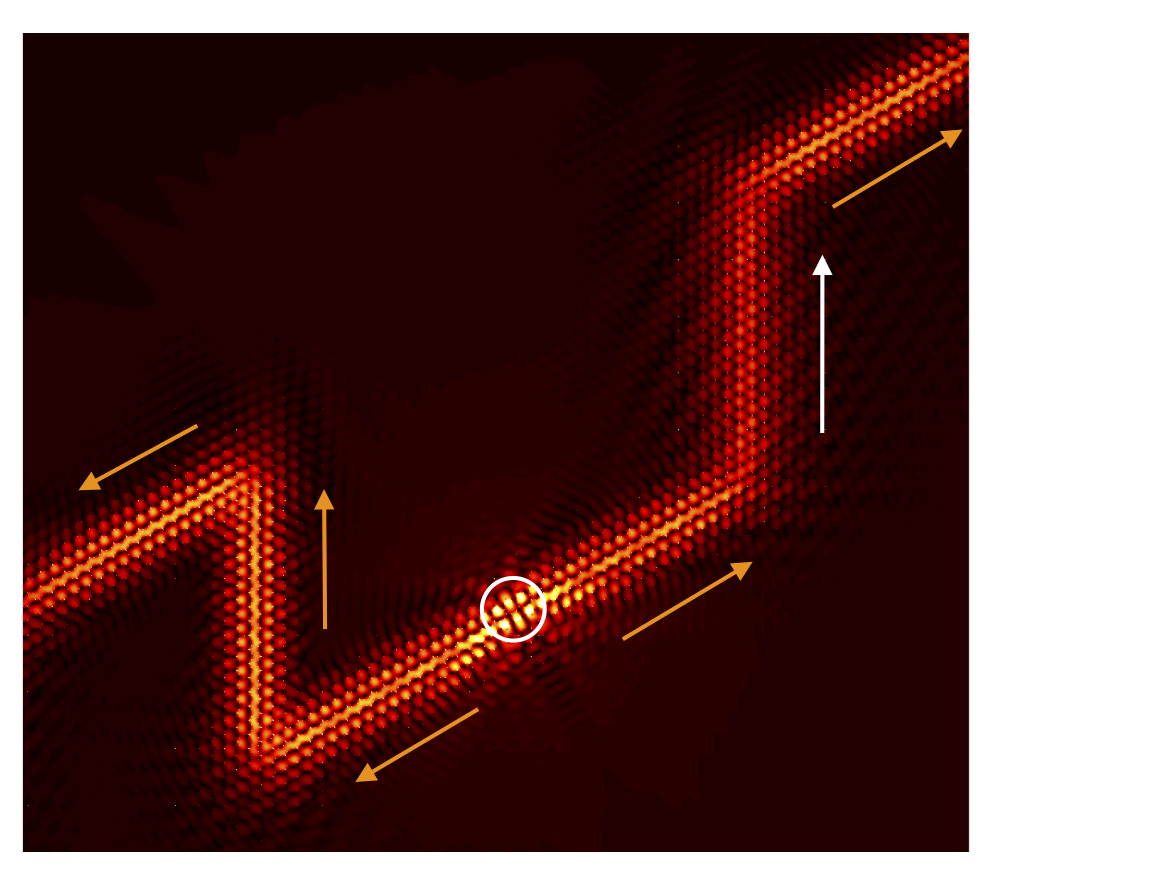}};
\end{tikzpicture}
\caption{The scattered field computed from the generalised Foldy scheme for the example shown in Fig.  \ref{SchematicDZwithEigenFoldy}. \textcolor{black}{A isotropic (monopole) incident source is }placed within the white circle igniting a leftward and rightward propagating even-parity edge mode. The white arrow indicates the sole interface that hosts an odd-parity state; again $\Omega = 3.73$.} 
\label{Gen-DZ-Foldy}
\end{figure}

For completeness, we also perform the {\color{black}conventional} Foldy scattering calculation (section \ref{Sec:solnInPhysGen}) with line source excitation which generates Fig.  \ref{Gen-DZ-Foldy}; this scattering solution mirrors the homogeneous-Foldy solution shown in Fig.  \ref{SchematicDZwithEigenFoldy}. The examples in this subsection demonstrate how our semianalytic expressions allow us to obtain highly resolved and precise edge states. The clarity of the solutions obtained is of paramount importance as they allow us to interpret the relative interface orientations with ease. 

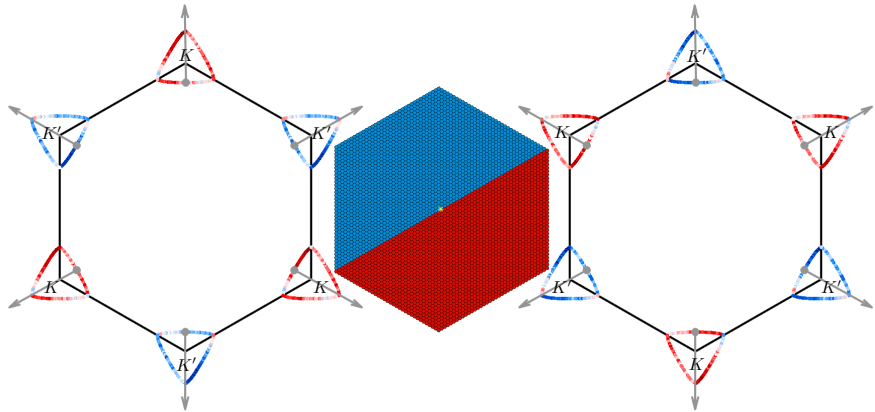
\begin{figure}[h!]
\centering
\begin{tikzpicture}[scale=0.65, transform shape]
    \begin{scope}[xshift=0, yshift=0]
    \newdimen\R
	\R=3.15cm 
    	\draw (1.9, -0.1) node[inner sep=0] {\includegraphics[scale=0.6]{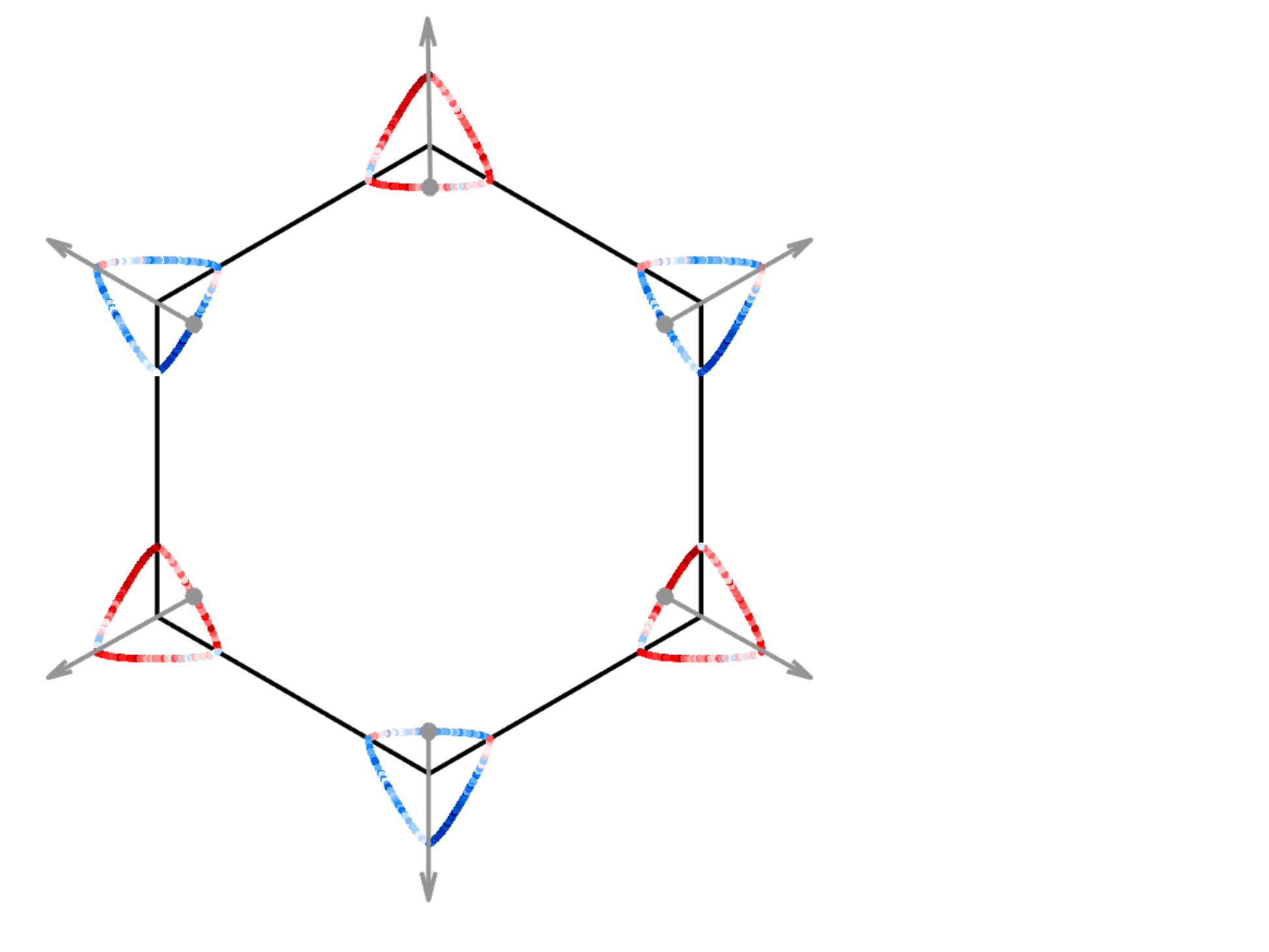}};
        \draw (30:\R) \foreach \x in {30,150,270} { 
                --cycle (\x:\R) node[] {$K'$} 
            } ;
            \draw (90:\R) \foreach \x in {90,210,330} { 
                --cycle (\x:\R) node[] {$K$} 
            } ;
     \end{scope}   
     
     \begin{scope}[xshift=10.33cm, yshift=0]
    \newdimen\R
	\R=3.15cm 
    	\draw (1.9, -0.1) node[inner sep=0] {\includegraphics[scale=0.6]{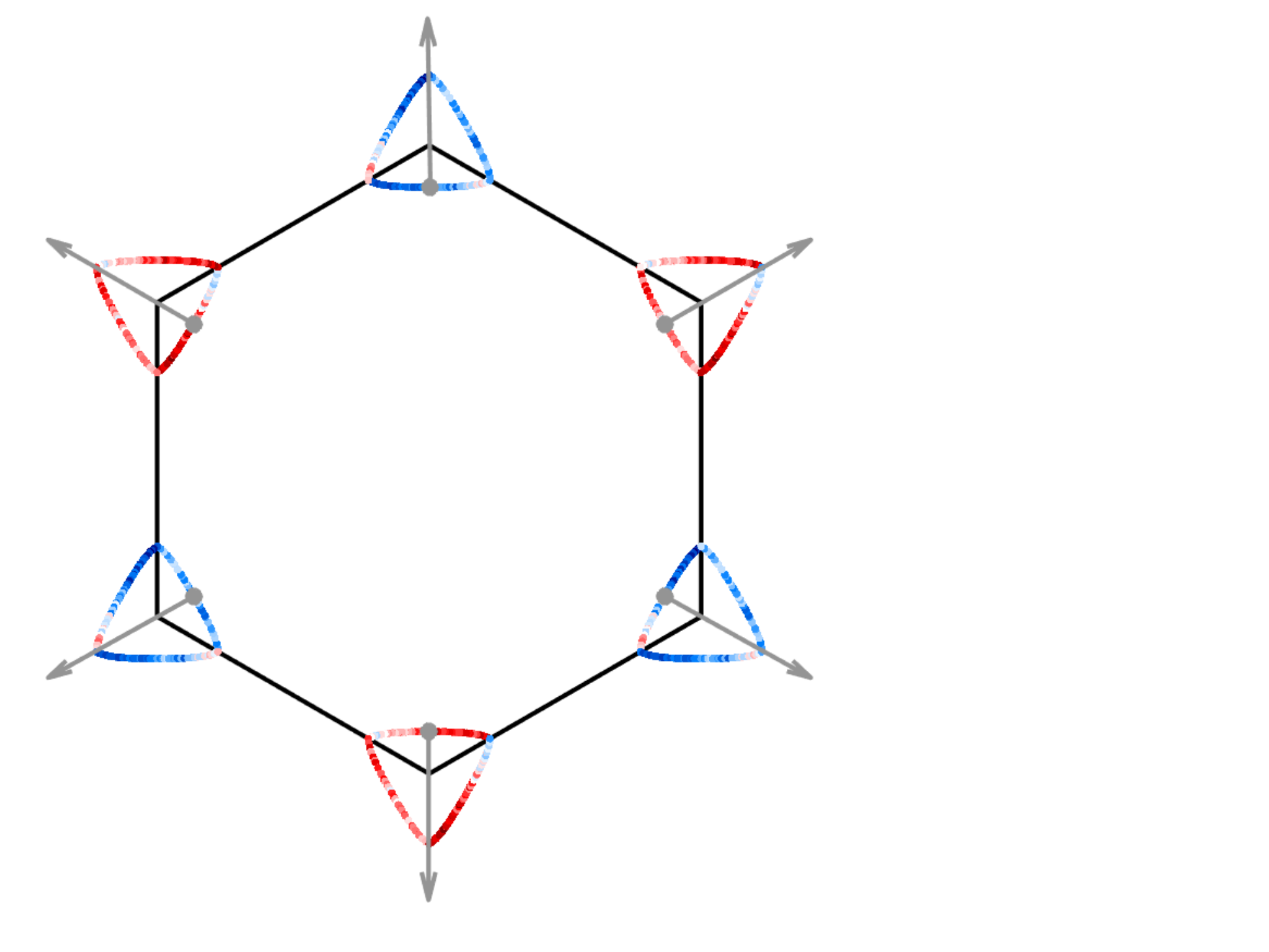}};
        \draw (30:\R) \foreach \x in {30,150,270} { 
                --cycle (\x:\R) node[] {$K$} 
            } ;
            \draw (90:\R) \foreach \x in {90,210,330} { 
                --cycle (\x:\R) node[] {$K'$} 
            } ;
     \end{scope} 
     
     \begin{scope}[xshift=5.15cm, yshift=0]
		\node[regular polygon, regular polygon sides=6,draw, inner sep=1.565cm,rotate=90,line width=0.0mm, white,
           path picture={
               \node[rotate=-90] at (0.24,-1.52){
                   \includegraphics[scale=0.38]{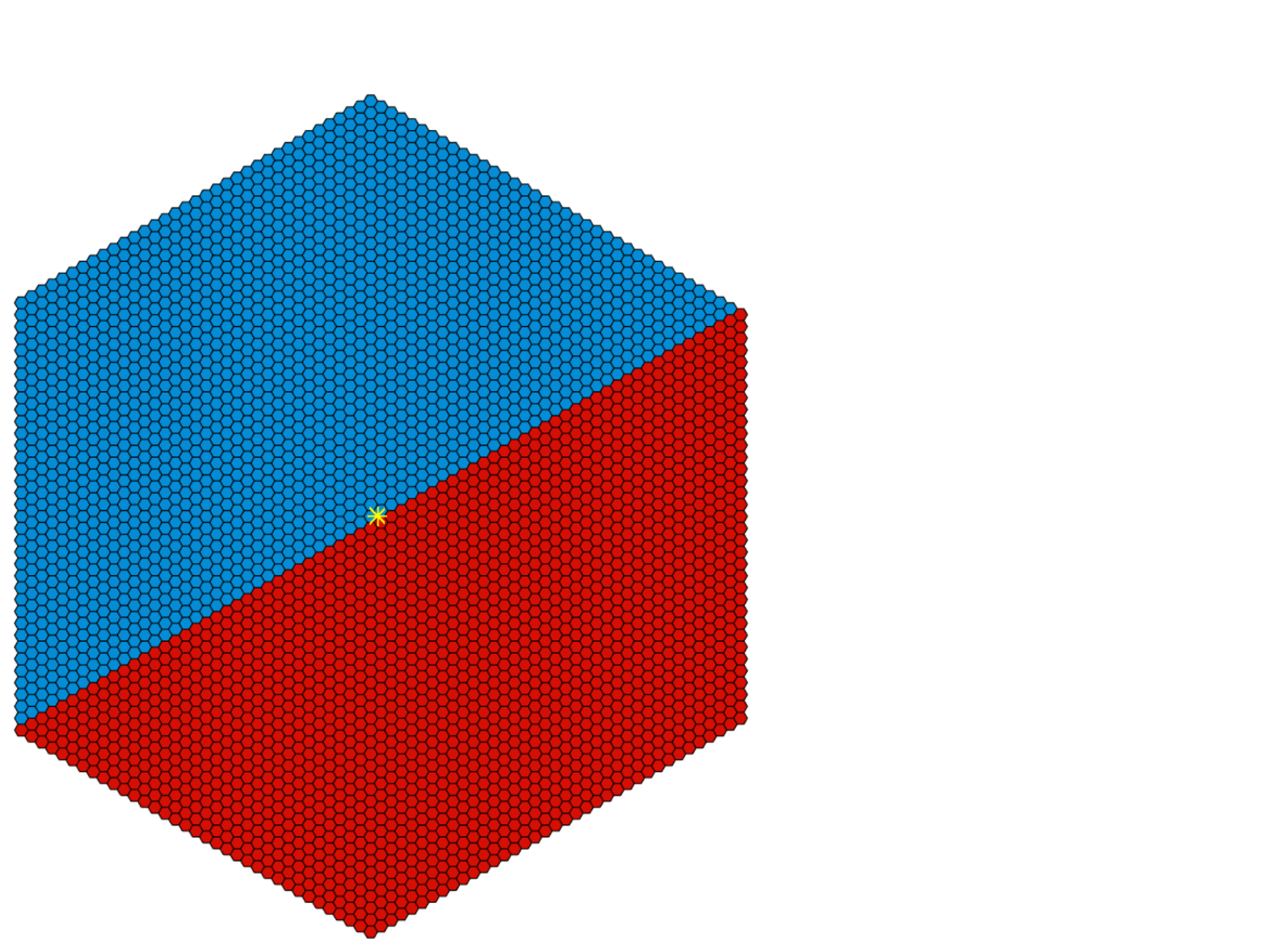}
               };
           }]{};
     \end{scope}         
\end{tikzpicture}
\caption{Isofrequency contours, {\color{black}from (\ref{HelmholtzHardScheme_1}),} at $\Omega = 3.06$ for the blue medium 2 (left) and the red medium 1 (right). The red and blue colours of the contours correspond to +ve and -ve $Q$-factor values respectively; the modulus of $Q$ is proportional to the intensity of the colour. The grey arrows represent the favoured group velocity of waves excited by an isotropic source. The schematic (middle) contains 3780 cells, half blue (medium 2) and half red (medium 1), forming a structure with a total of 15120 scatterers. An isotropic source is positioned at the center of the interface (yellow point) and excited at $\Omega = 3.06$ to yield the chiral beaming phenomena as shown in Fig.  \ref{ChiralBeamSchemFoldyIsoFrequencyScatter}.  } 
\label{ChiralBeamSchemFoldyIsoFrequencyContour}
\end{figure}

\begin{figure}[h!]
\centering
\begin{tikzpicture}[scale=0.5, transform shape]
\draw (-3.5, 0) node[inner sep=0] {\includegraphics[scale=0.5]{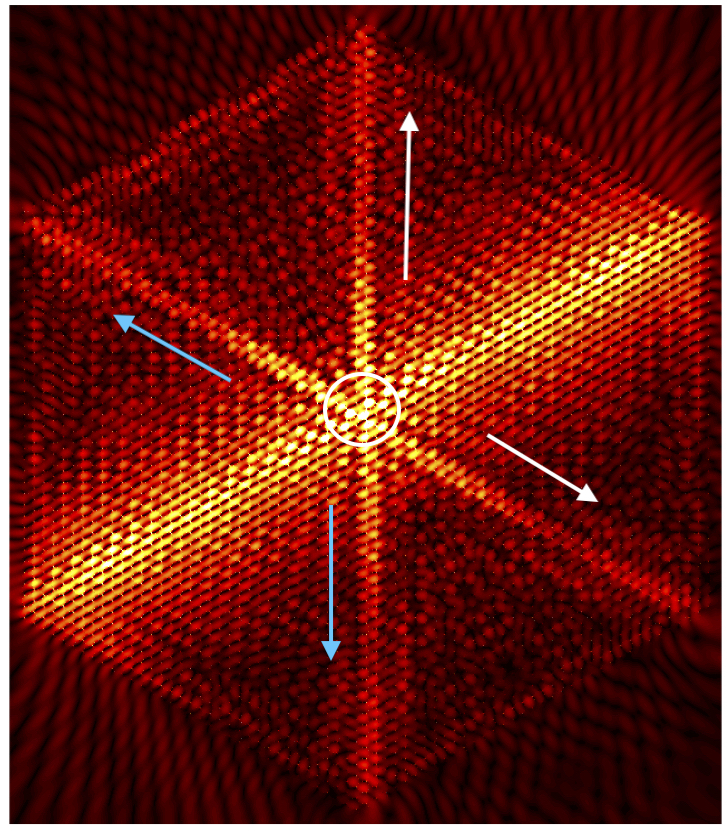}};
\end{tikzpicture}
\caption{The scattered field for the system whose schematic is shown in Fig.  \ref{ChiralBeamSchemFoldyIsoFrequencyContour}; calculated using the extended Foldy's method (section \ref{Sec:solnInPhysGen}). The anisotropic behaviour and favoured pseudospins are explained by examining the isofrequency contours and the $Q-$factors in the vicinity of the $KK'$ valleys (Fig.  \ref{ChiralBeamSchemFoldyIsoFrequencyContour}). 
} 
\label{ChiralBeamSchemFoldyIsoFrequencyScatter}
\end{figure}

\subsection{Chiral beaming in the propagating regime}
The modal conversion and preservation effect outlined in the preceding section occurs within the bulk band-gap frequency range (Fig.  \ref{RibbonDisp}). Contrastingly, in this subsection we operate within the propagating regime of the bulk. We use our succinct asymptotic formulae to show how opposite pseudospin modes are separated using a carefully placed isotropic source. This effect is more commonly referred to as chiral beaming and, ordinarily, the source is placed along the boundary between a topologically nontrivial domain and a homogeneous region \cite{lu_valley_2016}. Here, we opt to place our source along the interface between two topologically distinct domains, as shown by the schematic in Fig.  \ref{ChiralBeamSchemFoldyIsoFrequencyContour}. We clearly see two highly localised pulses beam into the upper (blue) and lower (red) domains in Fig.  \ref{ChiralBeamSchemFoldyIsoFrequencyScatter}. The angular difference between the pulses, within either the blue or red domain, is attributed to the $\pi/3$ rotational difference between the $K$ and $K'$ isofrequency contours; the contours for the upper (and lower) domain, superimposed onto the Brillouin zone, are shown in the leftmost and rightmost panels of Fig.  \ref{ChiralBeamSchemFoldyIsoFrequencyContour}. We also oust the favoured pseudospins by calculating the $Q-$factor \cite{lu_valley_2016, zhang_chiral_2017},
\begin{equation}
    Q \textbf{e}_{z} = \int_{\partial \Upsilon} \nabla \times \left< \textbf{F} \right> \, dS, 
\label{Q_factor}\end{equation}
where $\partial \Upsilon$ denotes the surface of a primitive cell and $Q$ represents the ``circulation" of the time-averaged flux. The $Q-$factor is indicated in Fig.  \ref{ChiralBeamSchemFoldyIsoFrequencyContour} by the colours of the contours. Interestingly, in the upper medium a right (left) pseudospin mode, associated with $K (K')$ beams off to the right (left) whilst in the lower region the pseudospin directions are switched. This is due to the $K$ ($K'$) mode for the upper medium being endowed with a positive (negative) pseudospin whilst for the lower medium the pseudospin polarisation is flipped. 
The modes with common chirality or pseudospin are indicated by identically coloured arrows in Fig.  \ref{ChiralBeamSchemFoldyIsoFrequencyScatter}. This chiral beaming phenomenon occurs near the standing wave frequency that demarcates the band gap and hence residual energy associated with the edge state is also shown to propagate along the interface.

\section{Concluding remarks}
\label{sec:conclude} 
{\color{black} We have designed semi-analytical schemes solving various propagation problems due to arrangements of small Neumann inclusions within a Helmholtz wavefield. Analytical solutions are singular approaching the center of each inclusion, however we apply the method of matched asymptotics to account for any spatial singularities present; two schemes follow the analysis, those of section \ref{sec:solnrecpsp} (matching inner to a divergent Fourier series) and section \ref{Sec:solnInPhysGen} (matching inner to singular Green's functions). The schemes work since the matched quantities in the neighbourhood of inclusions are equivalent, thus singularities in both must cancel.

The numerical schemes are highly efficient, accurate and rapid. Those in section \ref{sec:solnrecpsp} determine the eigensolutions and dispersion diagrams of the wavefield through a fundamental cell, containing an arbitrary arrangment of inclusions, within periodic media. The rapid scheme determines solutions in seconds, expediting the process of tailoring arrangements to create sought after dispersive properties. Subsequently we design photonic crystals giving us real control over how energy will propagate through the structure (examples were presented in section \ref{sec:topological}).   

The second scheme, in section \ref{Sec:solnInPhysGen}, allows us to test our designs. This scheme, a generalized Foldy approach, considers the interaction between a finite collection of scatterers and an incident source of energy (forced problems). The periodic assumption here is dropped in Foldy's method, however we can build photonic crystals from large (but finite) periodic collections of scatterers; this allows us to test our designs from section \ref{sec:solnrecpsp} by performing numerical experiments as in Figs. \ref{Gen-DZ-Foldy} and \ref{ChiralBeamSchemFoldyIsoFrequencyContour}.

We present a new twist to Foldy's method, applicable to similar numerical (FE) problems \footnote{Or any other analogous system of equations, going from inhomogenous to homogeneous systems of equations.}, by considering the solutions to the homogeneous (unforced) Foldy problem. Exact homogenous solutions generally do not exist, however we can construct approximate solutions through the singular value decomposition. The approximate solution will be a good approximation to the homogenous problem, provided it is built from a singular vector corresponding to a singular value of very small magnitude. The solution to the homogeneous Foldy problem visualizes dormant modes, in real space, awaiting excitation.} 


The accuracy of our asymptotically derived formulae against FE computations was quantitatively demonstrated in section \ref{sec:results}. Finally, we illustrated the efficacy of our asymptotic scheme by analysing, with a high-degree of precision, nontrivial phenomena associated with symmetry-induced topological edge states \textcolor{black}{associated with the photonic crystals design in section \ref{sec:topological}}. We anticipate that the versatile approach, justified via matched asymptotic expansions and presented herein, will allow large-scale computations to be performed with ease.

\section*{Acknowledgements}

The authors thank the UK EPSRC for their support through Programme grant EP/L024926/1, grant EP/T002654/1. R.W. acknowledges funding from the EPSRC Centre for Doctoral Training in Fluid Dynamics across Scales, reference EP/L016230/1. R.V.C acknowledges the support of the Leverhulme Trust and of the European Union FET Open, grant number 863179, Boheme. 

\begin{figure}[h!]
\centering
\begin{tikzpicture}
  \draw[->] (-1,0) -- (3,0) coordinate[label=below: $\displaystyle \xi_{1}$] (x);
\draw[->] (0,-1) -- (0,3) node[left] {$\displaystyle \xi_{2}$} ; 
\draw[thick,->] (0,0) -- (0.5,1.5);
\draw[thick,->] (0,0) -- (-0.75,0.75);
\draw[thick,->] (0,0) -- (2.0,0.55);
\draw[thick,->] (0,0) -- (2.5, 2.75 );
\draw[dashed] (2.0,0.55) -- (2.75, 0.55);
\node[above right] at (2.5,2.75) {$\displaystyle \textbf{x}$}; 
\node[below] at (2.0,0.55) {$\displaystyle \quad \textbf{X}_{IJ}$}; 
\node[left] at (2.25,1.6) {$\displaystyle \quad \quad r$}; 
\draw[-] (2.0,0.55) -- (2.5, 2.75 );

  \draw
  	(2.5,0) coordinate (a) node[above right] {}
     (0,0) coordinate (b) node[above right] {}
     (-0.75,0.75) coordinate (c) node[above left] {$\displaystyle \textbf{b}$}
    pic["$\displaystyle (2 \pi - \sigma) \quad \quad \quad \quad$", draw=black, <->, angle eccentricity=1.2, angle radius=0.5cm]
    {angle=c--b--a};
    
\draw
  	(2.5,0) coordinate (a) node[above right] {}
     (0,0) coordinate (b) node[above right] {}
     (0.5,1.5) coordinate (c) node[above] {$\displaystyle \, \, \, \boldsymbol{\xi}$}
    pic["$\displaystyle  \, \, \theta$", draw=black, <->, angle eccentricity=1.2, angle radius=0.750cm]
    {angle=a--b--c};    
    
\draw
  	(2.75,0.5) coordinate (a) node[above right] {}
     (2.0,0.55) coordinate (b) node[above right] {}
     (2.5,2.75) coordinate (c) node[above right] {}
    pic["$\, \, \displaystyle \varphi$", draw=black, <->, angle eccentricity=1.2, angle radius=0.50cm]
    {angle=a--b--c};  
    \node[below] at (0.75,-1.25) {$\displaystyle  (i)$}; 
\begin{scope}[shift = {(6,0)}]
 \draw[->] (-1,0) -- (3.5,0) coordinate[label=below: $\displaystyle G_{1}$] (x);
\draw[->] (0,-1) -- (0,3.5) node[left] {$\displaystyle G_{2}$} ; 
\draw[thick,->] (0,0) -- (0.5,1.5);
\draw[thick,->] (0,0) -- (-0.75,0.75);
\draw[thick,->] (0,0) -- (1.5,-0.75);
\draw[thick,->] (0,0) -- (2.0,0.55);
\draw[thick,->] (0,0) -- (2.5, 2.75 );
\draw[dashed] (2.0,0.55) -- (2.75, 0.55);
\node[above right] at (2.5,2.75) {$\displaystyle \textbf{x}$}; 
\node[below] at (2.0,0.55) {$\displaystyle \quad \textbf{X}_{11}$}; 
\node[left] at (2.25,1.6) {$\displaystyle \quad \quad r$}; 
\node[right] at (1.5,-0.75) {$\displaystyle \boldsymbol{\kappa}$}; 
\draw[-] (2.0,0.55) -- (2.5, 2.75 );

  \draw
  	(2.5,0) coordinate (a) node[above right] {}
     (0,0) coordinate (b) node[above right] {}
     (1.5,-0.75) coordinate (c) node[above left] {}
    pic["$\quad \quad \quad \, \, \displaystyle (2 \pi - \theta_{\kappa})$", draw=black, <->, angle eccentricity=1.2, angle radius=1.0cm]
    {angle=c--b--a};

 \draw
  	(2.5,0) coordinate (a) node[above right] {}
     (0,0) coordinate (b) node[above right] {}
     (-0.75,0.75) coordinate (c) node[above left] {$\displaystyle \textbf{b}$}
    pic["$\displaystyle (2 \pi - \sigma) \quad \quad \quad \quad$", draw=black, <->, angle eccentricity=1.2, angle radius=0.5cm]
    {angle=c--b--a};   
    
\draw
  	(2.5,0) coordinate (a) node[above right] {}
     (0,0) coordinate (b) node[above right] {}
     (0.5,1.5) coordinate (c) node[above] {$\displaystyle \, \, \, \textbf{G}$}
    pic["$\displaystyle  \, \, \, \, \theta_{G}$", draw=black, <->, angle eccentricity=1.2, angle radius=0.750cm]
    {angle=a--b--c};    
    
\draw
  	(2.75,0.5) coordinate (a) node[above right] {}
     (2.0,0.55) coordinate (b) node[above right] {}
     (2.5,2.75) coordinate (c) node[above right] {}
    pic["$\, \, \displaystyle \varphi$", draw=black, <->, angle eccentricity=1.2, angle radius=0.50cm]
    {angle=a--b--c}; 
        \node[below] at (0.75,-1.25) {$\displaystyle  (ii)$};  
\end{scope}          
\end{tikzpicture}
\caption{The required vector quantities in Fourier space to derive: \\
$(i) \, - $ The Green's functions (\ref{eqn::AssocGreens}), here $\boldsymbol{\xi}$ denotes the transform variable in Fourier space. \\ 
$(ii) - $ The residual portion of the Fourier series, that is $\phi_{\mathrm{res}}$, in (\ref{NAP::G>Rcont}). 
}
\label{Xixandb}
\end{figure}
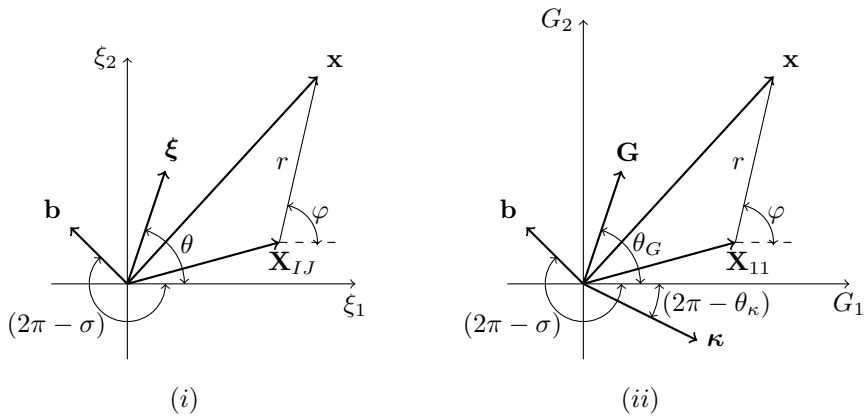

\appendix

\section{Determining the residual field $\phi_{\mathrm{res}}$} \label{sec:Residual}
The residual part of the sum, {\color{black}in \eqref{HelmCondConvSer}, is calculated with the aid of Fig. \ref{Xixandb} 
 $(ii)$. This is an extension of the analysis presented in Appendix B of \cite{schnitzer2017bloch}. Knowing $G>R' \gg 1$ in $\phi_{\mathrm{res}}$, we can utilize a Laurent series expansion with respect to $G$ at infinity. $\phi_{\mathrm{res}}$ is considered as follows
\begin{equation}
\begin{aligned}
\phi_{\mathrm{res}} = \frac{4}{i \mathscr{A}} \exp \left( i  \boldsymbol{\kappa} \cdot \textbf{r} \right) \sum_{G > R'} f(G, \theta_{G}), \label{eqn::B1}
\end{aligned}
\end{equation} 
where we denote $|\textbf{G}| = G$, $|\textbf{b}| = b$,  $|\boldsymbol{\kappa}| = \kappa$ and
\begin{equation}
\begin{split}
    f(G, \theta_{G}) = \epsilon^{2} \left( a - i \textbf{b} \cdot \textbf{K}_{\textbf{G}} \right) \exp \left[i G r \cos( \theta_{G} - \varphi ) \right]  \\
    \cdot \left[ \frac{1}{G^{2}} - \frac{2 \kappa \cos(\theta_{G} - \theta_{\kappa})}{G^{3}} + \frac{4 \kappa^{2} \cos^{2}(\theta_{G} - \theta_{\kappa}) + \Omega^{2} - \kappa^{2}}{G^{4}} + \mathcal{O}\left(\frac{1}{G^{5}}\right) \right] 
\end{split}
\end{equation}
The above sum is computed using the Euler–Maclaurin formula \cite{olver1997asymptotics} and is approximated by the following integral
\begin{equation}
\phi_{\mathrm{res}} = \frac{1}{i \pi^{2}} \exp \left( i  \boldsymbol{\kappa} \cdot \textbf{r} \right) \int_{R'}^{\infty} \int_{0}^{2 \pi} f(G, \theta_{G}) \cdot G \, d \theta_{G} \, dG + o(1). \label{eqn::integral::B1}
\end{equation}

The remainder ($o(1)$) term comes from approximating the summation as an integral and is negligible for $1 \ll R \ll \frac{1}{r}$ as $r \to 0$. Similar remainder terms are also negligible when considering $\nabla \phi_{\mathrm{res}}$ in the same asymptotic regime. Subsequent direct integration gives $\phi_{\mathrm{res}}$  as defined in Eq. \eqref{NAP::G>Rcont}.
}


To derive $\phi_{\mathrm{res}}$ we utilise Van der Pol's $Ji_{0}$ \cite{humbert1933bessel}. Interestingly $Ji_{0}$ can be used to evaluate integrals containing combinations of Bessel functions and powers, of the form \eqref{eqn::IntegralsThisForm}. Integrating by parts we find a recursive relation which can be exploited to find 
\begin{equation}
    \int_{x}^{\infty} \frac{J_{n}(\xi)}{\xi^{n+1}} d \xi = \left\lbrace \sum_{m=1}^{n} \frac{(m-1)! J_{m}(x)}{n!x^{m} 2^{(n-m+1)}} \right\rbrace  - \frac{1}{2^{n} n!} Ji_{0} (x), \quad \mbox{where $n \in \mathbb{N} : n \ne 0$.} \label{eqn::IntegralsThisForm}
\end{equation}  

\section{Matched asymptotics} \label{Appendix::matching}
The matching procedure is described within section 6.3 
 of \cite{crighton1992matched}. The outer solution to the wavefield is defined $\phi = \phi(r, \varphi ; \epsilon)$ and is a function of outer variable $r$. The inner solution to the wavefield is defined $\Phi = \Phi(R, \varphi ; \epsilon)$ and is a function of the inner variable $R$. Both $\phi$ and $\Phi$ are to be expressed as asymptotic expansions dependent upon the small parameter $\epsilon$, where
\begin{equation}
    r = \epsilon R.
\end{equation}
Denote $\phi^{(n)}(r, \varphi ; \epsilon)$ to represent $\phi$ correct up to and including $\mathcal{O}(\epsilon^{n})$. Moreover, the inner limit of $\phi^{(n)}(r, \varphi ; \epsilon)$ correct up to and including $\mathcal{O}(\epsilon^{m})$, is denoted $\phi^{(n,m)}$ {\color{black}and defined}
\begin{equation}
    \phi^{(n)}(r = \epsilon R, \varphi ; \epsilon) = \phi^{(n,m)} (R, \varphi; \epsilon) + o(\epsilon^{m}).
\end{equation}
Similarly for the outer solution, $\Phi$ correct up to $\mathcal{O}(\epsilon^{m})$ is denoted $\Phi^{(m)}(R, \varphi ; \epsilon)$, where it is appropriate to examine the outer limit of the inner solution correct up to and including $\mathcal{O}(\epsilon^{n})$ 
\begin{equation}
\Phi^{(m)}\left(R = \frac{r}{\epsilon}, \varphi ; \epsilon\right) = \Phi^{(m,n)} (r, \varphi; \epsilon) + o(\epsilon^{n}).    
\end{equation}
The matching procedure describes the {\color{black}equivalence} of $ \phi^{(n,m)}$ and $\Phi^{(m,n)}$ as follows 
\begin{equation}
    \phi^{(n,m)} \equiv \Phi^{(m,n)}
\end{equation}
The above procedure is naturally consistent with Van Dyke's matching rule \cite{van1964perturbation}, in which $\log \epsilon$ terms are regarded as order unity \cite{schnitzer2017bloch, crighton1992matched}. 

\subsection{Outer region}
The matching procedure is {\color{black}simplified considering outer and inner expansions for} one inclusion. Subsequently \eqref{HelmholtzOuterEase} is modified to give 
\begin{equation}
    (\nabla^{2} + \Omega^{2}) \phi = 4i \epsilon^{2} \left\lbrace a \delta(\textbf{x} - \textbf{X}) - \textbf{b} \cdot \nabla \left[ \delta( \textbf{x} - \textbf{X}) \right] \right\rbrace. \label{eqn::OuterForInnerHelmholtz}
\end{equation}
Motivated by \cite{schnitzer2017bloch} we consider the outer solution in two parts 
\begin{align}
\phi = {\color{black} \epsilon^{2} \Big\lbrace \chi + \psi \Big\rbrace,}
\end{align} where $\chi$ and $\psi$ denote the complimentary and particular solution of \eqref{eqn::OuterForInnerHelmholtz}. The form of the complementary solution is obtained via separation of variables,
\begin{equation}
\chi = \sum_{n} (B_{1 \, n} \cos n \varphi + B_{2 \, n} \sin n \varphi ) J_{n} (\Omega r) + (C_{1 \, n} \cos n \varphi + C_{2 \, n} \sin n \varphi ) Y_{n} (\Omega r). \label{HelmholtzCompliment}
\end{equation} $\chi$ should not be singular {\color{black}as $r \to 0$. $\psi$} is given by the Green's functions for the monopole and dipole line sources 
\begin{equation}
\psi = a H_{0} (\Omega r) + (\textbf{b} \cdot \hat{\textbf{r}}) \Omega H_{1} (\Omega r). \label{HelmholtzParticular}
\end{equation} 
{\color{black}
Applying the Neumann condition we set $\frac{\partial \phi}{\partial r} \Big|_{r = \epsilon} = 0$.   Subsequently the 
 outer solution is given by  
\begin{equation}
\phi^{(2)}(r, \varphi) = 
\epsilon^{2} \Big\lbrace \chi + a H_{0} (\Omega r) + (\textbf{b} \cdot \hat{\textbf{r}}) \Omega H_{1} (\Omega r) \Big\rbrace, \label{HelmholtzOuterAsymptotic} 
\end{equation} where, around the scatterer, $\chi$ must be of the following form
\begin{equation}
   \lim_{r \to 0} \chi(r) = - a \frac{H_{1} (\Omega \epsilon)}{J_{1} (\Omega \epsilon)} J_{0} (\Omega r) - \textbf{b} \cdot \hat{\textbf{r}} \Omega \frac{H_{0} (\Omega \epsilon) - H_{2}(\Omega \epsilon) }{J_{0}(\Omega \epsilon) - J_{2}(\Omega \epsilon)} J_{1} (\Omega r). \label{chi_Only}
\end{equation}
We only require $\chi$ approaching the inner region; thus, stating (\ref{chi_Only}) is adequate and analogous to adding some standing wave in the vicinity of the inclusion, forcing the Neumann condition at the inclusion. However, for $r$ increasing $\chi$ must tend to zero rapidly enough such that any field satisfies the Sommerfeld radiation condition, corresponding to outgoing cylindrical waves at infinity \cite{sommerfeld1949partial}.    

Expanding the above for small $\epsilon$ and $r$ gives the inner limit of the outer solution as }
\begin{equation}
\begin{split}
\phi^{(2,3)}(R, \varphi) =  \frac{4i}{\pi} \left\lbrace \frac{a}{\Omega^{2}} - \epsilon \frac{\textbf{b} \cdot \hat{\textbf{r}}}{\Omega}  \frac{\Omega}{2} \left[ R + \frac{1}{R} \right] \right\rbrace
 + \epsilon^{2} a \left[ \frac{2 i}{\pi} \left( \log R + \frac{3}{4} - \frac{R^{2}}{2} \right) \right] + \\
+ \epsilon^{3} \textbf{b} \cdot \hat{\textbf{r}}  \left[ \frac{i \Omega^{2} R}{\pi} \left( \log R  - \frac{7}{4} + \frac{R^{2}}{4} \right) \right] .  \label{InnerOuter} 
\end{split}
\end{equation} 

\subsection{Inner region}
From \eqref{InnerOuter}, the inner solution is asymptotically expanded as
\begin{equation}
\Phi(R, \varphi; \epsilon) = \Phi_{0} + \epsilon \Phi_{1} + \epsilon^{2} \Phi_{2} + \epsilon^{3} \Phi_{3} + \ldots , \label{HelmholtzInnerAsymptotic} 
\end{equation} 
and needs to satisfy the following problem
\begin{align}
\left( \nabla^{2} + \epsilon^{2} \Omega^{2} \right) \Phi = 0, \label{InnerForHelmholtz} \\ 
\frac{\partial \Phi}{\partial R} = 0, \quad \mbox{on $R = 1$}. \label{InnerForHelmholtzBC}
\end{align} 
Formally, we work within a low-frequency regime; the asymptotics developed will only hold if the $\epsilon^{2} \Omega^{2}$ term within (\ref{InnerForHelmholtz}) is of $o(1)$. The solution in the inner limit is found by considering (\ref{InnerForHelmholtz}) to the orders indicated in (\ref{HelmholtzInnerAsymptotic}). Solving each $\Phi_{i}$ in polar coordinates is not a difficult task, the dependence on $\varphi$ is known{\color{black}, since each term corresponds to a monopole or dipole like source}. Matching to (\ref{InnerOuter}) at various orders determines all unknown constants in the inner solution. Subsequently
\begin{equation}
\begin{split}
\Phi^{(3)} = \frac{4 i}{\pi} \frac{a}{\Omega^{2}} + \epsilon \frac{2}{i \pi}  \Big( \frac{1}{R} + R \Big) \Big[ \textbf{b} \cdot \hat{\textbf{r}} \Big]  + \epsilon^{2}  \frac{2a i}{\pi} \left\lbrace \log R  + \frac{3}{4} - \frac{1}{2} R^{2}  \right\rbrace + \\
 + \epsilon^{3}  \frac{i \textbf{b} \cdot \hat{\textbf{r}}}{\pi} \Omega^{2} \left\lbrace R \log R - \frac{7}{4} R + \frac{1}{4} R^{3}  \right\rbrace
\end{split} . \label{HelmholtzInnerLimINNER}
\end{equation} 
Therefore $\Phi^{(3,2)}$ may be determined, which yields the inner solution in the outer limit, that is
\begin{equation}
\begin{split}
\Phi^{(3,2)} = \frac{4 i}{\pi} \frac{a}{\Omega^{2}} \left[ 1 - \frac{\Omega^{2} r^{2}}{4} \right] + \frac{4}{i \pi} \frac{ \textbf{b} \cdot \hat{\textbf{r}} }{\Omega}  \left( \frac{r \Omega}{2} - \frac{\Omega^{3} r^{3}}{16} \right)  + \\
+ \epsilon^{2} \left\lbrace \frac{2ia}{\pi} \left[ \log \frac{r}{\epsilon} + \frac{3}{4} \right] + \frac{\textbf{b} \cdot \hat{\textbf{r}}}{i \pi} \left(\frac{2}{r} - \Omega^{2} r \left[ \log \frac{r}{\epsilon} - \frac{7}{4} \right] \right) \right\rbrace.
\end{split} \label{HelmholtzInnerLim}
\end{equation}

\section{The explicit components within equation (\ref{HelmholtzHardScheme_1})} \label{AllOfTheComponentsEigen}
The components of  $\mathcal{A}$, $\mathcal{B}$ and $\boldsymbol{\Phi}$  are 
\begin{equation}
\mathcal{A}_{[\mathrm{(N+3P)\times(N+3P)}]} = \begin{pmatrix} \circled{1}_{[N \times N]} & \circled{2}_{[N \times P]} & \circled{3}_{[N \times P]} & \circled{4}_{[N \times P]} \\
\circled{5}_{[P \times N]} & \circled{6}_{[P \times P]} & \circled{7}_{[P \times P]} & \circled{8}_{[P \times P]} \\
\circled{9}_{[P \times N]} & \circled{10}_{[P \times P]} & \circled{11}_{[P \times P]} & \circled{12}_{[P \times P]} \\
\circled{13}_{[P \times N]} & \circled{14}_{[P \times P]} & \circled{15}_{[P \times P]} & \circled{16}_{[P \times P]} 
\end{pmatrix}, \label{HelmA}
\end{equation}
\begin{equation}
\mathcal{B}_{[\mathrm{(N+3P)\times(N+3P)}]} =  \begin{pmatrix} \circled{1'}_{[N \times N]} & \circled{2'}_{[N \times P]} & \circled{3'}_{[N \times P]} & \circled{4'}_{[N \times P]} \\
\circled{5'}_{[P \times N]} & \circled{6'}_{[P \times P]} & \circled{7'}_{[P \times P]} & \circled{8'}_{[P \times P]} \\
\circled{9'}_{[P \times N]} & \circled{10'}_{[P \times P]} & \circled{11'}_{[P \times P]} & \circled{12'}_{[P \times P]} \\
\circled{13'}_{[P \times N]} & \circled{14'}_{[P \times P]} & \circled{15'}_{[P \times P]} & \circled{16'}_{[P \times P]} 
\end{pmatrix}. \label{HelmB}
\end{equation}
Here the subscript $[P \times Q]$ denotes the dimensions of a matrix with $P$ rows and $Q$ columns. We have factorised the eigenvalue problem with the eigenvector $\boldsymbol{\Phi}_{\mathrm{[(N+3P) \times 1}]}$, containing all of the unknowns, as follows 
\begin{equation}
\boldsymbol{\Phi}^{\dagger} = \begin{pmatrix} \Phi_{\textbf{G}_{1}} & \ldots & \Phi_{\textbf{G}_{N} }   & a_{1} & \ldots & a_{P} & b_{1 \, 1} & \ldots & b_{1 \, P}  & b_{2 \, 1} & \ldots & b_{2 \, P} \end{pmatrix}. \label{eqn::PhiDagger} 
\end{equation}
The block matrices forming \eqref{HelmA} and \eqref{HelmB} are
\begin{equation}
\circled{1}_{rc} = \textbf{K}_{\textbf{G}_{r}} \cdot \textbf{K}_{\textbf{G}_{r}}  \widetilde{\delta}_{rc} \quad r,c = 1, \ldots, N.
\end{equation}
\begin{equation}
\circled{2}_{rc} = \frac{4i \epsilon_{c}^{2}}{\mathscr{A}} \exp(-i \textbf{K}_{\textbf{G}_{r}} \cdot \textbf{X}_{1c} ) \quad r = 1, \ldots, N \quad c = 1, \ldots, P.
\end{equation}
\begin{equation}
\circled{3}_{rc} = \frac{4\epsilon_{c}^{2}}{\mathscr{A}} K_{1 \, \textbf{G}_{r}} \exp(-i \textbf{K}_{\textbf{G}_{r}} \cdot \textbf{X}_{1c} ) \quad r = 1, \ldots, N \quad c = 1, \ldots, P.
\end{equation}
\begin{equation}
\circled{4}_{rc} = \frac{4\epsilon_{c}^{2}}{\mathscr{A}} K_{2 \, \textbf{G}_{r}} \exp(-i \textbf{K}_{\textbf{G}_{r}} \cdot \textbf{X}_{1c} ) \quad r = 1, \ldots, N \quad c = 1, \ldots, P.
\end{equation}
\begin{equation}
\circled{6}_{rc} = \frac{4}{i \pi} \widetilde{\delta}_{rc} \quad r,c = 1, \ldots, P.
\end{equation}
\begin{equation}
\circled{9}_{rc} = i K_{1 \, \textbf{G}_{c}} \exp(i \textbf{K}_{\textbf{G}_{c}} \cdot \textbf{X}_{1r} ) \quad r = 1, \ldots, P \quad c = 1, \ldots, N.
\end{equation}
\begin{equation}
\circled{10}_{rc} = - \frac{\epsilon_{r}^{2} \kappa_{1}}{\pi} \widetilde{\delta}_{rc} \quad r,c = 1, \ldots, P.
\end{equation}
\begin{equation}
\circled{11}_{rc} = \frac{i}{\pi} \left[2 + \frac{\epsilon_{r}^{2} R'^{2}}{2} + \frac{\epsilon_{r}^{2}}{4}(\kappa_{1}^{2} - \kappa_{2}^{2}) \right] \widetilde{\delta}_{rc} \quad r,c = 1, \ldots, P.
\end{equation}
\begin{equation}
\circled{12}_{rc} = \frac{\epsilon_{r}^{2} i}{2 \pi} \kappa_{1} \kappa_{2} \widetilde{\delta}_{rc} \quad r,c = 1, \ldots, P.
\end{equation}
\begin{equation}
\circled{13}_{rc} = i K_{2 \, \textbf{G}_{c}} \exp(i \textbf{K}_{\textbf{G}_{c}} \cdot \textbf{X}_{1r} ) \quad r = 1, \ldots, P \quad c = 1, \ldots, N.
\end{equation}
\begin{equation}
\circled{14}_{rc}=  - \frac{\epsilon_{r}^{2} \kappa_{2}}{\pi}  \widetilde{\delta}_{rc} \quad r,c = 1, \ldots, P.
\end{equation}
\begin{equation}
\circled{15}_{rc} = \frac{\epsilon_{r}^{2}i}{2 \pi} \kappa_{1} \kappa_{2}  \widetilde{\delta}_{rc} \quad r,c = 1, \ldots, P.
\end{equation}
\begin{equation}
\circled{16}_{rc} = \frac{i}{\pi} \left[2 + \frac{\epsilon_{r}^{2} R'^{2}}{2} - \frac{\epsilon_{r}^{2}}{4}(\kappa_{1}^{2} - \kappa_{2}^{2}) \right]  \widetilde{\delta}_{rc} \quad r,c = 1, \ldots, P.
\end{equation}
\begin{equation}
\circled{1'}_{rc} = \widetilde{\delta}_{rc} \quad r,c = 1, \ldots, N.
\end{equation}
\begin{equation}
\circled{5'}_{rc} = - \exp(i \textbf{K}_{\textbf{G}_{c}} \cdot \textbf{X}_{1r} ) \quad r = 1, \ldots, P \quad c = 1, \ldots, N.
\end{equation}
\begin{equation}
\circled{6'}_{rc} = \epsilon_{r}^{2} \frac{2i}{\pi} \left[ \log \frac{2}{\epsilon_{r} R'} + \frac{3}{4} - \gamma_{E} \right] \widetilde{\delta}_{rc} \quad r,c = 1, \ldots, P.
\end{equation}
\begin{equation}
\circled{7'}_{rc} = - \frac{\epsilon_{r}^{2}}{\pi} \kappa_{1} \widetilde{\delta}_{rc} \quad r,c = 1, \ldots, P.
\end{equation}
\begin{equation}
\circled{8'}_{rc} = - \frac{\epsilon_{r}^{2}}{\pi} \kappa_{2} \widetilde{\delta}_{rc} \quad r,c = 1, \ldots, P.
\end{equation}
\begin{equation}
\circled{11'}_{rc} = \circled{16'}_{rc}  = \epsilon_{r}^{2} \frac{i}{\pi} \left[ \log \frac{2}{\epsilon_{r} R'} - \frac{5}{4} - \gamma_{E} \right] \widetilde{\delta}_{rc} \quad r,c = 1, \ldots, P. \label{Helm16'}
\end{equation}
In (\ref{eqn::PhiDagger}) the superscript $\dagger$ denotes the transpose operation. The components of any block matrices, forming $\mathcal{A}$ or $\mathcal{B}$, which are ``missing" from this list are all zero. Here $r$ and $c$ denotes the row and column number respectively, and $\widetilde{\delta}_{rc}$ denotes the Kronecker delta function. $\textbf{G}_{i}$ denotes the $i$th arbitrary reciprocal position vector for the $i = 1, \ldots, N$ Bloch modes considered within the radius $R'$ of truncation.

\section{The explicit components within the extended Foldy system} \label{AllOfTheComponentsFoldy}
The matrices forming $\circledast$ from equation (\ref{FoldyMatrix}) are
\begin{equation} 
\circled{1}_{\tilde{r}c} = \left( \frac{4i}{\pi \Omega^{2}} - \epsilon_{\tilde{r}}^{2} \left[ 1 - \frac{2i}{\pi} \left( \log \frac{2}{\epsilon_{\tilde{r}} \Omega} + \frac{3}{4} - \gamma_{E}  \right) \right]  \right) \widetilde{\delta}_{\tilde{r}c}  - \epsilon_{c}^{2} H_{0}(\Omega r_{\tilde{r} c}) \left\lbrace 1 - \widetilde{\delta}_{\tilde{r}c} \right\rbrace
\end{equation}
\begin{equation}
\circled{2}_{\tilde{r}c} = - \epsilon_{c}^{2} \Omega \cos \varphi_{\tilde{r}c} H_{1}(\Omega r_{\tilde{r} c}) \left\lbrace 1 - \widetilde{\delta}_{\tilde{r}c} \right\rbrace
\end{equation}
\begin{equation}
\circled{3}_{\tilde{r}c} = - \epsilon_{c}^{2} \Omega \sin \varphi_{\tilde{r}c} H_{1}(\Omega r_{\tilde{r} c}) \left\lbrace 1 - \widetilde{\delta}_{\tilde{r}c} \right\rbrace
\end{equation}
\begin{equation}
\circled{4}_{\tilde{r}c} = - \circled{2}_{\tilde{r}c}
\end{equation}
\begin{equation}
\begin{split}
& \circled{5}_{\tilde{r}c} = \left[ \frac{2}{i \pi} + \frac{\epsilon_{\tilde{r}}^{2} i \Omega^{2}}{\pi} \left( \log \frac{2}{\epsilon_{\tilde{r}} \Omega} - \frac{5}{4} - \gamma_{E} \right) - \frac{\epsilon_{\tilde{r}}^{2} \Omega^{2}}{2} \right] \widetilde{\delta}_{\tilde{r}c} -  \\
& - \epsilon_{c}^{2}  \left\lbrace\frac{\Omega^{2} \cos^{2} \varphi_{\tilde{r}c}}{2} [  H_{0}(\Omega r_{\tilde{r} c}) - H_{2}(\Omega r_{\tilde{r} c})] + \frac{\Omega}{ r_{\tilde{r} c}} \sin^{2} \varphi_{\tilde{r}c} H_{1}(\Omega r_{\tilde{r} c}) \right\rbrace \left\lbrace 1 - \widetilde{\delta}_{\tilde{r}c} \right\rbrace
\end{split}
\end{equation}
\begin{equation}
\begin{split}
\circled{6}_{\tilde{r}c} = - \epsilon_{c}^{2}  \cos \varphi_{\tilde{r}c} \sin \varphi_{\tilde{r}c} \left\lbrace\frac{\Omega^{2} }{2} [  H_{0}(\Omega r_{\tilde{r} c}) - H_{2}(\Omega r_{\tilde{r} c})] - \frac{\Omega}{ r_{\tilde{r} c}} H_{1}(\Omega r_{\tilde{r} c}) \right\rbrace \cdot \\
\cdot \left\lbrace 1 - \widetilde{\delta}_{\tilde{r}c} \right\rbrace
\end{split}
\end{equation}
\begin{equation}
\circled{7}_{\tilde{r}c}= - \circled{3}_{\tilde{r}c}
\end{equation}
\begin{equation}
\circled{8}_{\tilde{r}c} = \circled{6}_{\tilde{r}c}
\end{equation}
\begin{equation}
\begin{split}
& \circled{9}_{\tilde{r}c} = \left[ \frac{2}{i \pi} + \frac{\epsilon_{\tilde{r}}^{2} i \Omega^{2}}{\pi} \left( \log \frac{2}{\epsilon_{\tilde{r}} \Omega} - \frac{5}{4} - \gamma_{E} \right) - \frac{\epsilon_{\tilde{r}}^{2} \Omega^{2}}{2}  \right] \widetilde{\delta}_{\tilde{r}c} -  \\
& - \epsilon_{c}^{2}  \left\lbrace\frac{\Omega^{2} \sin^{2} \varphi_{\tilde{r}c}}{2} [  H_{0}(\Omega r_{\tilde{r} c}) - H_{2}(\Omega r_{\tilde{r} c})] + \frac{\Omega}{ r_{\tilde{r} c}} \cos^{2} \varphi_{\tilde{r}c} H_{1}(\Omega r_{\tilde{r} c}) \right\rbrace \left\lbrace 1 - \widetilde{\delta}_{\tilde{r}c} \right\rbrace
\end{split}
\end{equation}
In the above $\tilde{r},c = 1, \ldots m$ denotes the row and column of the block matrices assembling the scheme (\ref{FoldyMatrix}). The incident field is inserted into the scheme as follows.

\textcolor{black}{For a monopole incident source:}
\begin{equation}
\boldsymbol{\phi}_{\mathrm{inc \, s} \, \, \tilde{r}} = \frac{\epsilon_{\mathrm{min}}^{2} a_{\mathrm{inc}}}{4i} H_{0} (\Omega |\textbf{X}_{\tilde{r}} - \textbf{X}_{\mathrm{inc}}|), \quad \tilde{r} = 1, \ldots, m. \label{MonoFoldyIncident1}
\end{equation}
\begin{equation}
\nabla \boldsymbol{\phi}_{\mathrm{inc1 \, s} \, \, \tilde{r}} = \epsilon_{\mathrm{min}}^{2} \frac{i a_{\mathrm{inc}}}{4} \cos \varphi_{\tilde{r} \mathrm{inc}} \Omega H_{1} (\Omega |\textbf{X}_{\tilde{r}} - \textbf{X}_{\mathrm{inc}}|), \quad \tilde{r} = 1, \ldots, m.
\end{equation}
\begin{equation}
\nabla \boldsymbol{\phi}_{\mathrm{inc2 \, s} \, \, \tilde{r}} = \epsilon_{\mathrm{min}}^{2} \frac{i a_{\mathrm{inc}}}{4} \sin \varphi_{\tilde{r} \mathrm{inc}} \Omega H_{1} (\Omega |\textbf{X}_{\tilde{r}} - \textbf{X}_{\mathrm{inc}}|), \quad r = 1, \ldots, m. \label{MonoFoldyIncident3}
\end{equation}

\textcolor{black}{For a dipole incident source:}
\begin{equation}
\begin{split}
\boldsymbol{\phi}_{\mathrm{inc \, s} \, \,\tilde{r}} = \frac{\epsilon_{\mathrm{min}}^{2} i }{4} \Omega \left[b_{1 \mathrm{inc}} \cos \varphi_{\tilde{r} \mathrm{inc}} + b_{2 \mathrm{inc}} \sin \varphi_{\tilde{r} \mathrm{inc}}\right] H_{1} (\Omega |\textbf{X}_{\tilde{r}} - \textbf{X}_{\mathrm{inc}}|) & \\ 
 \tilde{r} = 1, & \ldots, m. 
\end{split}
\end{equation}
Similarly to (\ref{MonoFoldyIncident1})-(\ref{MonoFoldyIncident3}) the required gradient terms $\nabla \boldsymbol{\phi}_{\mathrm{inc1 \, s}}$ and $\nabla \boldsymbol{\phi}_{\mathrm{inc2 \, s}}$ can be determined for the dipole source term.

\bibliography{mybibfile,CompleteReferences_4}

\end{document}